\documentclass[12pt]{iopart}
\usepackage{amstext}
\usepackage{iopams}
\usepackage{setstack}
\usepackage{graphicx}
\usepackage{bm}
\usepackage{psfrag}
\usepackage{color}

\begin{document}

\title{Stochastic population dynamics in spatially extended predator-prey systems}

\author{Ulrich Dobramysl$^\dag$, Mauro Mobilia$^\ddag$, Michel Pleimling$^\P$$^*$,
  and Uwe C. T\"auber$^\P$}
\address{$^\dag$ Wellcome Trust / Cancer Research UK Gurdon Institute,
  University of Cambridge, Cambridge CB2 1QN, U.K.}
\address{$^\ddag$ Department of Applied Mathematics, School of Mathematics,
  University of Leeds, Leeds LS2 9JT, U.K.}
\address{$^\P$ Department of Physics (MC 0435) 
  and Center for Soft Matter and Biological Physics, Robeson Hall,
  850 West Campus Drive, Virginia Tech, Blacksburg, VA 24061, USA}
\address{$^*$ Academy of Integrated Science (MC 0405), 300 Turner Street NW,
  Virginia Tech, Blacksburg, VA 24061, USA}
\ead{u.dobramysl@gurdon.cam.uc.uk, m.mobilia@leeds.ac.uk, pleim@vt.edu,
	tauber@vt.edu}

% Topical Review, to appear in J. Phys. A: Math. Theor. (2017); arXiv: 1708.07055

\begin{abstract}
Spatially extended population dynamics models that incorporate demographic 
noise serve as case studies for the crucial role of fluctuations and 
correlations in biological systems. Numerical and analytic tools from 
non-equilibrium statistical physics capture the stochastic kinetics of these 
complex interacting many-particle systems beyond rate equation approximations. 
Including spatial structure and stochastic noise in models for predator-prey 
competition invalidates the neutral Lotka--Volterra population cycles. 
Stochastic models yield long-lived erratic oscillations stemming from a resonant 
amplification mechanism. Spatially extended predator-prey systems display 
noise-stabilized activity fronts that generate persistent correlations. 
Fluctuation-induced renormalizations of the oscillation parameters can be 
analyzed perturbatively via a Doi--Peliti field theory mapping of the master 
equation; related tools allow detailed characterization of extinction pathways. 
The critical steady-state and non-equilibrium relaxation dynamics at the predator 
extinction threshold are governed by the directed percolation universality class.
Spatial predation rate variability results in more localized clusters, enhancing 
both competing species' population densities. Affixing variable interaction rates
to individual particles and allowing for trait inheritance subject to mutations 
induces fast evolutionary dynamics for the rate distributions. Stochastic spatial 
variants of three-species competition with `rock-paper-scissors' interactions 
metaphorically describe cyclic dominance. These models illustrate intimate 
connections between population dynamics and evolutionary game theory, underscore 
the role of fluctuations to drive populations toward extinction, and demonstrate 
how space can support species diversity. Two-dimensional cyclic three-species 
May--Leonard models are characterized by the emergence of spiraling patterns 
whose properties are elucidated by a mapping onto a complex Ginzburg--Landau 
equation. Multiple-species extensions to general `food networks' can be 
classified on the mean-field level, providing both fundamental understanding of 
ensuing cooperativity and profound insight into the rich spatio-temporal 
features and coarsening kinetics in the corresponding spatially extended systems. 
Novel space-time patterns emerge as a result of the formation of competing 
alliances; e.g., coarsening domains that each incorporate rock-paper-scissors 
competition games.
\end{abstract}

% 87.23.Cc - Population dynamics and ecological pattern formation
% 02.50.Ey - Stochastic processes
% 05.40.-a - Fluctuation phenomena, random processes, noise, and motion
% 87.18.Tt - Noise in biological systems

\submitto{\JPA -- \today}

\section{Introduction and historical overview}
\label{sec:intro}

\subsection{Introduction: population dynamics}

For the purpose of this topical review, we shall view population dynamics as the 
study of (classical) interacting particle systems typically involving a number of 
different species, and their time 
evolution~\cite{May73,Maynard74,Hofbauer98,Murray02,Neal04}. Population dynamics is 
most deeply rooted in ecology, where biologists and mathematicians have been 
investigating the dynamics of interacting animal, plant, or microbial species for
centuries~\cite{Kingsland1985}, via observations, theoretical considerations, and 
more recently by means of experiments in engineered, controlled environments. 
However, its principles as well as basic mathematical and computational tools have 
been successfully applied to an extremely diverse range of fields, such as the 
study of (bio-)chemical reactions~\cite{Schuster08}, genetics~\cite{Hanski04}, 
laser physics~\cite{Kim05}, economics~\cite{Malthus1798}, epidemiology, and the 
analysis of cancerous growths~\cite{Vineis06}, to list but a few. It has hence 
become a foundational subfield of non-equilibrium statistical physics. In recent 
years, population dynamics has become a crucial tool to investigate the fundamental 
puzzle of the emergence and stabilization of biodiversity, and to seek means to 
preserve the latter in endangered and besieged ecosystems.

In this review, we chiefly focus our attention on the population dynamics of 
spatially extended systems. While non-spatial and well-mixed systems already 
exhibit intriguing properties, the explicit inclusion of spatial degrees of freedom 
that allow the propagation and mutual invasion of species may lead to the 
appearance of fascinating spatio-temporal patterns that include activity fronts, 
traveling waves, and spiral structures characteristic of excitable 
media~\cite{Cross93,Cross09}. While the deterministic nature of mean-field 
equations, with the possible extension to spatially extended systems via the 
inclusion of, {\em e.g.}, diffusive spreading, already yields important insights 
into the dynamics of competing populations, the explicit incorporation of 
stochasticity can fundamentally change and renormalize the behavior of a system of
interacting species. We will therefore devote a substantial part of this brief 
overview to the discussion of random fluctuations in addition to spatio-temporal 
correlations induced by the underlying stochastic kinetics, and their consequences
to species stability, model robustness, and dynamical pattern formation in coupled
non-linear population models.

In particular, the description of species interactions via stochastic chemical
reaction processes intrinsically includes internal reaction noise. In the case
of the basic Lotka--Volterra (LV) two-species predator-prey system, stochastic 
models exhibit long-lived but ultimately decaying oscillations, already 
contradicting the classical picture of neutral population cycles. These features 
transcend the mean-field type mass-action treatment, and their theoretical analysis
allows deep insights into the role of fluctuations on species survival as well as 
the formation of intriguing spatio-temporal patterns in spatially extended systems. 
Such stochastic interacting and reacting particle models can be elegantly studied 
via individual-based Monte Carlo simulations, {\em i.e.}, stochastic cellular 
automata, uncovering intriguing spatio-temporal features such as predator-prey 
activity fronts that are induced and stabilized by the intrinsic reaction noise. 
Going beyond simulations, the Doi--Peliti framework enables a mapping of the 
(chemical) master equation to a Liouville operator formulation that encodes the 
reaction processes by means of (bosonic) creation and annihilation operators. 
Employing well-established tools from quantum many-particle physics and quantum or 
statistical field theory in turn allows qualitative insights as well as 
quantitative treatments beyond straightforward linear approximations, {\em e.g.},
addressing the renormalization through non-linear stochastic fluctuations of
observables such as the population oscillation frequencies, their relaxations, and 
characteristics pattern wavelengths.

In order to set the scene and to put this review into broader context, we start
with a brief overview of the historical background and important experimental
work over the last two centuries. The subsequent section~\ref{sec:sllvm} focuses on
the classic LV model, the role of intrinsic `reaction' noise and its treatment via
stochastic chemical processes, its spatial extension via placement on a regular 
lattice, the emergence of spatio-temporal patterns, spatial heterogeneity, critical
properties near the predator extinction threshold, and the inclusion of 
evolutionary dynamics. This is followed by section~\ref{sec:cyclc} on cyclic 
three-species systems, where we discuss the paradigmatic rock-paper-scissors  
and May--Leonard models. In section~\ref{sec:mulsp}, we extend our attention 
to general systems with more than three species and extended food networks, 
discussing the formation of both spatio-temporal patterns and emerging species 
alliances. Finally, we end this overview in section~\ref{sec:concl} with our 
conclusions and outlook on the future evolution of this field.

\subsection{Historical overview}

The study of population dynamics looks back over two centuries of history in the
mathematical and ecological sciences. Malthus' growth law~\cite{Malthus1798} is
widely regarded as the `first law of population ecology'. In this work, he
debated that the exponential human population growth is incompatible with linear
growth of food resources and argued for population controls to be put in place.
However, unchecked infinite exponential growth is obviously unphysical, hence
the exponential law was later amended by Verhulst to construct the logistic growth
model~\cite{Verhulst1838}. Almost ninety years later, Pearl applied the logistic 
equation, which had been derived independently by Lotka~\cite{Lotka1925}, to model 
population growth in the US~\cite{Pearl1925}. In 1926, Volterra introduced a simple 
model describing the effects of two interacting species in close 
proximity~\cite{Volterra26}, and used this model to explain oscillations of fish 
populations (and the resulting catch volumes) in the Adriatic sea as observed by
D'Ancona~\cite{Kingsland1985}. Lotka independently introduced the same set of rate
equations in his work on chemical oscillations and physical 
biology~\cite{Lotka20,Murray02}.

Volterra argued that the growth rate of the prey population density $b$, given as 
$b^{-1} db / dt$, should be a decreasing function of the predator density $a$ and 
greater than zero when the predator density is zero. Conversely, the predator 
growth rate $a^{-1} da / dt$ should increase with the prey count, but be negative 
when $b = 0$. The simplest coupled set of non-linear differential equations 
following these arguments represents the LV competition model that describes the 
population densities of predators $a$ and prey $b$:
\begin{equation}
\label{eq:classiclv}
  \frac{d a(t)}{dt} = a(t) \left[ \lambda b(t) - \mu \right] , \quad
  \frac{d b(t)}{dt} = b(t) \left[ \sigma - \lambda a(t) \right],
\end{equation}
where $t$ denotes the (continuous) time.
The parameters $\lambda$, $\mu$, and $\sigma$ describe the phenomenolgical
predation, predator death, and prey reproduction rates, respectively. This set of
coupled ordinary differential equations gives rise to characteristic, undamped, 
non-linear oscillations, see figure~\ref{fig:lv_oscillations}. Very similar 
techniques were also used to model warfare~\cite{Kingsland1985}.

However, the LV equations~\eref{eq:classiclv} are rather simplistic, with the
most obvious flaw being the unchecked growth of the prey in the absence of
predators. In the 1930s, Gause proposed a generalized mathematical model in which 
the rate parameters effectively become response functions of the respective 
species, allowing more realistic control of populations compared to the original 
LV model~\cite{Gause1934,Royama71,Sigmund07}. In his 1936 note, Kolmogorov
published an even more general set of equations governing a system of two interacting
predator and prey species~\cite{Kolmogorov36}, 
\begin{equation}
\label{eq:kolmogorov-eqns}
  \frac{d a(t)}{dt} = \alpha(a,b) \, a(t) \, , \quad
  \frac{d b(t)}{dt} = \beta(a,b) \, b(t) \, .
\end{equation}
He argued that the response functions $\alpha$ and $\beta$ should decrease with
increasing predator density $a$, i.e,. $\partial_a \alpha(a,b) < 0$ and
$\partial_a \beta(a,b) < 0$, which is a broadly accepted fact in theoretical
ecology~\cite{Sigmund07}, and makes Kolmogorov's model much more realistic than the
LV model. Nevertheless, the LV model has remained quite popular owing to its
simplicity and the low number of free parameters. We will introduce the LV model
and discuss its spatially extended and stochastic counterpart in detail in
section~\ref{sec:sllvm}.

The models we discussed so far are concerned with the mean population densities 
only, neglecting the influence of random fluctuations on the populations. The study
of the stochastic aspects of population dynamics (or indeed epidem\-iology) likely 
began with McKendrick in 1926, where, in his seminal paper, he derived the 
probabilistic master equation for the Susceptible-Infected-Recovered (SIR)
model~\cite{McKendrick1926,Murray02,Bacaer2011}. The SIR system represents the
most basic model of epidemiology to describe the spread of an infection through
a susceptible population, and still serves as an extremely popular paradigmatic
model in the current literature.

In 1937, MacLulich analyzed time series data on the populations of lynx and
snowshoe hare in Canada~\cite{MacLulich1937}, first published by 
Hewitt~\cite{Hewitt1921}. These data were inferred from the volume of fur traded 
via the Hudson Bay Company going back to the early nineteenth century. Strikingly, 
both lynx and hare populations show multi-year recurrent spikes, and thus exhibit 
the signature non-linear oscillatory cycles of the LV system. Therefore this 
time series data is considered as one of the classic examples for LV oscillations 
and an early validation of this rather simplistic model.

In the 1950s, spatial aspects of predator-prey population dynamics were first
considered. Huffaker created an experimental spatially extended lattice for two
competing species of mites~\cite{Huffaker1958}. He arranged oranges in a
two-dimensional array such that mites can traverse between them and use the
oranges as a habitat. In this system, the mite species \textit{Eotetranychus
sexmaculatus} serves as the prey, while the mite \textit{Typhlodromus occidentalis}
plays the role of predator. Huffaker counted the number of mites of either species 
over time on each orange in the lattice arrangement. His data indeed also displayed
the characteristic oscillations associated with the LV model, in addition to the
striking spatio-temporal patterns formed by the mite population distribution.
Huffaker's mite universes also contained heterogeneity in the form of rubber balls
replacing oranges, thereby showing that spatial heterogeneity and species motility
stabilize the coexistence of the two competing species. The oranges and rubber 
balls can be considered to be interacting patches in a metapopulation model for 
spatially extended predator-prey dynamics. 

Based on an observation by Wright that in spread-out areas larger than the average
migration distance, populations occur in essentially isolated patches, with 
individuals migrating between neighboring patches, Kimura introduced the stepping 
stone model for population genetics~\cite{Kimura1953,Kimura1964}. This model, which
is equivalent to the well-known voter model, explained the local genetic
characteristics based on the size of population patches~\cite{Durrett99}. Durrett 
and co-workers later extended this to understand the effects of local 
mutations~\cite{Durrett99}.

In his seminal paper on evolutionary genetics, Moran proposed a process for
neutral drift and genetic fixation in populations of fixed
size~\cite{Moran58,Nowak06}. In this model, the population of alleles A or B can change
via the death of a randomly chosen individual and its replacement by the
offspring of a second randomly chosen individual. The Moran model can then be
used to predict the fixation times and probabilities of alleles A and B,
depending on population size and initial populations. It can be extended to
account for evolutionary selection when one of the two alleles is assigned a
higher reproduction probability consistent with a fitness advantage.

In the late 1960s and the 1970s, spatial aspects of population dynamics gained
traction in the theoretical ecology community. For example, Levins derived
principles for the control of invading insect pests using simple spatial
models~\cite{Levins1969}. Hastings considered a two-species predator-prey system in
which prey can migrate to empty areas and predators can subsequently invade prey 
patches~\cite{Hastings77}. This all-to-all connected system exhibits stable
predator-prey coexistence phases. Meanwhile, May showed that randomly assembled
species interaction networks become less stable with growing size and connectivity 
according to random matrix theory and linear stability analysis~\cite{May73}, an 
intriguing result as it seemingly contradicts the common belief that higher 
diversity generically leads to more stable ecosystems.

In their seminal 1973 essay, Maynard Smith and Price laid the groundwork for the
application of game theory concepts to population dynamics and to the study of
evolutionary theory~\cite{Maynard73}. Crucially, Maynard Smith also popularized the
`rock-paper-scissors' (RPS) game in which three species interact cyclically via LV 
terms~\cite{Maynard82}. May and Leonard published their important modified cyclic 
interaction model in 1975~\cite{May75}. The crucial difference between these two 
model variants is that in the RPS model consumed prey change identity and are 
converted into their respective predator species, with the total population number 
remaining conserved, while in the May--Leonard model (MLM) the competition is more
indirect, and unconstrained by net population conservation. Perhaps the most
prominent example of cyclic interaction in nature are the mating strategies of the 
Californian side-blotched lizard, observed to obey the rules of the RPS game by 
Sinervo and Lively~\cite{PatternsRPS}. We discuss cyclic games in detail in
section~\ref{sec:cyclc}.

In 1957, Kerner developed a statistical mechanics framework for interacting species
based on the LV model. He made the observation that the LV equations admit a 
Liouville theorem and exhibit a conserved quantity~\cite{Kerner57}. In the early 
1990s, more researchers began applying the methods of statistical physics to study 
population dynamics in spatially extended systems. Dunbar showed that in one 
dimension, the LV equations with diffusive spreading support travelling wave 
solutions~\cite{Dunbar83} akin to Fisher--Kolmogorov waves. Matsuda {\it et al.} 
developed a lattice model based on the LV interaction rules and analyzed its 
evolutionary stability and the emergence of altruism using a pair approximation 
method~\cite{Matsuda92}. Satulovsky and Tom\'e developed a stochastic model for 
predator-prey dynamics on a square lattice and investigated its phase diagram using
dynamical mean-field theory and computer simulations~\cite{Tome94}. They observed 
absorbing states as well as a coexistence regime with local oscillatory behavior. 
Boccara, Roblin, and Roger studied a similar stochastic lattice system and also 
found that overall species densities tended to stable values in the coexistence 
phase, while local oscillations still persisted~\cite{Boccara94}. Similar numerical
and analytical studies were also performed for cyclically competing
species~\cite{Frachebourg96,Frachebourg96b}.

\section{Stochastic lattice Lotka--Volterra predator-prey models}
\label{sec:sllvm}

\subsection{Classical mean-field rate equations}

We have already introduced the classical mean-field Lotka--Volterra rate 
equations~\eref{eq:classiclv} in section~\ref{sec:intro}, and discussed their most
important shortcomings. One may view these coupled ordinary non-linear differential
equations as the well-mixed, deterministic limit for time evolution of the mean 
populations of a system of two interacting predator and prey species. The 
microscopic and more general Lotka--Volterra reaction rules from which 
equations~\eref{eq:classiclv} derive are given by:
\begin{equation}
\label{eq:lvrules}
  B \stackrel{\sigma}{\rightarrow} B + B \, , \quad
  A \stackrel{\mu}{\rightarrow}\emptyset \, , \quad
  A + B \stackrel{\lambda'}{\rightarrow} A + A \, .
\end{equation}
The prey $B$ reproduce with rate $\sigma > 0$; the predators $A$ spontaneously die
with rate $\mu > 0$; and upon encountering each other in their immediate vicinity, 
both species may interact with (microscopic) predation rate $\lambda' > 0$,
whereupon the participating prey is consumed while the predator generates one 
offspring. On a hypercubic $d$-dimensional lattice with lattice constant $a_0$, the
mean-field continuum reaction rate in eqs.~\eref{eq:classiclv} is connected with 
its microscopic counterpart via $\lambda = a_0^d \lambda'$. The above three 
reaction rules are to be interpreted as continuous-time stochastic processes, and 
the discreteness in individual numbers is important near the absorbing states, 
where either the predator population $A$ goes extinct, leaving ever multiplying 
prey $B$, or even both species disappear. Neither of the processes \eref{eq:lvrules}
allows the system to leave these absorbing states, affirming the irreversibility 
and hence non-equilibrium character of this stochastic kinetics. We remark that 
host-pathogen systems are modeled essentially by the same set of non-linear 
reactions \cite{Rauch03,Aguiar03,Aguiar04}.

If implemented on a lattice, one may in addition allow for nearest-neighbor (or 
more long-range) particle hopping processes with rate $D'$ (with associated 
continuum diffusivity $D = a_0^d D'$); alternatively, random particle exchange has 
been implemented. If one imposes the restriction that each lattice site can at most
be occupied by a single individual, prey birth entails that the offspring particle
be placed on an adjacent position; likewise, the predation reaction must then 
involve two neighboring lattice sites. These processes then automatically generate
diffusive population spreading. As we shall see below, spatial as well as temporal 
correlations turn out to be important features in the LV system that often 
crucially influence the ensuing population dynamics. Nevertheless, much insight can
be gathered by beginning with an analysis of the mean-field description 
\eref{eq:classiclv} of the LV model. Note that these coupled rate equations entail 
a mass action factorization of a two-point correlation function that encodes the 
likelihood of predators and prey meeting each other at given location at the same 
time into a simple product of their average uniform densities $a$ and $b$; hence 
spatio-temporal correlations are manifestly ignored in this approximate and, in 
general, rather crude description.

\begin{figure}
  \centering
  \includegraphics[width=0.8\columnwidth]{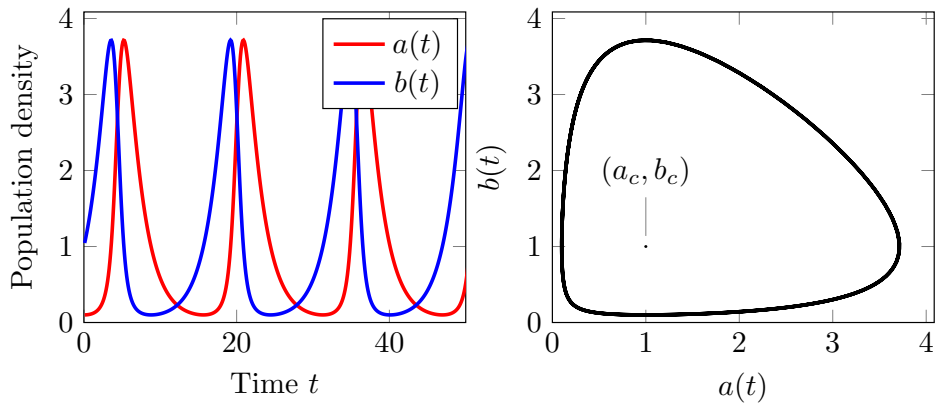}
  \caption{Left: Characteristic non-linear LV mean-field oscillations of predator
    (red) and prey (blue) populations over time obtained through numerical
    integration of the rate equations~\eref{eq:classiclv} with parameters 
	$\sigma = \mu = \lambda = 0.5$ and initial densities $a(0) = 0.1$, $b(0) = 1$.
	Right: The oscillatory dynamics implies closed orbits (neutral cycles) in
	population density phase space (reproduced with permission from 
	Ref.~\cite{Dobramysl13}).}
\label{fig:lv_oscillations}
\end{figure}

Straightforward linear stability analysis of the rate equations~\eref{eq:classiclv} 
yields three stationary states~\cite{Haken83,Murray02,Georgiev07}: (1) the linearly
unstable (for $\sigma > 0$) absorbing total population extinction state 
$(a = 0, b = 0)$; (2) a linearly unstable (for $\lambda > 0$) predator extinction 
and Malthusian prey explosion state $(a = 0, b \to \infty)$; and (3) the marginally
stable species coexistence fixed point $(a = \sigma / \lambda, b = \mu / \lambda)$.
Expanding in fluctuations about this fixed point, one obtains within the linear 
approximation neutral cycles with purely imaginary stability matrix eigenvalues, 
{\em i.e.}, undamped population oscillations with characteristic frequency 
$\omega_0 = \sqrt{\sigma \mu}$. Indeed, the full non-linear coupled differential LV
equations~(\ref{eq:classiclv}) give rise to a conserved first integral 
\begin{equation}
\label{eq:K}
  K = \lambda \left[ a(t) + b(t) \right] - \sigma \ln a(t) - \mu \ln b(t).
\end{equation}
This in turn implies closed orbits in phase space spanned by the two population 
densities, which therefore each display characteristic non-linear oscillations with 
the unrealistic feature that both predator and prey densities return precisely to 
their initial values after each cycle~\cite{Murray02,Georgiev07}, see 
figure~\ref{fig:lv_oscillations}.

The presence of neutral cycles in the deterministic LV model \eref{eq:classiclv}
moreover hints at the lack of robustness in this approximate mean-field description
with respect to even minor modifications. Indeed, model alterations typically 
induce a negative real part for the stability matrix eigenvalues in the species 
coexistence phase, representing attenuated kinetics that ultimately takes the 
populations to stationary fixed-point densities. One biologically relevant LV 
variant posits a finite carrying capacity $\rho$ for the prey~\cite{Murray02} 
mimicking either limited food resources or direct intra-species competition 
({\em e.g.}, $B + B \rightarrow B$ with rate $\sigma / \rho$~\cite{Tauber12}). On 
the mean-field level, the prey population rate equation in \eref{eq:classiclv} is 
then to be replaced with
\begin{equation}
\label{eq:fincaplv}
  \frac{d b(t)}{dt} = 
  \sigma \, b(t) \left[ 1 - \frac{b(t)}{\rho} \right] - \lambda a(t) b(t) \ . 
\end{equation}
This restricted LV model sustains again three stationary states~\cite{Georgiev07}:
(1) total extinction $(a = 0, b = 0)$; (2') predator extinction and prey saturation
at its carrying capacity $(a = 0, b = \rho)$; and (3') predator-prey coexistence 
$(a = \sigma (1 - \mu / \rho \lambda) / \lambda, b = \mu / \lambda)$. This 
coexistence state only exists, and then is linearly stable, if the predation rate
exceeds the threshold $\lambda > \lambda_c = \mu / \rho$. Otherwise, fixed point
(2') is stable, and the predator species is driven toward extinction. In the
two-species coexistence regime, the stability matrix eigenvalues become
\begin{equation}
\label{eq:lvrese}
  \epsilon_\pm = - \frac{\sigma \mu}{2 \lambda \rho} \left[ 1 \pm \sqrt{1 - 
  \frac{4 \lambda \rho}{\sigma} \left( \frac{\lambda \rho}{\mu} - 1 \right)} 
  \right] .
\end{equation}
Consequently, if the eigenvalues $\epsilon_\pm$ are both real, {\em i.e.}, 
$\sigma > \sigma_s = 4 \lambda \rho (\lambda \rho / \mu - 1) > 0$, or 
$\mu / \rho < \lambda < \lambda_s = \mu (1 + \sqrt{1 + \sigma / \mu}) / 2 \rho$,
the neutral cycles of the unrestricted model \eref{eq:classiclv} turn into a stable
node; alternatively, if $\sigma < \sigma_s$ or $\lambda > \lambda_s$, fixed point
(3') becomes a stable focus, and is approached in phase space via a spiraling
trajectory. In the former situation, both predator and prey densities approach 
their stationary values exponentially, in the latter case through damped population
oscillations.

Spatial structures can be accounted for within the mean-field framework through
extending the rate equations \eref{eq:classiclv}, \eref{eq:fincaplv} to a set of 
coupled partial differential equations for local predator and prey densities 
$a({\vec x},t)$ and $b({\vec x},t)$, and heuristically adding diffusive spreading
terms:
\begin{eqnarray}
\label{eq:recdiflv} 
  \frac{\partial a({\vec x},t)}{\partial t} &=& \left( D_A \nabla^2 - \mu \right) 
  a({\vec x},t) + \lambda \, a({\vec x},t) \, b({\vec x},t) \ , \nonumber \\
  \frac{\partial b({\vec x},t)}{\partial t} &=& \left( D_B \nabla^2 + \sigma\right) 
  b({\vec x},t) - \frac{\sigma}{\rho} \, b({\vec x},t)^2 - \lambda \, a({\vec x},t) 
  b({\vec x},t) \ ;
\end{eqnarray}
note that these reaction-diffusion equations still assume weak correlations as
encoded in the mass action factorization in the predation terms. In one dimension,
equations~\eref{eq:recdiflv} permit explicit travelling wave solutions of the
form $a(x,t) = {\bar a}(x - vt)$, $b(x,t) = {\bar b}(x - vt)$ that describe a
predator invasion front from the coexistence phase described by homogeneous fixed
point (3') into a region occupied only by prey~\cite{Dunbar83,Sherratt97,Murray02}.
An established lower bound for the front propagation speed is 
$v > \sqrt{4 D_A (\lambda \rho - \mu)}$~\cite{Hosono98}.

\subsection{Non-spatial stochastic LV systems}

The dominant role of stochastic fluctuations in the LV model that are due to the
internal reaction noise, is already apparent in non-spatial or zero-dimensional
(purely on-site) `urn model' realizations. The system is then fully described by a
(local) stochastic master equation that governs the time evolution of the 
configurational probability $P(n,m;t)$ to find $n$ predators $A$ and $m$ prey $B$ at time 
$t$ through a balance of gain and loss terms. For the set of reactions 
\eref{eq:lvrules}, the master equation reads~\cite{McKane05,Georgiev07,Tauber12}
\begin{eqnarray} 
\label{eq:masteqlv}
  \frac{\partial P(n,m;t)}{\partial t} &=&
  \sigma \left[ (m-1) \, P(n,m-1;t) - m \, P(n,m;t) \right] \nonumber \\
  &+& \mu \left[ (n+1) \, P(n+1,m;t) - n \, P(n,m;t) \right] \\
  &+& \lambda' \left[ (n-1)(m+1) \, P(n-1,m+1;t) - n m \, P(n,m;t) \right] \ . 
  \nonumber
\end{eqnarray} 
It should first be noted that as $t \to \infty$, the system will inevitably reach
the completely empty absorbing state, where both species have become extinct. For
reasonably abundant populations, however, both direct and refined continuous-time
Monte Carlo simulations utilizing Gillespie's algorithm~\cite{Gillespie76} display
erratic but persistent population oscillations over effectively many generations;
as one would expect, their remarkably large amplitude scales roughly like the 
square-root of the mean total particle number in the system~\cite{McKane05}. McKane
and Newman elucidated the physical mechanism behind these persistent random 
oscillations by means of a van Kampen system size expansion~\cite{VanKampen92}, 
which essentially maps the stochastic LV system with finite prey carrying capacity
$\rho$ to a non-linear oscillator driven by white noise with frequency-independent
correlation spectrum. On occasion, these random kicks will occur at the resonance
frequency ($\omega_0 = \sqrt{\sigma \mu}$ in linear approximation) and thus incite
high-amplitude population density excursions away from the mean-field coexistence
fixed point (3'). For sufficiently large predation rate $\lambda'$, the subsequent
relaxation follows a spiral trajectory in phase space, {\em i.e.}, damped 
oscillatory kinetics. 

The discrete small number fluctuations or demographic noise also strongly impacts
the survival time near an absorbing extinction state~\cite{Parker09,Dobrinevski12}.
These can be treated in the framework of large-deviation theory based on the master
equation or, more directly, through the equivalent Doi--Hamiltonian that encodes 
the associated generating function~\cite{Tauber14}, for example by means of a
`semi-classical' WKB-type approach; Ref.~\cite{Assaf17} provides an excellent 
up-to-date review. The mean extinction time in the stochastic non-spatial LV model 
(\ref{eq:masteqlv}) is a relevant example of the intriguing influence of 
demographic noise in a mean-field setting: By exploiting the existence of the  
mean-field constant of motion \eref{eq:K}, Parker and Kamenev performed a 
semi-classical analysis of the Fokker--Planck equation derived from 
\eref{eq:masteqlv} and devised a method to average demographic fluctuations around 
the LV mean-field orbits~\cite{Parker09,Parker10}. They thus showed that the mean 
number of cycles prior to extinction scales as $N_S^{3/2}/N_L^{1/2}$, where $N_S$ 
and $N_L$ respectively denote the characteristic sizes of the smaller and larger of 
the predator or prey sub-populations, and determined the mean extinction time in 
terms of the number of cycles. This result means that extinction typically occurs 
faster when the number $N_L$ of individuals of the most abundant species increases.

The reverse problem of understanding the large spikes in the populations of
predators and prey can be described by turning the classic LV equations 
\eref{eq:classiclv} into stochastic differential equations via the introduction of 
multiplicative noise in the prey birth. This problem can be rewritten in the form 
of a Fokker--Planck--Kolmogorov equation for the probability distribution for the 
sizes of the predator and prey populations, which can be solved exactly. The 
solutions indicate strong intermittent behavior with small population means and 
rare, large excursions of the population sizes when the noise levels are high, 
which are also observed in Monte Carlo simulations~\cite{Dimentberg02}.

\subsection{Coexistence in spatially extended stochastic Lotka--Volterra systems}

The stochastic LV model \eref{eq:lvrules} can be implemented on a $d$-dimensional 
lattice, usually with periodic boundary conditions to minimize edge effects, in a 
straightforward manner via individual-based Monte Carlo update rules. One may 
either allow arbitrarily many predator or prey particles per site, or restrict the 
site occupancy representing a finite local carrying capacity. For example, one 
detailed Monte Carlo algorithm on a two-dimensional square lattice with site 
restrictions (at most a single particle allowed per site) proceeds as 
follows~\cite{Georgiev07,Chen16}:
\begin{itemize}
\item Select a lattice occupant at random and generate a random number $r$ 
      uniformly distributed in the range $[0,1]$ to perform either of the following
 	  four possible reactions (with probabilities $D'$, $\sigma$, $\mu$, and
	  $\lambda'$ in the range $[0,1]$):
\item If $r < 1/4$, select one of the four sites adjacent to this occupant, and 
      move the occupant there with probability $D'$, provided the selected 
      neighboring site is empty (nearest-neighbor hopping).
\item If $1/4 \leq r < 1/2$ and if the occupant is an $A$ particle, then with 
      probability $\mu$ the site will become empty (predator death, 
      $A \rightarrow \emptyset$).
\item If $1/2 \leq r < 3/4$ and if the occupant is an $A$ particle, choose a 
      neighboring site at random; if that selected neighboring site holds a $B$ 
      particle, then with probability $\lambda'$ it becomes replaced with an 
      $A$ particle (predation reaction, $A + B \rightarrow A + A$).
\item If $3/4 \leq r < 1$ and if the occupant is a $B$ particle, randomly select
      a neighboring site; if that site is empty, then with probability $\sigma$ 
	  place a new $B$ particle on this neighboring site (prey offspring production,
	  $B \rightarrow B + B$).
\end{itemize}
Notice that even for $D' = 0$, the particle production processes on neighboring
lattice sites effectively induce spatial population spreading. One Monte Carlo step
is considered completed when on average each particle present in the system has 
been picked once for the above processes. If arbitrarily many individuals of either
species are allowed on each lattice site, all reactions can be performed locally,
but then hopping processe need to be implemented explicitly to allow diffusive
propagation (an explicit Monte Carlo simulation algorithm is, {\em e.g.}, listed in 
Ref.~\cite{Washenberger07}). 

Alternatively, one can use event-driven simulation schemes such as the Gillespie
or kinetic Monte Carlo algorithm~\cite{Gillespie76}. A Gillespie algorithm
equivalent to the sequential update Monte Carlo scheme described above would be:
\begin{enumerate}
\item Set time $t=0$ and choose initial state.
\item List all possible reactions that change the state of the system:
  \begin{itemize}
  \item the number of possible prey reproduction events $N_\sigma$, i.e. the
    number of $B$ particles with empty neighboring sites;
  \item the number of possible predator death events $N_\mu$, i.e. the number of
    $A$ particles on the lattice;
  \item the number of possible predation events $N_\lambda$, i.e. the number of
    neighboring pairs of $A$ and $B$ particles;
  \item the number of possible diffusion events $N_D=\sum_{i=1}^{N}n_i$, where
    $n_i=0,\ldots,4$ is the number of empty adjacent sites of particle $i$, and $N$
    the total number of particles on the lattice.
  \end{itemize}
\item Calculate the reaction propensities of each event
  $R_\sigma=\sigma N_\sigma$, $R_\mu=\mu N_\mu$, $R_\lambda=\lambda' N_\lambda$
  and $R_D=D' N_D$. Choose an event type $i$ at random with the probability
  weighed according to the respective propensities.
\item Carry out a randomly chosen event of the given type in the list assembled
  in step (ii).
\item Advance time by $\Delta t=-\ln(r)/R_i$ where $r\in(0,1]$ is a uniformly
  distributed random number.
\item Continue with step (ii).
\end{enumerate}
The Gillespie algorithm requires no choice of time step length and thus provides
exact trajectory samples from the solution of the master equation. It is also
rejection-free because every event results in a reaction and is thus more
efficient than a sequential update algorithm. However, the Gillespie algorithm
(and its variants, see Ref.~\cite{Cao04} for a good overview) can be challenging to
implement due to the complexity involved in calculating the reaction
propensities. It has been successfully applied to study the LV model and the RPS
model in both spatial~\cite{SMR,SMR2,BS,MRS16} and non-spatial
contexts~\cite{RMFnonspatial1,Parker09,Dobrinevski12}.

\begin{figure} 
  \centering
  \includegraphics[width = 7.4cm]{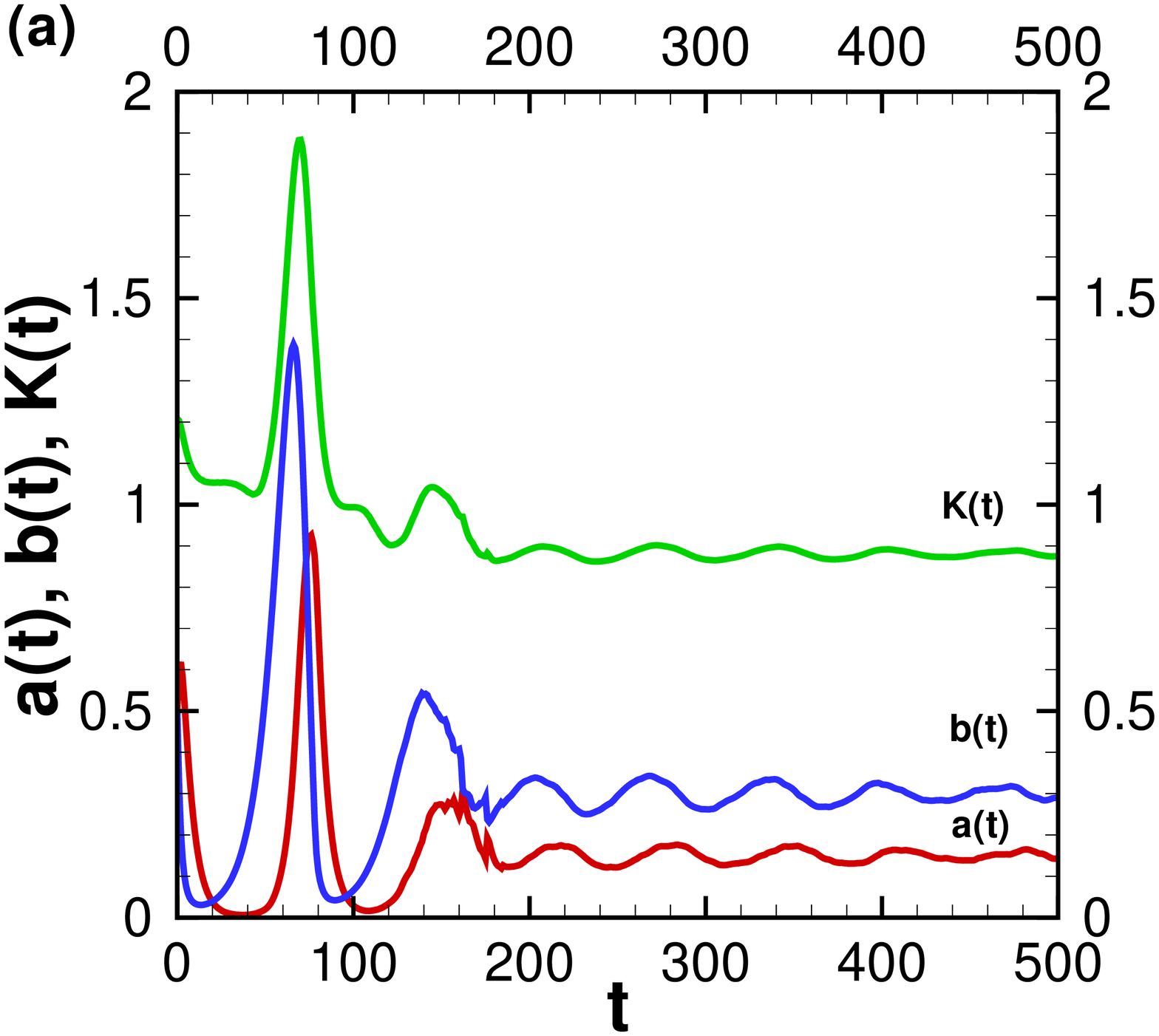} \
  \includegraphics[width = 7.4cm]{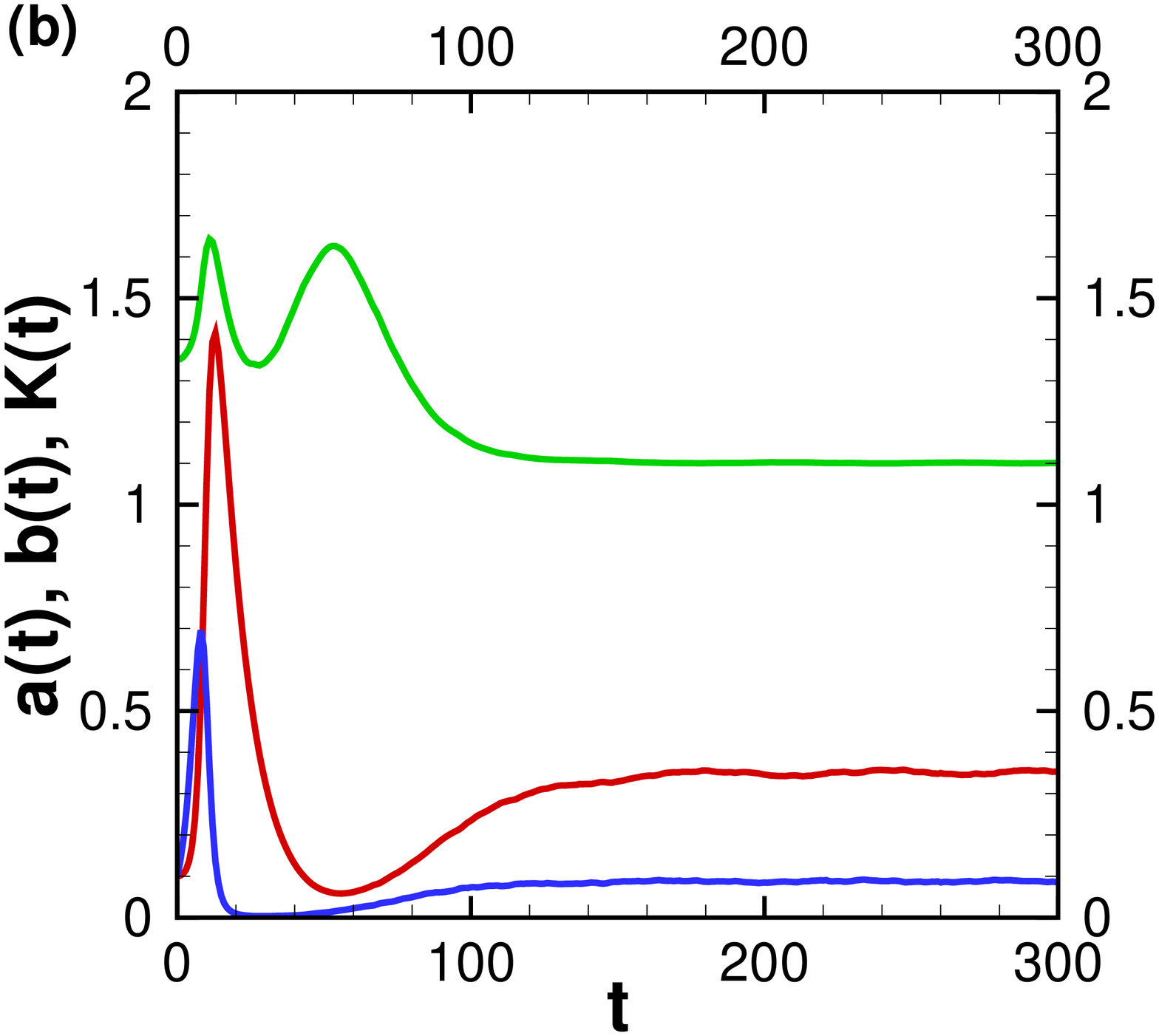}
  \caption{Early time evolution for the population density of predators $a(t)$ 
    (red), prey $b(t)$ (blue), and the mean-field first integral $K(t)$ (green) in
	a stochastic two-dimensional lattice LV model with $1024 \times 1024$ sites 
	(periodic boundary conditions, no occupation number restrictions) from two 
	single runs that both started with a random particle distribution with reaction
	rates (a) $\sigma = 0.1$, $\mu = 0.2$, and $\lambda = 1.0$, and 
	(b) $\sigma = 0.4$, $\mu = 0.1$, and $\lambda = 1.0$ 
	(reproduced with permission from Ref.~\cite{Washenberger07}).}
\label{fig:lv_trajectories} 
\end{figure}
Computer simulations based on similar Monte Carlo algorithms of sufficiently large 
stochastic LV systems, thus avoiding full population extinction, invariably yield 
long-lived erratic population oscillations and a stable predator-prey coexistence
regime~\cite{Tome94,Boccara94,Provata99,Rozenfeld99,Lipowski99,Lipowska00,
Monetti00,Droz01,Antal01,Tsekouras01,Kowalik02,Georgiev07,Washenberger07,Chen16}.
Two typical simulation runs on a two-dimensional square lattice without site 
occupation restrictions, starting with particles that are randomly distributed in 
the system, are shown in figure~\ref{fig:lv_trajectories}. Both predator and prey 
densities display damped oscillatory kinetics, with the oscillation frequency and 
the attenuation depending on the prescribed reaction rates. In stark contrast with 
the mean-field rate equation solutions (see figure~\ref{fig:lv_oscillations}), 
there are no neutral cycles, and instead one clearly observes relaxation towards 
(quasi-)stationary population densities representing a stable coexistence state.
The quantity $K(t)$ (\ref{eq:K}), representing the conserved rate equations' first 
integral, becomes time-dependent and oscillates along with the population densities.
The oscillation amplitude scales as the inverse square-root of the lattice size,
hinting at the presence of spatially separated and largely independent (non-linear)
oscillators in the system~\cite{Provata99,Lipowski99,Lipowska00,Droz01,Tsekouras01,
Kowalik02,Georgiev07}, each subject to local noise-induced resonant 
amplification~\cite{McKane05}.

\begin{figure}
  \centering
  \includegraphics[width=0.9\columnwidth]{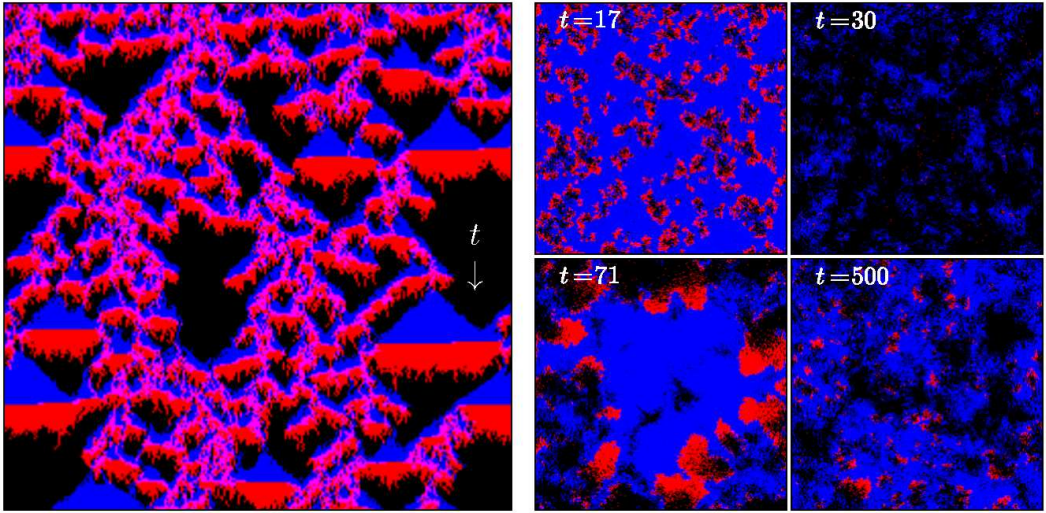}
  \caption{Left: space-time plot for a stochastic LV model simulation on a chain 
    with $250$ sites (and periodic boundary conditions), initial densities
    $a(0) = 1 = b(0)$, and rates $\sigma = 0.5$, $\mu = 0.5$, and $\lambda=0.3$; 
	the temporal evolution is displayed vertically downward, the colors blue and 
	red indicate the presence of prey and predator particles, respectively, purple 
	marks sites occupied by both species, and black pixels denote empty sites. At 
	$t = 0$ the system is well-mixed; over time, clusters of prey particles form 
	and grow; predators subsequently invade prey clusters, often removing them 
	completely.
	Right: snapshots from a two-dimensional stochastic lattice LV simulation with 
	$250 \times 250$ sites (periodic boundary conditions, multiple occupations
    allowed), $a(0) = 0.01$, $b(0) = 1$, $\sigma = 0.1$, $\mu = 0.9$, and 
    $\lambda = 1$. Here, the prey community survives an early predator invasion 
	(at $t = 17$ MCS), followed by prey recovery and proliferation due to predator 
    scarcity ($t = 30$ MCS). New predator fronts later invade a large prey cluster 
    ($t = 71$ MCS). Following transient oscillations, the system reaches a
    quasi-steady species coexistence state ($t = 500$ MCS) characterized by smaller
	prey clusters and recurring predator invasions (reproduced with permission from 
	Ref.~\cite{Dobramysl13}).}
\label{fig:lv_snapshots}
\end{figure}
Inspection of successive temporal snapshots of the Monte Carlo data for the spatial
population distributions and full movie visualizations~\cite{Mobilia06,Georgiev07,
Washenberger07,Dobramysl08,Dobramysl12,Dobramysl13,Chen16} further illuminate the
origin of the local population oscillations incited by the system-immanent
stochasticity or internal reaction noise~\cite{Movies}. As shown in the space-time
plot on the left panel of figure~\ref{fig:lv_snapshots}, quite complex temporal 
behavior emerges already in one dimension, with growing prey clusters suffering
predator invasion from their boundaries and subsequent near-elimination, with the
few surviving prey particles forming the nuclei for renewed growth spurts. This
repetitive spatially correlated dynamics induces striking spatio-temporal patterns,
and even more so in two dimensions; lattice snapshots for one example are depicted 
in the right panels of figure~\ref{fig:lv_snapshots}. Localized prey clusters are
invaded and devoured by predators, who then starve and come close to extinction,
until scarce survivors become the sources for radially spreading prey-predator
fronts that subsequently merge and interact, giving rise to spatially separated 
local population oscillations. The characteristic LV activity fronts represent a
proto-typical example for the more general non-equilibrium phenomenon of formation
of noise-induced and -stabilized spatio-temporal patterns~\cite{Butler09,Butler11}.
For the LV system, these recurring fluctuating structures and the associated 
erratic oscillations are in fact quite robust against modifications of the 
microscopic reaction scheme, rendering the features of spatially extended 
stochastic LV systems remarkably universal. They only become suppressed in lattice
simulations in $d > 4$ spatial dimensions, or upon implementation of effective 
species mixing through fast particle exchange (`swapping') processes, whereupon the
dynamics attains the characteristic signatures of the corresponding mean-field rate
equation solutions~\cite{Mobilia06}.
 
\begin{figure} 
  \centering
  \includegraphics[width = 5.1cm]{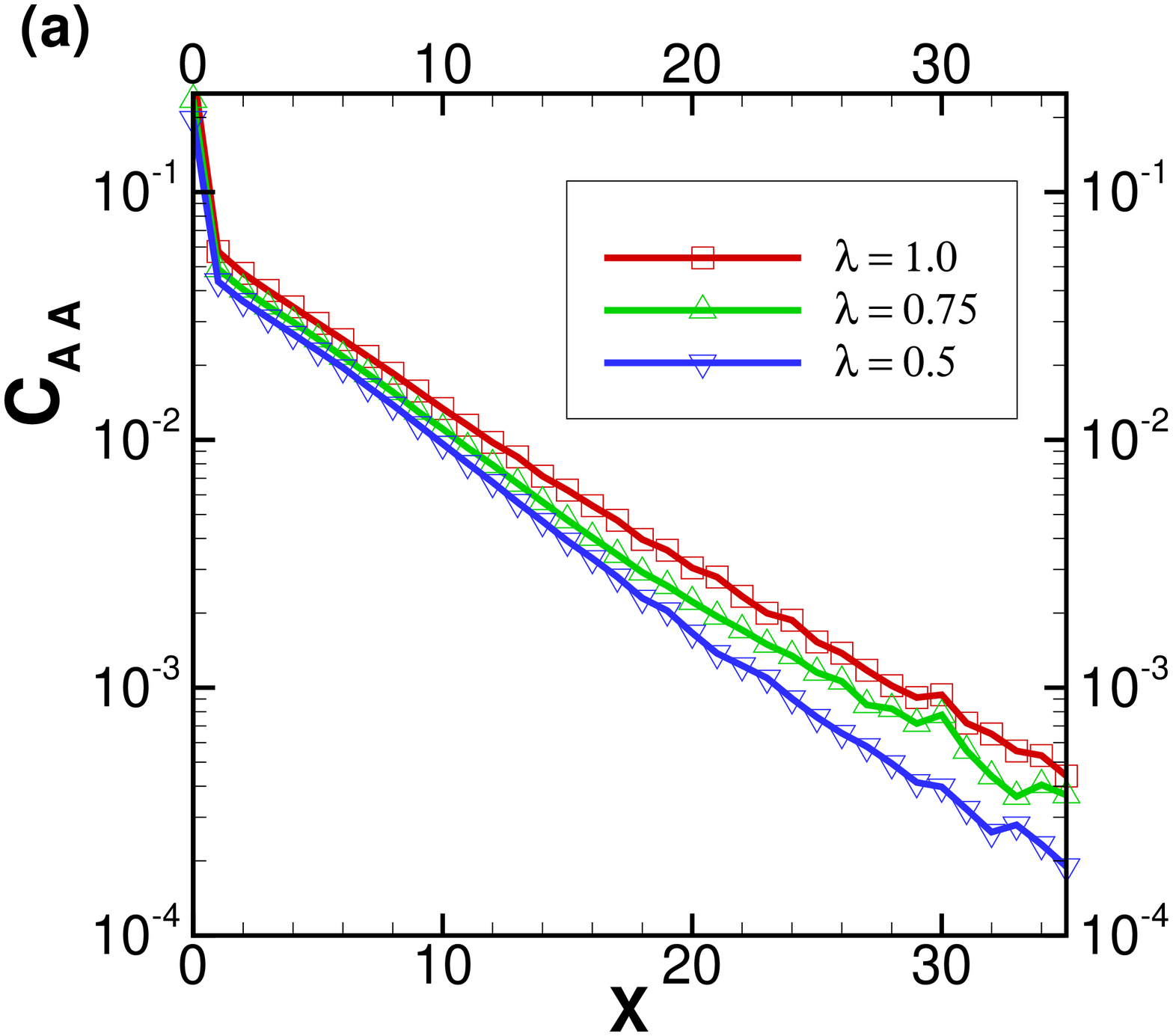} 
  \includegraphics[width = 5.1cm]{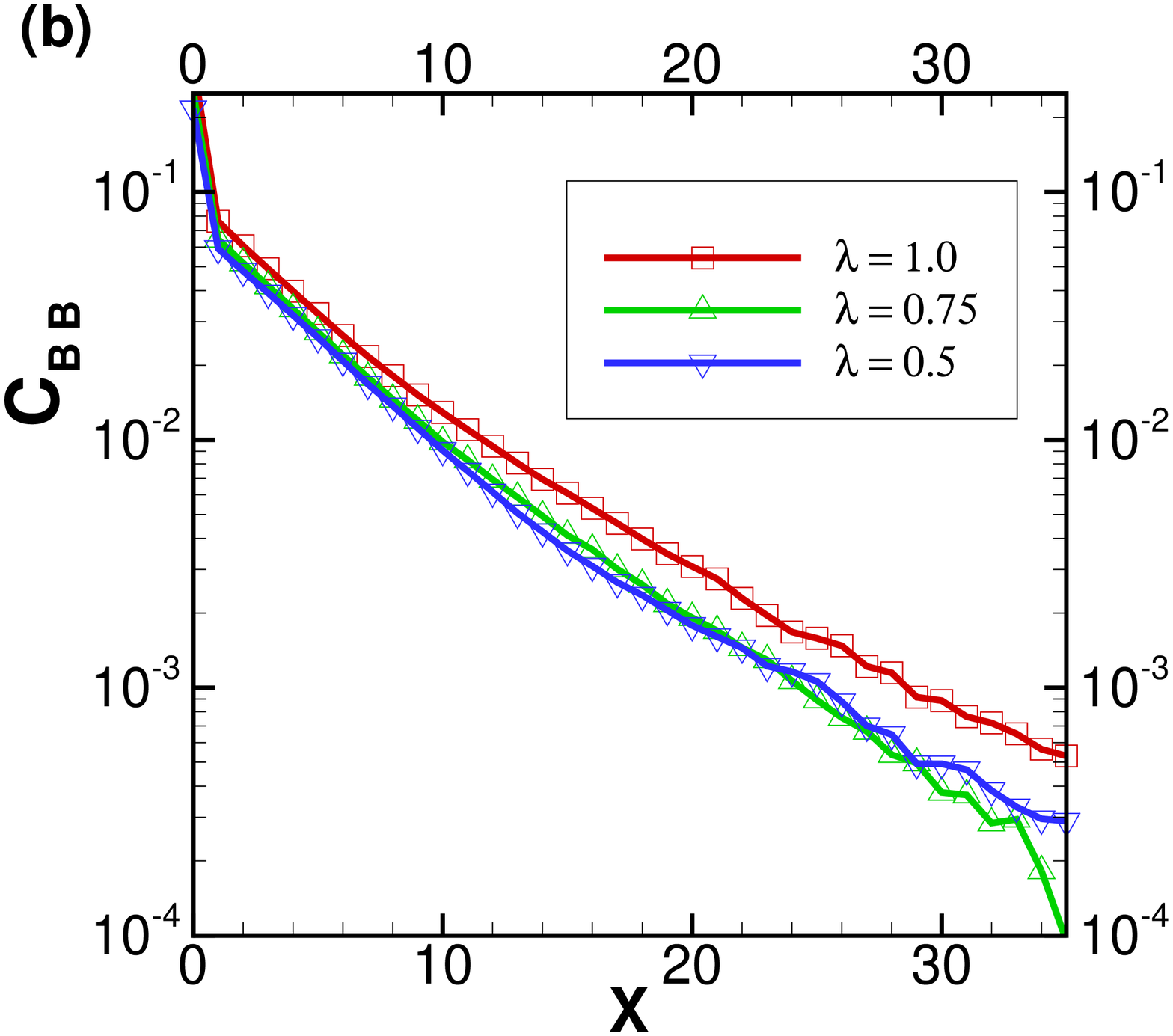} 
  \includegraphics[width = 5.1cm]{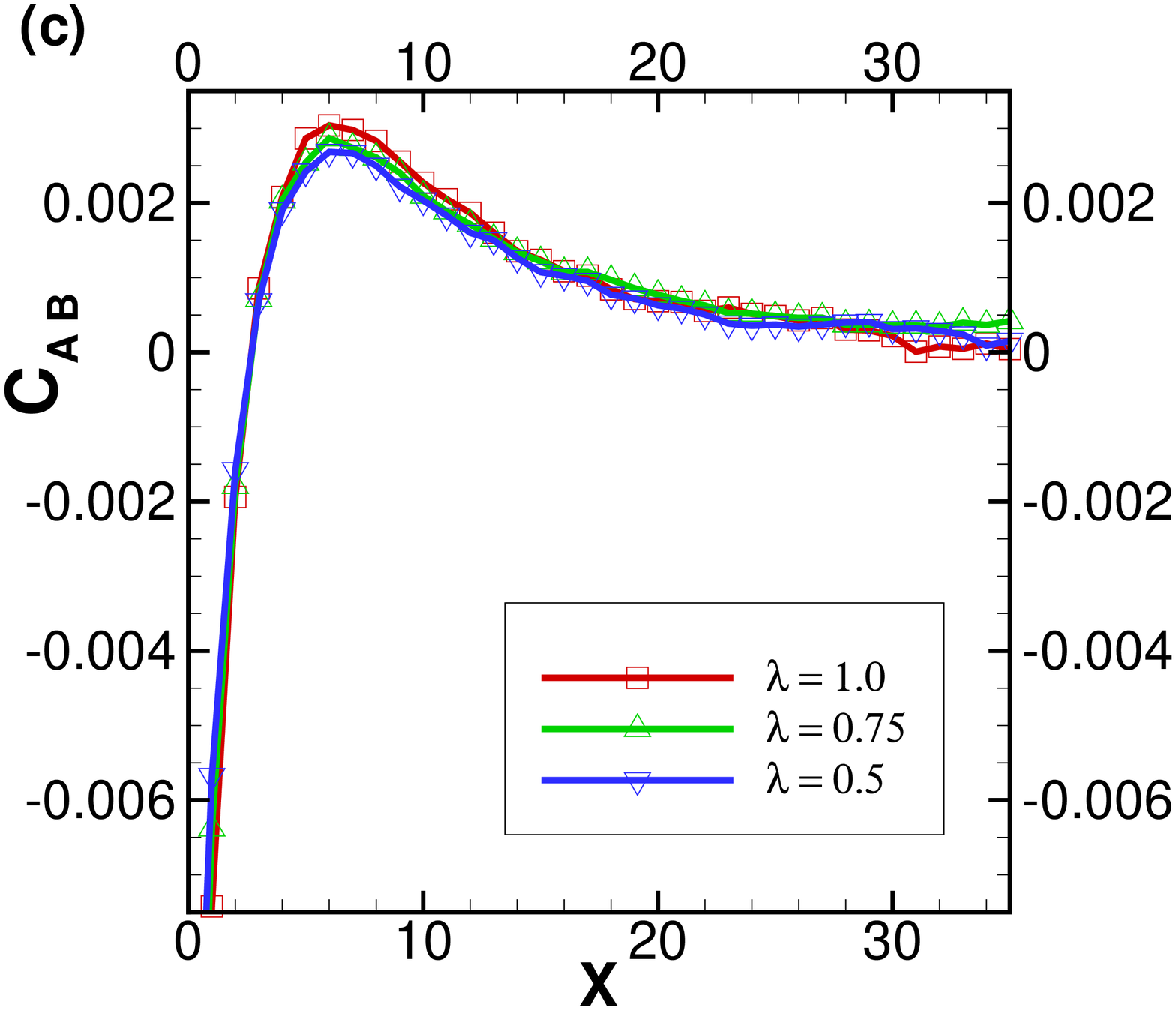}
  \caption{equational-time correlation functions (a) $C_{AA}(x)$, (b) $C_{BB}(x)$, and 
    (c) $C_{AB}(x)$, measured in stochastic LV simulations on a $1024 \times 1024$ 
	lattice, with rates $\sigma = 0.1$, $\mu = 0.1$, and different values of the
	predation rate $\lambda = 0.5$ (blue), $0.75$ (green), and $1.0$ (red)
	(reproduced with permission from Ref.~\cite{Washenberger07}).}
\label{fig:lv_correlations}
\end{figure}
The Monte Carlo simulation data are of course amenable to a quantitative analysis 
of the emerging spatial and temporal correlations in stochastic lattice LV 
systems~\cite{Georgiev07,Washenberger07}. Figure~\ref{fig:lv_correlations} displays
the three static (equal-time) correlation functions $C_{\alpha \beta}(x) = 
\langle n_\alpha(x) \, n_\beta(0) \rangle - \langle n_\alpha(x) \rangle \, 
\langle n_\beta(0) \rangle$, with local occupation numbers $n_\alpha$ and species 
indices $\alpha,\beta = A,B$, {\em i.e.}, the predator-predator and prey-prey 
correlators $C_{AA}(x)$ and $C_{BB}(x)$, as well as the two-species 
cross-correlations $C_{AB}(x)$, as function of distance $x$ in the predator-prey 
coexistence phase of a two-dimensional lattice with $1024 \times 1024$ sites for 
fixed rates $\sigma, \mu$, but different values of $\lambda$. As is apparent from 
figures~\ref{fig:lv_correlations}(a) and (b), both $C_{AA}(x)$ and $C_{BB}(x)$ are 
positive and decay exponentially $\sim e^{-|x|/\xi}$ with correlation lengths $\xi$
on the scale of a few lattice constants, which represents the typical width of the 
spreading population fronts. The predator-prey cross-correlations $C_{AB}(x)$ 
similarly decay at large distances from positive values, but naturally are negative
(anti-correlated) at short $x$, since prey that approach too closely to predators 
may not survive the encounter. The maximum in figure~\ref{fig:lv_correlations}(c) 
located at about six lattice constants indicates the mean spacing between the prey
and following predator waves.

\begin{figure} 
\centering
  \includegraphics[width = 7.5cm]{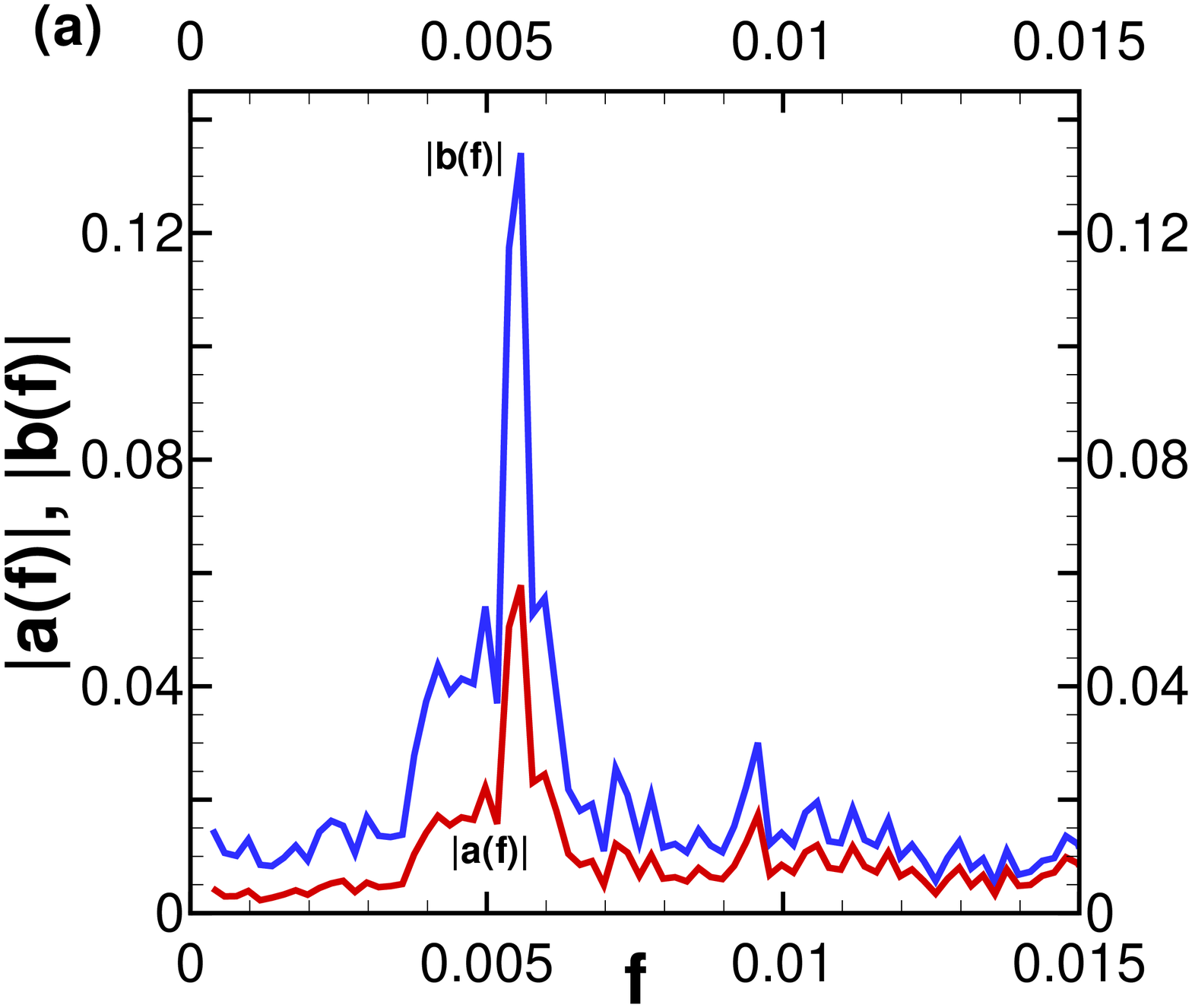} \
  \includegraphics[width = 7.5cm]{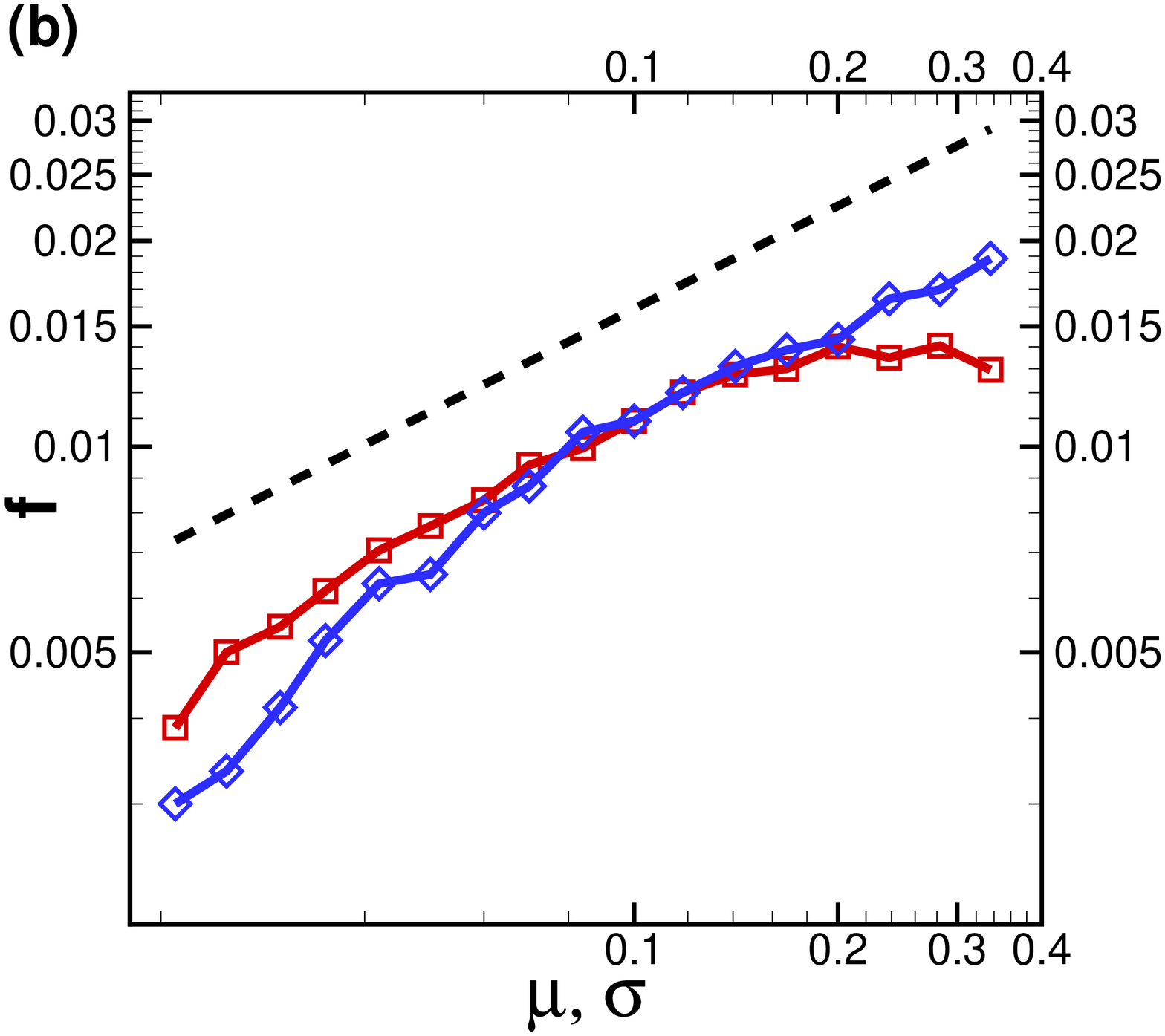}
  \caption{(a) Fourier transforms $|a(f)|$ and $|b(f)|$ of the predator (red) and 
	prey (blue) population density time series for a stochastic LV simulation run 
	on a $1024 \times 1024$ lattice with rates $\sigma = 0.03$, $\mu = 0.1$, and 
	$\lambda = 1.0$, as functions of frequency $f$.
	(b) Measured dependence of the characteristic peak frequencies in $|a(f)|$ and
	$|b(f)|$ on the rates $\sigma$ (red squares) and $\mu$ (blue diamonds), with 
	the respective other rate held fixed at the value $0.1$ and $\lambda = 1.0$, as
	obtained from simulation data on $1024 \times 1024$ lattices up to time 
	$t = 20,000$; for comparison, the dashed black line shows the linearized 
	mean-field oscillation frequency $f_0 = \sqrt{\sigma \mu} / 2 \pi$
	(reproduced with permission from Ref.~\cite{Washenberger07}).} 
\label{fig:lv_ftpeaks} 
\end{figure}

The characteristic oscillation frequency and damping are most efficiently and
reliably determined through Fourier analysis of the density time series, 
$a(f) = \int e^{2 \pi i f t} \, a(t) \, dt$ for the predators (and similarly for
the prey density $b$). As demonstrated in figure~\ref{fig:lv_ftpeaks}(a), the
population density Fourier amplitudes display a marked peak, whose location can be
identified with the oscillation frequency, while its half-width at half-maximum
gives the attenuation rate or inverse relaxation time. The double-logarithmic plot
of ensuing data measured for two-dimensional stochastic LV systems with various
reaction rates shows that the functional dependence of the oscillation frequency
on $\mu$ and $\sigma$ roughly follows the mean-field square-root behavior, yet with
noticeable deviations as the ratio $\sigma / \mu$ deviates from $1$. However, the
characteristic population oscillations in spatially extended stochastic LV models 
clearly occur at markedly lower frequencies, here reduced by a factor $\sim 2$ as
compared with the linearized rate equation prediction.

The algorithm described above can be modified to introduce more complex
interaction patterns. For example, Rozenfeld and Albano introduced the ability for 
prey to forego reproduction if predators are within a range $V_H$, and allowed for 
escape-pursuit via an interaction potential~\cite{Rozenfeld01,Rozenfeld04}. This
yields a phase in which self-sustained oscillatory behavior of the overall
populations can be observed -- even in the thermodynamic limit -- based on dynamic
percolation. This finding is in contrast to the above model in which oscillations 
are always decaying towards the coexistence fixed point.

\subsection{Doi--Peliti field theory and perturbative analysis}

The remarkably strong renormalization of the characteristic population oscillation
features through intrinsic stochastic fluctuations in spatially extended LV systems
can be understood at least qualitatively through a perturbative computation based
on a field theory representation of the master equation \eref{eq:masteqlv} as
afforded by the powerful Doi--Peliti formalism (recent reviews are, {\em e.g.},
provided in Refs.~\cite{Tauber05,Cardy08,Tauber14}). To this end, we
permit an arbitrary number of predator and prey particles
per site, $n_i, m_i = 0,1,\ldots,\infty$, but 
implement a growth-limiting pair annihilation reaction for the prey species, 
$B + B \rightarrow B$ with rate $\nu'$. Since all stochastic reactions locally 
alter occupation numbers by integer values, it is convenient to introduce a bosonic
ladder operator algebra $[ a_i , a_j ] = 0$, $[ a_i , a_j^\dagger ] = \delta_{ij}$ 
for the predator species $A$, from which their particle number eigenstates 
$| n_i \rangle$ can be constructed, satisfying 
$a_i \, |n_i \rangle = n_i \, |n_i-1 \rangle$, 
$a_i^\dagger \, |n_i \rangle = |n_i + 1 \rangle$, 
$a_i^\dagger \, a_i \, |n_i \rangle = n_i \, |n_i \rangle$. In the same manner,
one proceeds with bosonic prey operators, and imposes 
$[a_i , b_j ] = 0 = [a_i , b_j^\dagger ]$. A general state vector is then 
defined as a linear combination of all particle number eigenstates, weighted with 
their configurational probabilities: $| \Phi(t) \rangle = \sum_{\{ n_i \}, 
\{ m_i \}} P(\{ n_i \},\{ m_i \};t) \, | \{ n_i \}, \{ m_i \} \rangle$. The master
equation is then transformed into the (linear) time evolution equation
$\partial | \Phi(t) \rangle / \partial t = - H \, | \Phi(t) \rangle$ with local 
reaction pseudo-Hamiltonian (Liouville operator)
\begin{equation}  
\label{eq:lovham}  
  \fl \ H_{\rm reac} = - \sum_i \left[ \mu \, \bigl( 1 - a_i^\dagger \bigr) \, 
  a_i + \sigma \, \bigl( b_i^\dagger - 1 \bigr) \, b_i^\dagger\,  b_i 
  + \nu' \, \bigl( 1 - b_i^\dagger \bigr) \, b_i^\dagger \, b_i^2 + \lambda' \,
  \bigl( a_i^\dagger - b_i^\dagger \bigr) \, a_i^\dagger \, a_i \, b_i \right] . \ 
\end{equation}
Nearest-neighbor hopping processes are similarly represented by
\begin{equation} 
\label{eq:difham} 
  H_{\rm diff} = \sum_{<ij>} \left[ D'_A \, \bigl( a_i^\dagger - a_j^\dagger
  \bigr) \, \bigl( a_i - a_j \bigr) + D_B' \, \bigl( b_i^\dagger - b_j^\dagger
  \bigr) \, \bigl( b_i - b_j \bigr) \right] \, .
\end{equation} 

In order to arrive at a continuum field theory representation, one follows the
standard route in quantum many-particle physics to utilize coherent states,
{\em i.e.}, eigenstates of the annihilation operators with complex eigenvalues
$\alpha_i$ and $\beta_i$: $a_i \, | \alpha_i \rangle = \alpha_i \, | \alpha_i 
\rangle$ and $b_i \, | \beta_i \rangle = \beta_i \, | \beta_i \rangle$, 
to construct a path integral for the time evolution of arbitrary observables
in this basis,
\begin{equation}
\label{eq:copain} 
  \langle {\cal O}(t) \rangle \! \propto \! \int \! \prod_i \! d\alpha_i \, 
  d\alpha_i^* \, d\beta_i \, d\beta_i^* \, {\cal O}(\{ \alpha_i \} , \{ 
  \beta_i \}) \, \exp (- S[\alpha_i^*, \beta_i^*; \alpha_i, \beta_i;t]) \, , \
\end{equation}
with a statistical weight that is determined by the action
\begin{equation}
\label{eq:cohact} 
  S[\alpha_i^*,\beta_i^*; \alpha_i,\beta_i] = \sum_i \int dt \, \left[ \alpha_i^* 
  \, \frac{\partial \alpha_i}{\partial t} + \beta_i^* \, 
  \frac{\partial \beta_i}{\partial t} + H(\alpha_i^*,\beta_i^*;\alpha_i,\beta_i) 
  \right] \ .
\end{equation}
Finally, the continuum limit is taken: $\sum_i \to a_0^{-d} \int d^dx$; where $a_0$
denotes the lattice constant, $\alpha_i(t) \to a_0^d \, a(\vec{x},t)$, 
$\beta_i(t) \to a_0^d \, b(\vec{x},t)$, $\alpha_i^*(t) \to {\hat a}(\vec{x},t)$, 
and $\beta_i^*(t) \to {\hat b}(\vec{x},t)$. Upon performing the field shifts
${\hat a}({\vec x},t) = 1 + {\tilde a}({\vec x},t)$, 
${\hat b}({\vec x},t) = 1 + {\tilde b}({\vec x},t)$, and setting 
$\nu = a_0^d \nu' = \sigma / \rho$, the action becomes explicitly
\begin{eqnarray}
\label{eq:slvact} 
  &&S[{\tilde a},{\tilde b};a,b] = \int \! d^dx \int \! dt \, \Biggl[ 
  {\tilde a} \, \Biggl( \frac{\partial}{\partial t} - D_A \, \nabla^2 
  + \mu \Biggr) a + {\tilde b} \, \Biggl( \frac{\partial}{\partial t} 
  - D_B \, \nabla^2 - \sigma \Biggr) b \nonumber \\
  &&\quad\qquad\qquad\qquad\quad\ - \sigma \, {\tilde b}^2 \, b + 
  \frac{\sigma}{\rho} \, (1 + {\tilde b}) \, {\tilde b} \, b^2 - \lambda
  \, (1 + {\tilde a}) \, ({\tilde a} - {\tilde b}) \, a \, b \, \Biggr] .
\end{eqnarray}
If it is interpreted as a Janssen--De~Dominicis functional in a path integral
representation for stochastic partial differential equations~\cite{Tauber14}, it
may be viewed equivalent to two coupled Langevin equations for complex fields
\begin{eqnarray}
\label{eq:langeq}
  \frac{\partial a({\vec x},t)}{\partial t} &=& (D_A \nabla^2 - \mu) \,
  a({\vec x},t) + \lambda \, a({\vec x},t) \, b({\vec x},t) + \zeta({\vec x},t)
  \ , \nonumber \\
  \frac{\partial b({\vec x},t)}{\partial t} &=& (D_B \nabla^2  + \sigma) \,
  b({\vec x},t) - \frac{\sigma}{\rho} \, b({\vec x},t)^2 
  - \lambda \, a({\vec x},t) \, b({\vec x},t) + \eta({\vec x},t) \, . \ 
\end{eqnarray}
These resemble the reaction-diffusion equations (\ref{eq:recdiflv}) for (real) local particle 
densities, with additional Gaussian stochastic forcing with zero mean, 
$\langle \zeta \rangle = 0 = \langle \eta \rangle$, and the noise 
(cross-)correlations
\begin{eqnarray}
\label{eq:noicor}
  &&\langle \zeta({\vec x},t) \, \zeta({\vec x}',t') \rangle = 2 \lambda \, 
  a({\vec x},t) \, b({\vec x},t) \, \delta({\vec x}-{\vec x}') \, \delta(t-t') \ , 
  \nonumber \\
  &&\langle \zeta({\vec x},t) \, \eta({\vec x}',t') \rangle = - \lambda \, 
  a({\vec x},t) \, b({\vec x},t) \, \delta({\vec x}-{\vec x}') \, \delta(t-t') \ ,
  \nonumber \\
  &&\langle \eta({\vec x},t) \, \eta({\vec x}',t') \rangle = 2 \sigma \, 
  b({\vec x},t) \, \Bigl[1 - b({\vec x},t) / \rho \Bigr] \, 
  \delta({\vec x}-{\vec x}') \, \delta(t-t'), \, 
\end{eqnarray}
which describe multiplicative noise terms that vanish with the particle densities, 
as appropriate for the presence of a fully absorbing state at $a = 0 = b$; similar 
Langevin equations were derived in Ref.~\cite{Butler09} by means of a van~Kampen
system size expansion~\cite{VanKampen92}.

In the predator-prey coexistence regime, one proceeds by expanding the fields about
their stationary values. Subsequent diagonalization of the resulting Gaussian 
(bilinear) action yields circularly polarized eigenmodes with dispersion 
$i \omega({\vec q}) = \pm i \omega_0 + \gamma_0 + D_0 q^2$ (for equal predator and
prey diffusivities $D_A = D_0 = D_B$) with `bare' oscillation frequency 
$\omega_0^2 = \sigma \mu (1 - \mu / \lambda \rho) - \gamma_0^2$ and damping 
$\gamma_0 = \sigma \mu / 2 \lambda \rho$, see eq.~\eref{eq:lvrese}; note that 
$\gamma_0 \to 0$ in the absence of site occupation restrictions or infinite 
carrying capacity $\rho$. One may then compute fluctuation corrections to the
oscillation frequency, diffusion constant, and attenuation by means of a systematic
perturbation expansion with respect to the non-linear predation rate $\lambda$; in
fact, the effective expansion parameter turns out to be 
$(\lambda / \omega_0) (\omega_0 / D_0)^{d/2}$~\cite{Tauber12}. As observed in the
computer simulations, first-order perturbation theory gives a downward 
renormalization for the characteristic frequency, which is particularly strong in
dimensions $d \leq 2$ owing to the very weak (for large $\rho$) mean-field 
attenuation. The fluctuation corrections moreover appear largely symmetric in the
rates $\sigma$ and $\mu$ and become enhanced as $\sigma \ll \mu$ or 
$\sigma \gg \mu$; both these features are also in accord with the Monte Carlo data,
see figure~\ref{fig:lv_ftpeaks}(b). The diffusion rate is shifted upwards by the
fluctuations, indicating faster front propagation. In contrast, the damping rate $\gamma$
is reduced; both these renormalizations facilitate instabilities towards 
spontaneous pattern formation that occur for $\gamma < 0$ at wavenumbers 
$q < \sqrt{|\gamma| / D}$.

\subsection{Predator species extinction threshold}

\begin{figure}
\centering
  \includegraphics[width=0.4\columnwidth,height=0.4\columnwidth,clip=untrue]{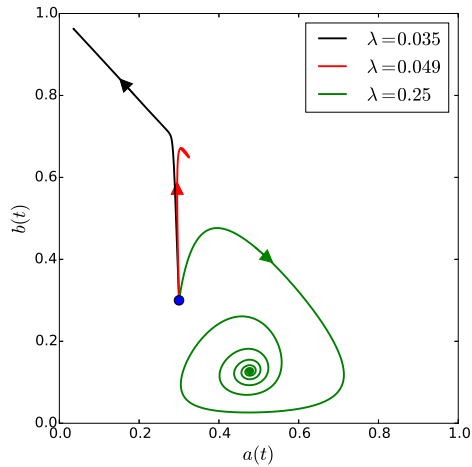} 
  \caption{Monte Carlo simulation trajectories for a stochastic LV model on a 
    $1024 \times 1024$ square lattice with periodic boundary conditions and 
	restricted site occupancy (at most one particle allowed per site) in the 
	predator-prey density phase plane ($a(t) + b(t) \leq 1$) with initial values 
	$a(0) = b(0) = 0.3$ (blue dot), fixed rates $\sigma = 1.0$, 
	$\mu = 0.025$, and predation rates $\lambda = 0.035$ (black): predator 
	extinction phase; $\lambda = 0.049$ (red): exponential relaxation to the 
	quasi-stationary state just beyond the extinction threshold in the active 
	coexistence phase; and $\lambda = 0.250$ (green): deep in the two-species 
	coexistence phase, with spiraling trajectories representing a damped oscillatory 
	relaxation (adapted with permission from Ref.~\cite{Chen16}).}
\label{fig:lv_phases}
\end{figure}
In the presence of site occupation number restrictions, stochastic lattice LV
models feature a sharp continuous non-equilibrium phase transition that separates
the two-species coexistence state from a prey-only phase wherein the predator
species is driven to extinction~\cite{Lipowska00,Antal01}. Different simulation
trajectories in the population density phase plane obtained for varying values of
the predation rate $\lambda$ are displayed in figure~\ref{fig:lv_phases}. At least
qualitatively, these follow remarkably well the fixed-point analysis following
eq.~\eref{eq:lvrese}: For small predation efficiency, the predators die out, and
the system reaches an absorbing state with prey proliferation. Just beyond this
extinction threshold in the coexistence phase, both species quickly reach their 
asymptotic stationary densities through exponential relaxation; simulation 
snapshots or movies show localized predator clusters immersed in a `sea' of 
abundant prey. For large predation rates, persistent damped population oscillations 
emerge, reflected in spiraling trajectories in the phase plane; in this situation, 
one observes complex spatio-temporal patterns induced by spreading and colliding 
prey-predator activity fronts~\cite{Georgiev07,Movies}.

On quite general grounds, one expects the critical properties of non-equilibrium
phase transitions from an active phase to an absorbing state to be described by
the universality class of critical directed percolation, with directionality set
along the time `direction'~\cite{Hinrichsen00,Odor04,Janssen05,Henkel08,Tauber14}.
In the absence of additional conservation laws and quenched spatial disorder, one
expects this statement to be true even for multi-species systems~\cite{Janssen01}.
It was thus surmised early on that the predator extinction transition in spatially 
extended LV models with restricted local carrying capacities is governed by the 
directed-percolation scaling exponents, and there now exists ample numerical
evidence to support this statement~\cite{Tome94,Boccara94,Rozenfeld99,Lipowski99,
Lipowska00,Monetti00,Antal01,Kowalik02,Georgiev07,Chen16}. Indeed, once the 
predator species becomes sparse and the prey abundant, essentially uniformly 
filling the lattice, the predation reaction $A + B \rightarrow A + A$ may just 
happen everywhere and is effectively replaced by a spontaneous branching process
$A \rightarrow A + A$; in conjunction with the decay processes 
$A \rightarrow \emptyset$ and $A + A \rightarrow A$ (reflecting again a locally
restricted carrying capacity), one hence arrives at the basic stochastic processes
defining directed percolation. Formally, considering fluctuations of the predator 
density near zero and of the prey density about $\rho$, one may directly map the 
Doi--Peliti action \eref{eq:slvact} for the near-threshold stochastic LV model onto
Reggeon field theory
\begin{equation}
\label{eq:dprft}  
  S[{\widetilde {\psi}},\psi] = \int d^dx \int dt \left[ {\widetilde \psi} \left( 
  \frac{\partial}{\partial t} + D \left( \tau - \nabla^2 \right) \right) \psi 
  - u \, {\widetilde \psi} \left( {\widetilde \psi} - \psi \right) \psi \right] 
\end{equation}
that captures the universal scaling properties of critical directed 
percolation~\cite{Cardy80,Janssen05,Tauber14}.

\begin{figure}
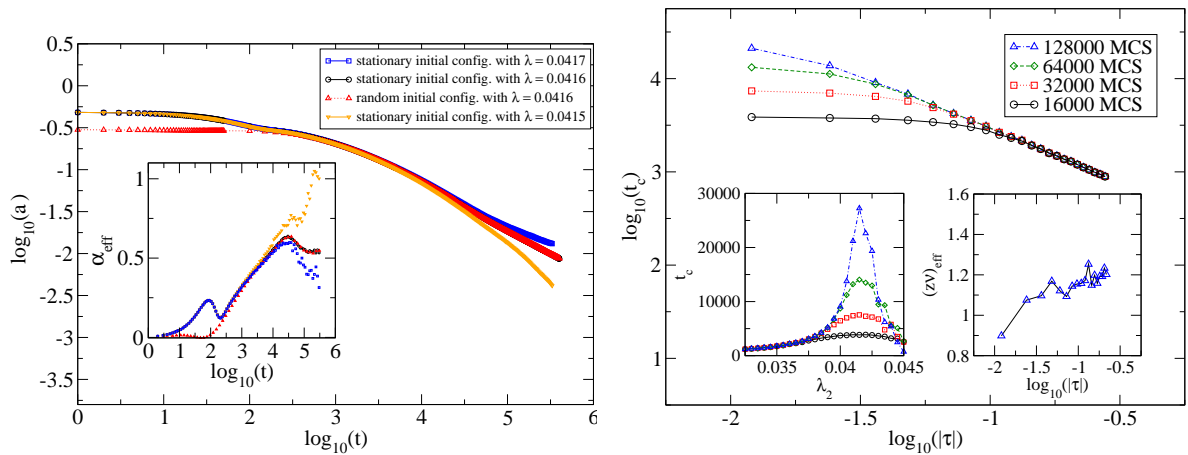

  \centering
  \includegraphics[width=0.5\columnwidth,clip=untrue]{figure7a.eps} \
  \includegraphics[width=0.48\columnwidth,clip=untrue]{figure7b.eps}
  \caption{Left: Decay of the mean predator density (double-logarithmic plots) for a
    stochastic Lotka--Volterra model on a $1024 \times 1024$ square lattice at the
    extinction threshold $\lambda_c = 0.0416$ for $\sigma = 1.0$ and $\mu = 0.025$
	and for both quasi-stationary (top, black) and random (lower curve, red dotted)
	initial configurations (data averaged over $2000$ independent simulation runs).
	For comparison, the predator density decay data are shown as well for 
	$\lambda = 0.0417$ (blue, active coexistence phase) and $\lambda = 0.0415$ 
	(orange, predator extinction phase). The inset shows the local effective decay 
	exponent $\alpha_{\rm eff}(t)$.
	Right: Characteristic relaxation time $t_c$ near the critical point. Left 
	inset: $t_c(\lambda_2)$ after the system is quenched from a quasi-steady state 
	at $\lambda_1 = 0.25 > \lambda_2$ near $\lambda_c = 0.0416$. (The different 
	graphs indicate $t_c$ when $128000$, $64000$, $32000$, and $16000$ MCS elapsed
	after the quench; data averaged over $500$ runs.) Main panel: same data in 
	double-logarithmic form. For $|\tau| = |(\lambda_2 / \lambda_c) - 1| > 0.1$, 
	the different graphs collapse, yielding $z \, \nu = -1.208 \pm 0.167$. Right 
	inset: associated effective exponent $(z \, \nu)_{\rm eff}(\tau)$ that tends 
	towards $z \, \nu \approx 1.3$ as $|\tau| \to 0$ (reproduced with permission 
	from Ref.~\cite{Chen16}).} 
\label{fig:lv_critdyn}
\end{figure}
Examples for numerically determined dynamical critical properties at the LV 
predator extinction threshold $\lambda_c$ in a square lattice with 
$1024 \times 1024$ sites are shown in figure~\ref{fig:lv_critdyn}~\cite{Chen16}: 
The predator density should decay algebraically according to 
$a(t) \sim t^{- \alpha}$, and upon approaching the transition, 
$\tau \sim \lambda - \lambda_c \to 0$, the characteristic relaxation time should 
display critical slowing down $t_c(\tau) \sim |\tau|^{- z \, \nu}$, with critical 
exponents $\alpha \approx 0.45$ and $z \nu \approx 1.295$ for directed percolation 
in two dimensions. The left panel in figure~\ref{fig:lv_critdyn} depicts rather
extensive simulation data for the critical predator density decay, starting from 
either random particle distributions, or following equilibration at higher 
predation rate, corresponding to a (quasi-)steady state in the coexistence region. 
The best estimate for the decay exponent here is $\alpha \approx 0.54$, but the 
asymptotic critical regime is only reached after $\sim 10^5$ MCS. As evident in the
right panel of figure~\ref{fig:lv_critdyn}, data collapse for different time 
durations after the quench from the coexistence state ensues only for 
$|\tau| \geq 0.1$ in this system, giving $z \, \nu \approx 1.2$, but apparently 
extrapolating towards the directed-percolation value as $|\tau| \to 0$. Other
characteristic signatures that can be examined in a straightforward manner in
Monte Carlo simulations on reasonably small lattices include the expected power 
laws in the decay of the survival probability and the growth of the active-site
number directly at the critical point~\cite{Georgiev07}.

\begin{figure}
  \centering
  \includegraphics[width=0.6\columnwidth,clip=untrue]{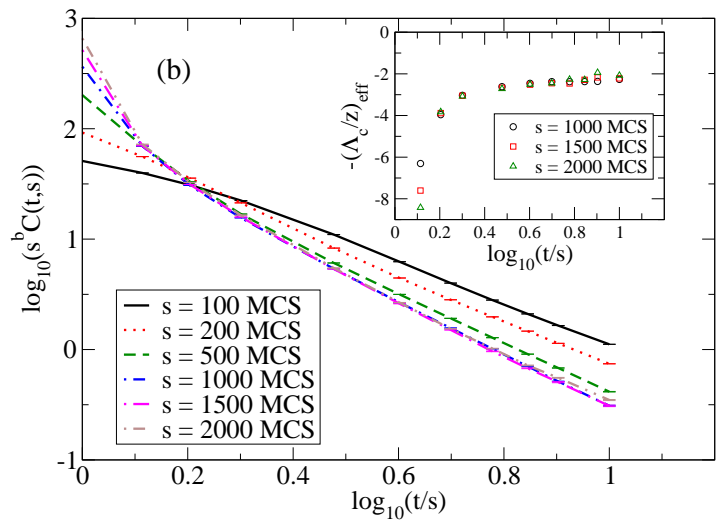}
  \caption{Aging scaling plot (double-logarithmic) for the scaled predator density 
    autocorrelation function $s^b \, C(t,s)$ as a function of the time ratio $t/s$ 
	for various waiting times $s = 100, 200, 500, 1000, 1500, 2000$ MCS at the 
	predator extinction critical point $\lambda_c = 0.0416$ (data averaged over 
	$1000$ independent runs for each value of $s$). The straight-slope section of 
	the curves with waiting times $s \geq 1000$ MCS yields 
	$\Lambda_c / z =  2.37 \pm 0.19$; the aging scaling exponent is found to be 
	$b = 0.879 \pm 0.005$. Inset: effective exponent $-(\Lambda_c /z)_{\rm eff}(t)$
	(reproduced with permission from Ref.~\cite{Chen16}).}
\label{fig:lv_aging}	
\end{figure}
Universal dynamical critical behavior may also be accessed in studies of 
out-of-equilibrium relaxation. The system is then quenched from a fully disordered
initial configuration to the critical point. Since the relaxation time diverges
there, stationarity cannot be reached, and even in finite systems time translation
invariance is broken during an extended time period. In this physical aging region,
two-time autocorrelation functions satisfy the simple-aging dynamical scaling 
form~\cite{Henkel10}
\begin{equation}
\label{eq:agscal}
  C(t,s) = s^{-b} \, {\hat C}(t / s) \, , \quad 
  {\hat C}(x) \sim x^{- \Lambda_c / z} \ .
\end{equation}
As a consequence of the so-called rapidity reversal symmetry 
$\psi({\vec x},t) \leftrightarrow - {\widetilde \psi}({\vec x},-t)$ encoded in
the Reggeon field theory action \eref{eq:dprft}, the aging scaling exponents are
linked to stationary dynamical critical exponents through the scaling relations
$b = 2 \, \alpha$, $\Lambda_c / z = 1 + \alpha + d / z$~\cite{Janssen05,Henkel10},
giving $b \approx 0.9$ and $\Lambda_c /z \approx 2.8$ in $d = 2$ dimensions.
Corresponding scaling plots for the predator density autocorrelation function for
stochastic LV models at the critical point indeed obey equation~(\ref{eq:agscal}) with
$b \approx 0.88$ and $\Lambda_c / z \approx 2.37$~\cite{Chen16}, as demonstrated in
figure~\ref{fig:lv_aging}. Interestingly, these data require simulation runs for 
only $10^4$ MCS, an order of magnitude less than needed to (marginally) establish
stationary dynamic scaling. Critical aging might thus provide a faster indicator
for impending population collapse than critical slowing-down. 

\subsection{Random environmental influences versus demographic variability}

\begin{figure}
  \centering
  \includegraphics[width=0.7\columnwidth]{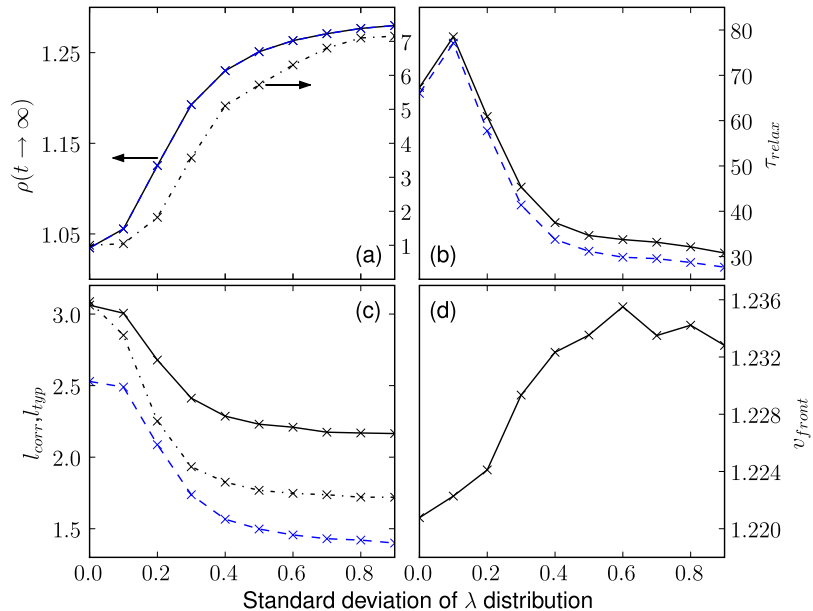}
  \caption{Effect of spatial heterogeneity (environmental variability) governed by
    the width of the standard deviation of the predation rate distribution, 
	determined for fixed $\sigma = 0.5$ and $\mu = 0.5$, on (a) the asymptotic 
	mean population densities of predators (solid lines) and prey (dashed lines) 
	compared to the mean-field prediction (dash-dotted line); (b) the relaxation 
	time towards the (quasi-)steady state; (c) the intra-species correlation
	lengths as well as the typical separation distance between predators and 
	prey; and (d) the front speed of spreading activity rings, measured for 
	$\sigma = 1$ and $\mu = 0.2$ (reproduced with permission from 
	Ref.~\cite{Dobramysl08}).}
\label{fig:lv_variability}
\end{figure}
In spatially extended predator-prey models, an interesting question can be
asked: How does spatial heterogeneity influence the population dynamics? In real
ecosystems, there tend to exist spatial regions in which it is easier for prey to 
hide, while other parts of the system might be beneficial hunting grounds for
predators. There might also be preferred breeding environments in which species
profileration is enhanced, or more hazardous places in which the probability
for species death is higher. Cantrell and Cosner looked at this question by
linearizing a deterministic diffusive logistic equation and using the principal
eigenvalue as a measure of environmental favorability~\cite{Cantrell91,Cantrell98}.
Spatial heterogeneity can be implemented by varying the rates governing the
reaction processes between sites on the simulation lattice.
For example, interesting boundary effects are found in the vicinity of interfaces 
separating active predator-prey coexistence regions from absorbing regions wherein
the predators go extinct. The net predator flux across such a boundary induces a 
local enhancement of the population oscillation amplitude as well as the 
attenuation rate~\cite{Heiba17}.

On the other hand, when the predation rates are treated as quenched random 
variables, affixed to the lattice sites and chosen from a truncated Gaussian 
distribution, remarkably the population densities of both predator and prey 
species are enhanced significantly beyond the baseline densities with homogeneous
rates~\cite{Dobramysl08}, see figure~\ref{fig:lv_variability}. The relaxation time
into the steady state, as well as the inter- and intra-species correlation lengths 
decrease with growing rate variability, which is controlled by the standard 
deviation of the Gaussian distribution. The underlying microscopic mechanism 
behind these effects is the presence of lattice sites with particularly low and 
hence favorable reaction rates which act as prey proliferation sites. In contrast,
heterogeneity in the prey reproduction and predator death rates does not 
significantly affect the species populations~\cite{Dobramysl08}.

\begin{figure}
  \centering
  \includegraphics[width=0.6\columnwidth]{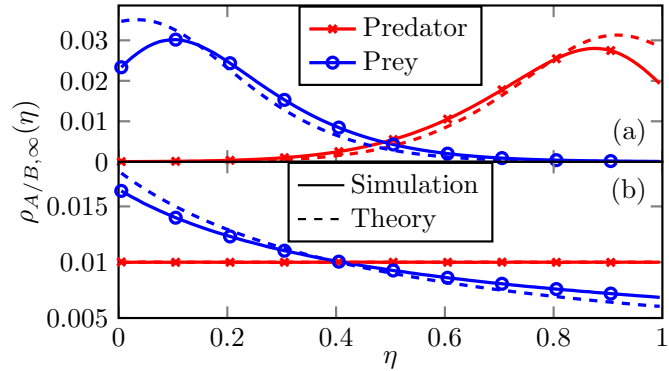}
  \caption{Distribution of predation efficencies at steady state,
    $\rho_{A,\infty}$ (predators) and $\rho_{B,\infty}$ (prey), in a
    predator-prey system with demographic variability and evolutionary dynamics
    for (a) finite correlation between the parent and offspring efficiencies, and 
	(b) uniformly distributed efficiencies. The densities of both species do not 
	fixate at extreme predation efficiencies (reproduced with permission from 
	Ref.~\cite{Dobramysl12}).}
\label{fig:lv_efficacy}
\end{figure}
Compared with spatial heterogeneity and quenched randomness, demographic 
variability plays a different role: Reaction efficiencies of predators and prey,
$\eta_A$ and $\eta_B$ respectively, become traits associated with individuals of
both species~\cite{Dobramysl12} as opposed to fixed, population-based properties. 
During an inter-species predation reaction, these efficiencies are then used to 
construct an instantaneous reaction rate $\lambda$ from the arithmetic mean of the 
individuals' $\eta_A$ and $\eta_B$, selecting for prey with low $\eta_B$ and 
predators with high $\eta_A$. Furthermore, individuals are assigned their
efficiencies at birth, drawn from a truncated Gaussian distribution centered 
around the parent's value of $\eta$. The ensuing coupled population and 
evolutionary dynamics of this system leads to an intriguing optimization of the 
efficiency distributions, shown in figure~\ref{fig:lv_efficacy}, which can also be 
approximated using an adapted multi-quasi-species mean-field 
approach~\cite{Dobramysl13}. Interestingly, the net effect on population densities 
of the evolutionary efficiency optimization is actually essentially neutral. 
Crucially, however, the mean extinction time in small systems is increased more 
than fourfold in the presence of such demographic 
variability~\cite{Dobramysl12,Dobramysl13}. The optimization of efficiency 
distributions is reminiscent of co-evolutionary arms race scenarios: Yoshida 
{\em et al.} studied the consequences of rapid evolution on the predator-prey 
dynamics of a rotifer-algae system, using experiments and simulations via coupled 
non-linear differential equations~\cite{Yoshida03,Yoshida07}.

\section{Cyclic dominance of three-species populations}
\label{sec:cyclc}
Unraveling what underpins the coexistence of species is of fundamental importance to understand and model
the biodiversity that characterizes ecosystems~\cite{biodiversity}. In this context,  the
cyclic dominance between competing species has been proposed as a possible mechanism to explain the persistent species
coexistence often observed in Nature, see, {\it e.g.}
Refs.~\cite{PatternsRPS,Kerr,Nahum,LizardMutation,Jackson}. In the last two decades, these observations have motivated a 
large body of work aiming at studying the dynamics of populations exhibiting cyclic dominance.
The simplest and, arguably,  most intuitive form of cyclic dominance consists of three species
in cyclic competition, as in the paradigmatic rock-paper-scissors game (RPS) - in which rock crushes scissors, scissors cut paper, and paper wraps rock.
Not surprisingly therefore, models exhibiting RPS interactions have been proposed as
paradigmatic models for the cyclic competition between three species
and have been the subject of a vast literature that we are reviewing in this section.

\subsection{Rock-paper-scissors competition as a metaphor of cyclic dominance in Nature}
As examples of populations governed by RPS-like dynamics, we can mention some
 communities of {\it E.coli}~\cite{PatternsRPS,Kerr,Nahum}, {\it Uta stansburiana} lizards \cite{LizardMutation}, as well
as coral reef invertebrates~\cite{Jackson}. In the absence of spatial degrees of freedom and mutations, the presence of
demographic fluctuations in finite populations leads to the loss of biodiversity with the extinction of two species in a
finite time, see, {\it e.g.},~\cite{Frean01,Ifti03,RMFnonspatial1,RMFnonspatial2,Gallas10,He10,He11}.
However, in Nature, organisms typically interact with a finite number of individuals in their neighborhood and are able to migrate. It is by now
well established both theoretically and experimentally that space and mobility greatly influence how species evolve and
how ecosystems self-organize, see {\it e.g.}~\cite{Turing,Murray02,Koch,patterns1,patterns2,patterns3,Weber14}.
The in vitro experiments with \textit{Escherichia coli} of Refs~\cite{PatternsRPS,Kerr,Nahum,Kerr2} have attracted particular
attention because they highlighted the importance of spatial degrees of freedom and local interactions. The authors of Ref.~\cite{Kerr}
showed that, when arranged on a Petri dish, three strains of bacteria in cyclic competition coexist for a long time while
two of the species go extinct  when the interactions take place in well-shaken flasks. Furthermore, in the in vivo experiments of
Ref.~\cite{Kirkup}, species coexistence is maintained when bacteria are allowed to migrate, which demonstrates the evolutionary
role of migration. These findings  have motivated a series of studies aiming at investigating the relevance of fluctuations,
space and movement on the properties of systems exhibiting cyclic dominance.
A popular class of three-species models exhibiting cyclic dominance are those with zero-sum RPS
interactions, where each predator replaces its prey in
turn ~\cite{Tainaka88,Tainaka89,Tainaka94,Frachebourg96,Frachebourg96b,Frachebourg98,Szolnoki02,Berr09,Parker09,He10,Ni10,Venkat10,Winkler10,Dobrinevski12,Juul13,Mitarai16,Avelino17}
and variants of the model introduced by May and
Leonard~\cite{May75}, characterized by cyclic `dominance removal' in which  each predator
`removes' its prey in turn (see below)~\cite{He11,Durrett97,Durrett98,RMF1,RMF2,RMF3,Matti,Jiang09,Wang11,Rulands11}. Particular interest has been drawn to questions concerning the survival statistics
(survival probability, extinction time) and in characterizing the spatio-temporal arrangements of the species.

Here, we first introduce the main models of population dynamics between three species in cyclic competition,
and then review their main properties in
well-mixed and spatially-structured settings.

\subsection{Models of three species in cyclic competition}

In the context of population dynamics, systems exhibiting cyclic dominance are often introduced at an individual-based level
as lattice models (often in two dimensions). Such an approach is the starting point for further analysis and coarse-grained descriptions.
Here, for the sake of concreteness we introduce a class of models exhibiting  RPS-like interactions between three species
by considering a periodic square lattice consisting of  $L\times L$ nodes ($L$ is the linear size of the
lattice) in which individuals of three species, $S_i$ ($i=1,2,3$), are in cyclic competition\footnote{See
Sec.~3.4.3 for a brief
discussion of the dynamics on other topologies, especially the interesting case of one-dimensional lattices.}.  Each node of the lattice is labeled by
a vector ${\bm \ell}=(\ell_1,\ell_2)$ and, depending on the details of the
model formulation, each node is either (i) a boolean random variable; (ii) a patch with a certain carrying capacity, (iii) an island that can accommodate
an unlimited number of individuals. More specifically, these distinct but related formulations correspond to the following settings
\begin{enumerate}
 \item[(i)] Each node  can be  empty or occupied at most by one individual, {\em i.e.}, if
$N_{S_i}({\bm \ell})$ denotes the number of individuals of species $S_i$ at ${\bm \ell}$, we have
$N_{S_i}({\bm \ell})=0$ or $1$ as well as $\sum\limits_i N_{S_i}(\bm \ell) = 0$ or $1$. This formulation  corresponds to a site-restricted model with 
volume exclusion and is sometimes referred to as being `fermionic',
see {\it e.g.} Refs.~\cite{Durrett97,Durrett98,RMF1,RMF2,RMF3,Matti,Jiang09,He11,RF08}
\item[(ii)] Each node is a  patch consisting of a  well-mixed population of species
$S_1, S_2, S_3$ and empty spaces $\emptyset$, with a finite carrying capacity $N$. In this case, we deal with a 
metapopulation model~\cite{Hanski99} and in each patch ${\bm \ell}$ there are
$N_{S_i}({\bm \ell})\leq N$ individuals of species $S_i$  and also $N_{\emptyset}({\bm \ell})=
N-N_{S_1}({\bm \ell})-N_{S_2}({\bm \ell})- N_{S_3}({\bm \ell})$
empty spaces, see, {\it e.g.}, Refs.~\cite{SMR,FigshareMovies,Rulands13,SMR2,cycl-rev,BS,MRS16}.
\item[(iii)] Each lattice site can accommodate  an unlimited number  of 
individuals of each species $N_{S_i}({\bm \ell})=0,1,\dots$ 
(In computer simulations, $N_{S_i}$ is practically capped to a large number),
see, {\it e.g.}, \cite{He10,Rulands11,Labavic16,Serrao17}. This formulation  corresponds to a site-unrestricted model 
(no volume exclusion) and is sometimes referred to as being `bosonic',  see Sec. 2.4.
\end{enumerate}

In most models, cyclic dominance between the species  $S_i$ ($i=1,2,3$) is implemented through
one or both of the following binary reactions among nearest-neighbors:
\begin{eqnarray}
\label{dom-rem}
  &S_i + S_{i+1} \to  S_i + \emptyset \ , \quad &{\rm \; at \; rate} \ \sigma_i   \\
\label{dom-repl}
  &S_i + S_{i+1} \to S_i + S_i \ , \quad &{\rm \;  at \; rate} \ \zeta_i \ ,
\label{lvreac}
\end{eqnarray}
where the  index $i \in \{1,2,3\}$ is ordered cyclically such that $S_{3+1}
\equiv S_1$ and $S_{1-1} \equiv S_3$. The reactions (\ref{dom-rem}) account
for {\it dominance--removal} with rate $\sigma_i$ while the scheme  (\ref{dom-repl}) accounts for
{\it dominance--replacement} with rate $\zeta_i$. While many works have focused exclusively either on the reactions
(\ref{dom-rem}), see {\it e.g.}
Refs.~\cite{Tainaka89,Tainaka94,Frachebourg96,Frean01,Szolnoki02,Ifti03,Berr09,He10,Winkler10,Dobrinevski12,Juul13,Mitarai16},
or only on (\ref{dom-repl}) as in Refs.~\cite{Durrett97,Durrett98,RMF1,RMF2,RMF3,Rulands11,He11}, it is convenient for the purpose of
this review to consider
the generic approach of Refs.~\cite{RF08,Jiang11,SMR,FigshareMovies,Rulands13,SMR2,BS,MRS16,Alastair16} and discuss
cyclic dominance in terms of the joint presence of the independent removal (\ref{dom-rem}) and replacement (\ref{dom-repl}) processes. In addition to cyclic dominance, we also consider
the processes of reproduction (with rate $\beta_i$) and mutation (with rate $\mu_i$) according to the schemes:
	\begin{eqnarray}
	\label{rep}
		&S_i + \emptyset \to S_i + S_i \ , \quad &{\rm \; at \; rate} \ \beta_i \\
	\label{mut}
		&S_i \to S_{i\pm1},  \quad &{\rm \; at \; rate} \ \mu_i \ .
	\end{eqnarray}
In principle, cyclic dominance, reproduction and mutation occur with different rates $(\sigma_i, \zeta_i, \beta_i, \mu_i)$
for each species, but here as in the vast majority of other works, and unless stated otherwise, we simply assume
 $(\sigma_i, \zeta_i, \beta_i, \mu_i)= (\sigma, \zeta, \beta, \mu)$, {\em i.e.}, the same reaction rates for each species.

Furthermore, to account for the fact that individuals can move, the models are endowed with spatial degrees of
freedom by allowing individuals to migrate from one node ${\bm \ell}$ to a neighboring site ${\bm \ell}'$
according to pair-exchange and hopping processes:
	\begin{eqnarray}
	\label{exch}
	&\big[X \big]_{{\bm \ell}} \big[Y \big]_{{\bm \ell}'} \to \big[Y	\big]_{{\bm \ell}} \big[X \big]_{{\bm \ell}'}\ ,
	\quad &{\rm  \; at \; rate} \ \delta_E\\
	\label{hop}
	&\big[X \big]_{{\bm \ell}} \big[\emptyset \big]_{{\bm \ell}'} \to \big[\emptyset	\big]_{{\bm \ell}} \big[X \big]_{{\bm \ell}'}\ ,
	\quad &{\rm \; at \; rate} \ \delta_D \ ,
	\end{eqnarray}
where $X\neq Y\in \{S_1, S_2, S_3\}$. It is worth noting that $\delta_E=\delta_D$ corresponds to the simplest form of
movement in which an individual (or a void $\emptyset$)  in ${\bm \ell}$ is
swapped with any other individual (or $\emptyset$) in  ${\bm \ell}'$~\cite{RMF1,RMF2,RMF3,RF08,SMR,FigshareMovies,Rulands13,SMR2,BS,MRS16}. When $\delta_E\neq \delta_D$ the pair-exchange and hopping processes
are divorced yielding non-linear diffusion effects~\cite{He10,SMR,SMR2,BS}. This mimics the fact that
 organisms rarely move purely diffusively, but rather sense and respond to their environment~\cite{Kearns}. By divorcing
 (\ref{exch})-(\ref{hop}), we can  discriminate between the movement in crowded regions, where mobility is
 dominated by pair-exchange, and mobility in diluted regions.
The individual-based models (\ref{dom-rem})-(\ref{hop}) are defined by the corresponding Markov processes
and the dynamics is governed by the underlying master equation~\cite{Gardiner,VanKampen92,PaulEliSid,Tauber14}. However, solving the
master equation is a formidable task, and one needs to rely on a combination of analytical approximations and
simulation techniques to make progress.

%%%%%%%%%%%%%%%%%
\subsection{Cyclic dominance in well-mixed populations}
%************

In the absence of spatial structure, the population is `homogeneous' or `well-mixed' and the analysis
is greatly simplified by the fact that in this case all individuals are nearest-neighbors and therefore interact
with each other. When the population size is infinitely large, the dynamics is aptly described by its deterministic
rate equations, whose predictions are dramatically  altered by demographic fluctuations when the population size is finite.
\subsubsection{Mean-field analysis \& relations with evolutionary game theory.}
In the limit of very large and spatially unstructured populations, the species densities can be treated as continuous variables
and any random fluctuations and correlations can be neglected. In such a mean-field 
setting, the main cyclic dominance scenarios are covered
by the rate equations of the generic model (\ref{dom-rem})-(\ref{mut}):
\begin{eqnarray}
	\label{MF}
	\hspace{-0.75cm}
	\frac{d}{dt} s_i &=& s_i[\beta(1-\rho) - \sigma s_{i-1}] + \zeta s_i[s_{i+1} - s_{i-1}]+
		\mu \left[s_{i-1} + s_{i+1} - 2 s_i\right],
	\end{eqnarray}
where $s_i$ denotes the density of species $S_i$,  $\rho=s_1+s_2+s_3$, and the indices are ordered cyclically.
It is  worth noting that the processes (\ref{exch})-(\ref{hop}) are obviously absent from the non-spatial rate equations (\ref{MF}). 
The deterministic description in terms of (\ref{MF}) is  widely used because of its simplicity. 
Here, it allows us to establish a connection with evolutionary game theory
~\cite{Maynard82,Hofbauer98,Nowak06,Szabo07,Frey10},
see below. It is also worth noting that equations~(\ref{MF}) are characterized by a steady state ${\bm s}^*=(s_1^*,s_2^*,s_3^*)$ 
at which all species coexist with the same density $s_i^*=\beta/(\sigma + 3\beta)$ and, when there are no mutations ($\mu=0$),  equations~(\ref{MF}) admit 
also three absorbing fixed points
${\bm s}=\{(1,0,0), (0,1,0), (0,0,1)\}$ corresponding to a population consisting of only one species.
The rate equations (\ref{MF}) encompass three main types of oscillatory dynamics around the coexistence state ${\bm s}^*$:\\
\vspace{0.1cm}
(a) The case $(\sigma, \zeta, \beta, \mu)=(0, \zeta, 0,0)$ corresponds
to the so-called {\it cyclic Lotka--Volterra model} (CLVM), see, {\it e.g.}, \cite{Tainaka89,Tainaka94,Frachebourg96,RMFnonspatial1,RMFnonspatial2}, which coincides with the {\it zero-sum
RPS game}, arguably the most popular version of this game. The latter is
a symmetric two-player game with three pure strategies, $S_1, S_2$ and $S_3$, corresponding to playing, say,
paper, rock, and scissors
using the payoff matrix~\cite{Maynard82,Hofbauer98,Nowak06,Szabo07,Frey10}
\begin{eqnarray}
\label{payoffM}
{\bm \Pi}=
\begin{tabular}{c|c c c}
vs & $S_1$ (paper) & $S_2$ (rock) & $S_3$ (scissors) \\
 \hline
$S_1$ (paper) &  $0$ & $+1$ & $-1$ \\
$S_2$ (rock) & $-1$ & $0$  & $+1$\\
$S_3$ (scissors) & $+1$ & $-1$  & $0$.\\
\end{tabular}
\end{eqnarray}
Without loss of generality, we have set $\zeta=1$.
The payoff matrix ${\bm \Pi}$ prescribes that $S_1$ (paper) dominates $S_2$ (rock), and gets a payoff $+1$ when it plays 
against it, while it is dominated by $S_3$ (scissors)
and get a payoff $-1$ by playing against it, etc. Here, ${\bm \Pi}$ is antisymmetric and we have a zero-sum game, {\em i.e.},
what a dominating strategy gains is exactly what the dominated  one loses; a fact that has far-reaching consequences such as
the existence of a nontrivial conserved quantity~\cite{Hofbauer98}.
According to the general tenets of evolutionary game theory, the dynamics of this evolutionary game is
formulated in terms of the {\it replicator equations}~\cite{Maynard82,Hofbauer98,Nowak06,Szabo07,Frey10,Broom13}. These are obtained by considering that each
strategy $S_i$ is played with a frequency $s_i$, and by computing the expected payoff $\Pi_i$ of a $S_i$-player:
$\Pi_i=({\bm \Pi}{\bm s})_i=s_{i+1}-s_{i-1}$ (cyclically ordered indices), while
the average payoff $\bar{\Pi}=\sum_i s_i \Pi_i=0$ vanishes (zero-sum game). Hence, the underlying
replicator equations are $\dot{s_i}=s_i[\Pi_i -\bar{\Pi}]=s_i[s_{i+1}-s_{i-1}]$ and coincide with (\ref{MF}).
In this case, the rate (replicator) equations (\ref{MF}) admit two constants of motion: the total density
$\rho=s_1+s_2+s_3$, here assumed to be set to one~\footnote{In this case,  state $\emptyset$ plays no role in the dynamics and can
simply assumed to be non-existent.},
and $s_1(t)s_2(t)s_3(t)$ are conserved. This leads to
{\it neutrally stable closed orbits} around  ${\bm s}^*$
 which is a marginally stable center. The three absorbing states ${\bm s}=\{(1,0,0), (0,1,0), (0,0,1)\}$
  are unstable, but, as discussed below, play an important role
 when the population size is finite. Interestingly, the zero-sum RPS dynamics has recently been generalized to a
 multi-species class of zero-sum games described by so-called
 antisymmetric Lotka--Volterra equations whose long-term (deterministic)
behavior has been classified in terms of the properties of the network's interaction matrix~\cite{Knebel13}. These
findings that are relevant to describe the phenomenon of condensation in some driven-dissipative quantum systems~\cite{Knebel15}.
 \\
 (b) When $\beta, \sigma>0$, $\zeta\geq 0$ and $\mu= 0$, equations~(\ref{MF}) can be recast as the rate equations of the
 {\it May--Leonard model} (MLM)~\cite{May75}. The coexistence steady state ${\bm s}^*$ is thus unstable  and the trajectories form
 {\it heteroclinic cycles}.
 The mean-field dynamics is thus similar to that of a generalized {\it non-zero-sum} RPS game where predators lose more than what
 their prey gain ({\it e.g.} `paper' obtains a payoff $+1$ against `rock', but `rock' gets a payoff
 less than $-1$ against `paper'
 \footnote{An example of payoff matrix for such a game is obtained by replacing the entries $-1$ of (\ref{payoffM})
 by $-\epsilon$, with $\epsilon>1$~\cite{Mobilia10}}.). It is noteworthy that in the  case without
 dominance-replacement ($\zeta=0$) the heteroclinic cycles are degenerate~\cite{May75}.\\
 (c) When $\beta,\sigma>0$, $\zeta\geq 0$ and $\mu>0$, with the mutation rate used as bifurcation parameter,
 a supercritical Hopf bifurcation arises at  $\mu_H=\beta \sigma/(6(3\beta +\sigma))$.
 The mean-field dynamics is thus characterized by a stable limit cycle around ${\bm s}^*$ when $\mu<\mu_H$, whereas the
 coexistence state ${\bm s}^*$ is asymptotically stable when $\mu>\mu_H$~\cite{Mobilia10,Andrae10,SMR,FigshareMovies,SMR2,BS,cycl-rev,MRS16,Toupo15,Yang17}.
 It is also worth noting that the mean-field dynamics of the `bosonic' models of Refs.~\cite{Labavic16,Serrao17}, where the schemes
 (\ref{dom-rem}), (\ref{rep}), (\ref{exch}) are supplemented by the   reactions $S_i+S_i \to S_i$~\footnote{In the `bosonic' formulation of 
  Refs.~\cite{Labavic16,Serrao17}, the coagulation  reactions $S_i+S_i \to S_i$ are introduced  to effectively limit the population size growth 
  in a way that is more suitable for 
  a perturbative analysis than the strict (`fermionic') site restrictions, see Sec. 2.5.}, is also 
  characterized by a Hopf bifurcation, see Sec.~2.5.
\subsubsection{Cyclic dominance in finite well-mixed populations.}
\begin{figure}
	\centering
	\includegraphics[width=0.55\linewidth,clip=untrue]{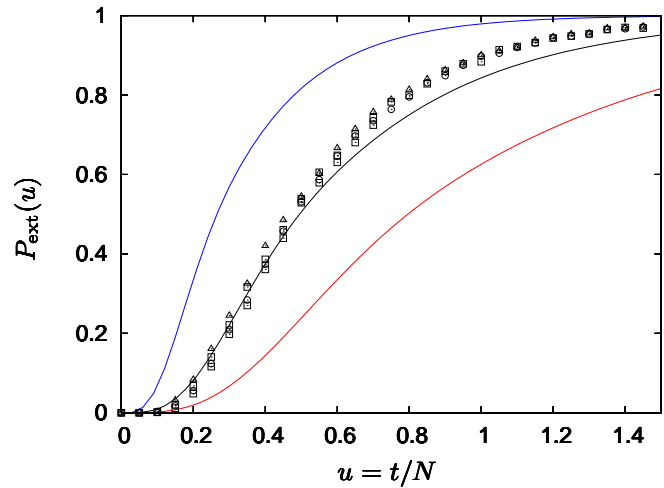}
	\caption{Extinction probability  $P_{\rm {ext}}(t)$ (starting at ${\bm s}^*$) in the CLVM 
	with $\zeta_i=\zeta$ as a function of
the rescaled time $u=t/N$: Symbols are stochastic simulation results  for different system sizes ($N=100$:
triangles; $N=200$: boxes; $N=500$: circles) and solid blue (dark gray) and red (light gray)
lines are analytical upper and lower bounds of $P_{\rm {ext}}(t)$, while the black line is the average of these upper and lower
bounds. For details, see Ref.~\cite{RMFnonspatial1}
from which this figure is reproduced with permission.}
	\label{Pext}
\end{figure}

The evolution in well-mixed populations of finite size  $N <\infty$ is usually formulated in terms of
birth-and-death processes, see, {\em e.g.},~\cite{Gardiner}, describing agents formally interacting on a complete
graph. The  pairwise interactions between a finite number of discrete individuals lead to fluctuations that
 alter the mean-field predictions. In particular, the Markov chains associated with the CLVM and MLM admit absorbing states
and these are unavoidably reached causing the extinction of all but one species, with the surviving species that takes over and 
`fixates' the entire population.
Hence,  the above deterministic scenarios (a) and (b) are dramatically
modified by demographic fluctuations since the dynamics in a finite population always leads to
the survival of one species and the extinction of the two others~\cite{Frean01,Ifti03,RMFnonspatial1,RMFnonspatial2}.
Questions of great importance, that have been studied in detail, concern the survival or fixation probability. The former
refers to the probability that, starting from a certain initial population composition,
a given species survives after an infinitely long time. In the presence of absorbing states in the CLVM and MLM,
the only surviving species fixates the population. A related question of great interest is the (unconditional) 
mean extinction time which is the average time that is necessary for two of the 
species to go extinct while the third survives (and fixates the population).
The question of  survival probability  is particularly interesting  when the species have different reaction rates:
For the CLVM with asymmetric rates,  when $N$ is large but finite, it has been shown that the species with the 
smallest dominance rate (the ``weakest species'') is the most likely to
fixate the population by helping the predator of its own predator~\cite{Frean01,Berr09}, a phenomenon referred to as the `law of 
the weakest'. A similar phenomenon
has also been found in other three-species models exhibiting cyclic dominance, see, {\it e.g.}, Ref.~\cite{Gallas10} and 
Ref.~\cite{Tainaka93} where a version of the `law of the weakest' was found in a two-dimensional CLVM with mutation. However,
 it has also been shown that no law  of the weakest holds when the number of species in cyclic competition is more 
than three~\cite{Case10,Durney11,Knebel13}, see also
Sec.~\ref{sec:mulsp}, and that this law is generally not followed when the rates are subject to external fluctuations~\cite{West17}.
The existence of quantities conserved by the CLVM  rate equations has been exploited to compute the
mean extinction time,  found to scale linearly with the system size~\cite{Ifti03,
RMFnonspatial1,RMFnonspatial2}, and the extinction  probability $P_{\rm {ext}}(t)$ giving the probability that,
starting at ${\bm s}={\bm s}^*$, two of the three species go extinct after a time
$t$~\cite{RMFnonspatial1,RMFnonspatial2,Parker09,Dobrinevski12},
see Figure~\ref{Pext}.
 Some other aspects of species extinction in well-mixed
 three-species models exhibiting cyclic dominance have been considered for instance
 in Refs.~\cite{Claussen08,Galla11},
 and the quasi-cycles arising in these systems have been studied in Refs.~\cite{RMFnonspatial1,Mobilia10}.
 In particular, the quasi-cyclic behavior around the  coexistence fixed point of the three-species
 cyclic model  with mutations of Ref.~\cite{Mobilia10} was investigated by computing the power-spectrum
 and the mean escape time from the coexistence fixed point. In  related models, the authors of Ref.~\cite{Andrae10}
 studied  the entropy production in the nonequilibrium steady state   while 
in Ref.~\cite{Yang17}  it is shown that demographic noise slows down  the quasi cycles of dominance. In Ref.~\cite{Knebel13}, it is shown that for a class of multi-species 
 zero-sum systems the mean-time for the extinction of one species scales linearly with the population size $N$ when the mean-field dynamics
 predicts the coexistence of all species (and logarithmically with $N$ otherwise), see also Sec.~\ref{sec:mulsp}.

\subsection{Cyclic three-species competition in structured populations}
Most ecosystems are spatially extended and populated by individuals that move and interact locally. It
is therefore natural to consider spatially-extended models of populations in cyclic competition.
When spatial degrees of freedom are taken into account, the  interactions between individuals are limited to their 
neighborhood and this restriction has far-reaching consequences that have been observed experimentally. In fact, since prey may avoid to encounter their
predators, species can coexist over long periods. Furthermore, long-term species coexistence is often accompanied by the
formation of spatio-temporal
patterns such as  propagating fronts or spiral waves, see, 
{\it e.g.},~\cite{RMF1,RMF2,RMF3,He11,SMR,Lamouroux12,Rulands13,SMR2}, that, for instance, have been
experimentally observed for Dictyostelium mounds and in Myxobacteria ~\cite{dictyo,myxo}.

Spatially-extended models in which immobile agents of three species  occupy the sites of a lattice and interact according to the schemes
(\ref{dom-rem})-(\ref{rep}) have received significant interest, see,
{\it e.g.},~\cite{Tainaka88,Tainaka89,Tainaka94,Frachebourg96,Frachebourg96b,Durrett97,Durrett98,Frachebourg98,Szolnoki02,Rulands11}.
For instance, the authors of Refs.~\cite{Frachebourg96,Frachebourg96b} considered the CLVM with immobile individuals ({\em i.e.},
with $\zeta>0$ and $\sigma=\beta=\mu=\delta_{D/E}=0$) and showed that spatial inhomogeneities develop
on a one-dimensional lattice: A coarsening phenomenon occurs with the formation of a mosaic of single-species domains with algebraically
growing size\footnote{The authors of Ref.~\cite{Frachebourg98} also considered the multi-species CLVM with immobile individuals on a hypercubic lattice and
determined the number of species above which a frozen state is attained.}. 
Spatial degrees of freedom also allow us to consider elementary processes associated with species' movement such as (\ref{exch}) and (\ref{hop}).
This  is particularly important in biology where  migration has been found to have a profound impact on the maintenance of biodiversity, see, {\it e.g.},
Refs~\cite{Kerr2,Kirkup}. In the last two decades, much effort has been dedicated to investigating various aspects of spatially-extended models combining
reactions like (\ref{dom-rem})-(\ref{mut}) and individual's movement typically modeled by   (\ref{exch}) and/or (\ref{hop}).
In this context, important issues, both from theoretical and biological perspectives, are:
\begin{enumerate}
{\it
 \item[-] What is the influence of mobility on the survival/coexistence of the species?
 \item[-]  Is the well-mixed scenario recovered when the individuals are highly mobile?
 \item[-]  How does mobility affect the spatio-temporal organization of the population?
}
\end{enumerate}
These questions have been addressed with many variants of the RPS-like systems (\ref{dom-rem})-(\ref{hop}),
and in particular in the framework of the MLM and CLVM. Some of the main findings are reviewed below.

A spatially-extended version of population dynamics with RPS-like interactions is generally formulated
at an individual-based level on a regular lattice, most often in two dimensions, which is the natural
 biologically-relevant choice for the interaction network.
In such a setting, a spatially-extended RPS-like model is defined by the processes
(\ref{dom-rem})-(\ref{hop})  implemented on the appropriate lattice (grid or array of patches of finite/infinite
carrying capacity). The model's dynamics is thus formally described by the
underlying master equation that appears to be intractable on the face of it.
Yet, when fluctuations can be neglected and the linear size $L$ of the lattice is large, the
 spatial population dynamics of RPS-like systems on a square domain
can often be well described in the continuum limit ($L\to \infty$)  by  the following set of partial differential
equations for the local densities  $s_i\equiv s_i({\bm x},t)$:
	\begin{eqnarray}
	\label{RPSpde}
		\partial_t s_i
		&=& D_E \Delta s_i + (D_E-D_D) (\rho\Delta s_i -s_i \Delta \rho) \nonumber\\
		&+&s_i[\beta(1-\rho) - \sigma s_{i-1}] + \zeta s_i[s_{i+1} - s_{i-1}] \nonumber\\ &+&
		\mu \left[s_{i-1} + s_{i+1} - 2 s_i\right],
	\end{eqnarray}
with periodic boundary conditions. Here, the density
$s_i\equiv s_i({\bm x},t)$ of species $S_i$
is a continuous variable, with $\rho=s_1+s_2+s_3$
and position vector
${\bm x}=(x_1,\dots,x_d)$.
The $d-$dimensional Laplacian is
$\Delta=\sum_{i=1}^{d} \partial_{x_i}^2$. As in most studies we focus  on the two-dimensional case, $d=2$, but see also Sec.~3.4.3.
The diffusion coefficients $D_{E/D}$ in (\ref{RPSpde}) and the migration rates $\delta_{E/D}$ of (\ref{exch}), (\ref{hop})
are simply related by $D_{E/D}=\delta_{E/D}/L^2$.
In the first line on the right-hand-side of (\ref{RPSpde}), we recognize the diffusion terms
while the second and third lines respectively correspond to the processes of cyclic dominance (\ref{dom-rem}),(\ref{dom-repl}) and mutation (\ref{mut}). It is worth
 noting  that non-linear diffusive terms arise when the pair-exchange and hopping processes (\ref{exch}), (\ref{hop}) are divorced and
  $\delta_E\neq \delta_D$~\cite{SMR,SMR2,BS}, whereas regular (linear) diffusion occurs when any neighboring pairs
  are exchanged with rate $\delta_E= \delta_D$~\cite{RMF1,RMF2,RMF3,RF08,Matti,Jiang09,Alastair16}. It is also noteworthy
  that within the metapopulation formulation
  of Refs.~\cite{SMR,FigshareMovies,SMR2,cycl-rev,BS,MRS16},
equations~(\ref{RPSpde}) arise at lowest order in an expansion in the inverse of the carrying capacity.

\subsubsection{Mobility promotes and jeopardizes biodiversity in models with rock-paper-scissors interactions.}
\begin{figure}
	\centering
	\includegraphics[width=0.8\linewidth]{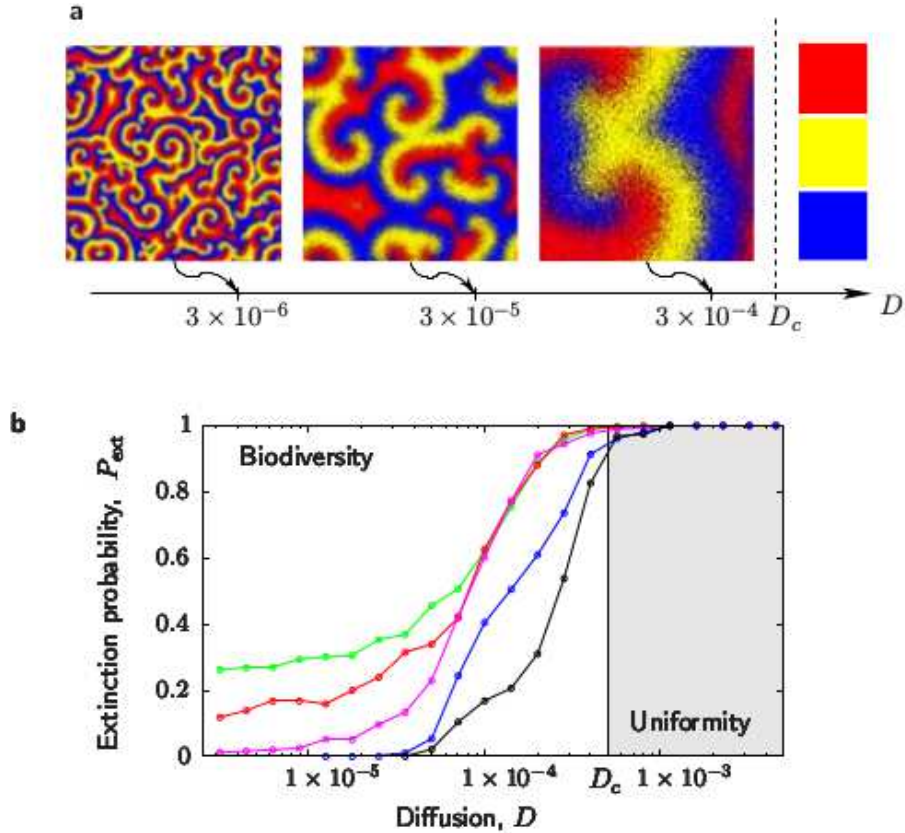}
	\caption{The diffusion coefficient $D$ in the two-dimensional MLM with
	rates $\sigma= \beta=1,\zeta=\mu=0$ and
	$\delta_D=\delta_C=\delta$. $D_c$ is the critical value, see text. Initially, individuals are randomly distributed.
	{\bf a}, Snapshots from lattice simulations of typical states of the system after
	long temporal development ($t\sim L^2$) and for different values of $D$
(each color represents one  species, black dots indicate empty spots). Increasing $D$  (from left
to right), the spiraling patterns grow, and  outgrow the system size when the diffusion coefficient
exceeds $D_c$: biodiversity is  lost above $D_c$, see text.
{\bf b}, The extinction probability $P_{\rm ext}$  from stochastic simulations after a waiting time $t=L^2$
as function of the diffusion $D$, computed  for different system sizes:
$N=20\times 20$ (green), $N=30\times 30$ (red), $N=40\times 40$  (purple), $N=100\times 100$ (blue), and $N=200\times 200$ (black),
from top to bottom (left side of the figure).
 %. As the system size increases, the  transition from stable coexistence ($P_\text{ext}=0$) to extinction ($P_\text{ext}=1$) sharpens  at a critical mobility  $M_c\approx(4.5\pm0.5)\times 10^{-4}$. \label{critical_D}
 Adapted from Ref.~\cite{RMF1}
}
	\label{Fig1RPS}
\end{figure}

The intriguing role of migration, is well exemplified by a series of {\it in vitro}
and {\it in vivo}
experiments:
The authors of Refs.~\cite{PatternsRPS,Kerr,Kerr2,Nahum} showed that when arranged on a Petri dish, three
strains of bacteria in cyclic  competition coexist for a long time while two  species go extinct
when the interactions take place in well-shaken flasks. On the other hand, in Ref.~\cite{Kirkup} it was shown
that mobility allows the bacterial  colonies in the intestines of co-caged mice  to migrate between mice which help
maintain the coexistence of bacterial species. Theoretical aspects related to the above questions  have been addressed in
a series of works~\cite{RMF1,RMF2,RMF3} on the two-dimensional MLM with at most one individual per lattice site 
(i.e. $N_{S_i}({\bm \ell})=0$ or 
$1$) and symmetric rates, {\em i.e.}, defined by
 (\ref{dom-rem})-(\ref{hop}) with $(\sigma_i, \zeta_i, \beta_i, \mu_i,\delta_E,\delta_D)=(\sigma,0,\beta,0,\delta,\delta)$,
that have demonstrated the critical impact of mobility (or migration) on biodiversity. In fact, by considering
the above spatially-extended model defined with only dominance-removal, no mutation  and
exchange between any pairs of neighbors,
it was shown  that below a critical mobility threshold $D_c$ all species coexist in a long-lived quasi-stationary state
and form spiraling patterns, whereas biodiversity is lost above the mobility threshold with only one surviving
species~\cite{RMF1,RMF2,RMF3}.
This phenomenon was analyzed by combining lattice simulations, with a description in terms of stochastic partial
differential equations and a complex Ginzburg--Landau equation \cite{CGLE} derived from equations~(\ref{RPSpde}) by approximating heteroclinic orbits
with limit cycles.  By exploiting the properties of the complex Ginzburg--Landau equation,
it has been  shown that the extinction probability in the MLM is  $P_{{\rm ext}}(t)\approx 0$ at  $t=L^2 \gg 1$
when $D=\delta/L^2<D_c$. In this case
all species coexist and form spiral waves, whereas  only one species survives when
$D>D_c$ and $P_{{\rm ext}}(t=L^2)\approx 1$, see Fig.~\ref{Fig1RPS}. Upon estimating  the wavelength $\lambda$ of the
spiraling patterns, the critical diffusion coefficient $D_c$ and the diagram allowing to identify the species coexistence
phase were obtained~\cite{RMF1,RMF2,RMF3}\footnote{When $\sigma=\beta=1$,
$D_c\approx (2.25 \pm 0.25)\times 10^{-4}$~\cite{RMF1,RMF2,RMF3}.}.
A similar analysis was then extended to the case of cyclic
dominance-removal and dominance-replacement  with linear mobility and no mutations, {\em i.e.}, for the scheme (\ref{dom-rem})-(\ref{hop}) with  rates
$(\sigma_i,\zeta_i,\beta_i, \mu_i, \delta_D,\delta_E)=(\sigma,\zeta,\beta, 0, \delta,\delta)$~\cite{RF08,Rulands13}.

The experimental findings of Refs.~\cite{Kerr,Kirkup} and theoretical results of Refs.~\cite{RMF1,RMF2,RMF3} suggest
that {\it mobility can both promote and  jeopardize biodiversity in systems with RPS interactions}.
Interestingly, recent experiments and agent-based simulations of the range expansion
of the {\it E.coli} communities used  in Ref.~\cite{Kerr}
showed that cyclic competition alone is not sufficient to guarantee species coexistence
in a two-dimensional expanding population: In this case, coexistence depends strongly on the diffusion of the toxin, the
composition of the inoculum, and the relative strain growth rates~\cite{Weber14}.

The oscillatory dynamics characterizing the metastable quasi-stationary state of the two-dimensional MLM
has also been studied by computing the species density correlation functions and the Fourier transform of the
densities~\cite{RMF1,RMF2,RMF3,He11}. In Ref.~\cite{He11}, it was shown that for the two-dimensional MLM
the above results are robust against quenched disorder in either the reaction rates or mobility rates.
Furthermore, the mean extinction time
$T_{\rm ex}$ (as the mean time for the first species to go extinct) in the  two-dimensional MLM with linear diffusion
($\delta_D=\delta_E=\delta$) was found to grow exponentially with the lattice size, {\em i.e.}, $\ln{(T_{\rm ex})} \sim L$, when
$D<D_c$, whereas  $T_{\rm ex} \sim L^2$ when $D>D_c$~\cite{He11}. Further properties of the spiral waves characterizing
the species coexistence in the MLM have been investigated. For instance, it has been shown that in the two-dimensional MLM
(with $\delta_D=\delta_E$) a pacemaker (localized periodic input of the three species) is able to maintain
target waves that spread across the entire population~\cite{Jiang09} (see also Ref.~\cite{Wang11}). For the same two-dimensional
MLM,  it has also  been found that when the rate of cyclic dominance $\sigma$ exceeds a critical value
(the other rates being kept fixed) the spiral waves become unstable Ref.~\cite{Jiang11} (see
also Refs.~\cite{SMR2,BS} and Sec.~3.4.2).

The spatially-extended CLVM with mobile individuals and reaction rates
$(\sigma_i, \zeta_i, \beta_i, \mu_i)=(0,\zeta,0,0)$
has also been extensively studied, both in the case of
linear diffusion ($\delta_D=\delta_E$) and non-linear mobility ($\delta_D>0, \delta_E=0$),
and while all species were still found to coexist in a long-lived quasi-stationary state, they
do not form coherent patterns, see, {\it e.g.},\cite{Matti,He10}. In particular, no spiraling patterns
have been observed in the CLVM at either high or low mobility rate:
By means of  an approximate mapping onto a complex Ginzburg-Landau equation,  it has been argued that
CLVM cannot sustain spiral waves, which is in stark contrast with the properties
of the MLM~\cite{Matti}. Yet, the effect of the range of the cyclic dominance and migration on the dynamics of the two-dimensional CLVM 
was investigated in Ref.~\cite{Ni10}, where spatio-temporal patterns (spiral and plane waves) were found 
in regimes characterized by interactions of sufficiently large range (see also Ref.~\cite{Avelino17}).
The oscillatory dynamics in the metastable coexistence state of the two-dimensional CLVM was
studied by computing the time-dependent densities, their Fourier transform and the two-point correlation
functions, with results that were found to be  robust against quenched disorder in the
reaction rates and in the presence/absence of site restriction~\cite{He10}. In that work,
the mean extinction time in the spatial CVLM was found to grow exponentially with the system size in two dimensions, {\em i.e.},
$\ln{T_{\rm ex}}\sim L$. For extreme choices of the reaction rates, where one species pair reacts much faster than
the other two pairs, such that the system is effectively set in one `corner' of parameter space,
it has been demonstrated that the resulting stochastic dynamics reduces to the two-species
Lotka--Volterra predator-prey model~\cite{He12}.

In Ref.~\cite{RF08}, it was found that a two-dimensional model
combining dominance-removal and dominance-replacement with linear diffusion
(and site restriction) can lead to stable spiraling patterns, as well as to convectively and absolutely unstable spiral waves. This picture was
complemented and unified in Refs.~\cite{SMR,SMR2,cycl-rev,BS}, as reviewed below.

\subsubsection{Spiraling patterns and the complex Ginzburg--Landau equation.}
\label{SpiralsCGLE}
\begin{figure}
	\centering
	\includegraphics[width=0.75\linewidth]{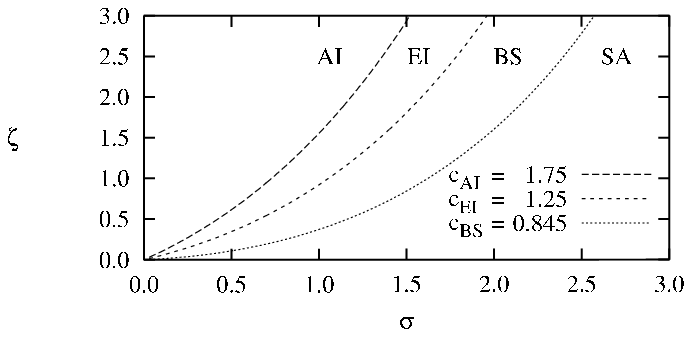}\\
	\includegraphics[width=0.225\linewidth]{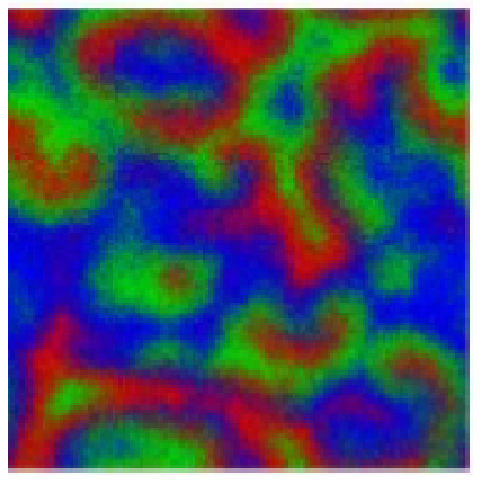}
	\includegraphics[width=0.225\linewidth]{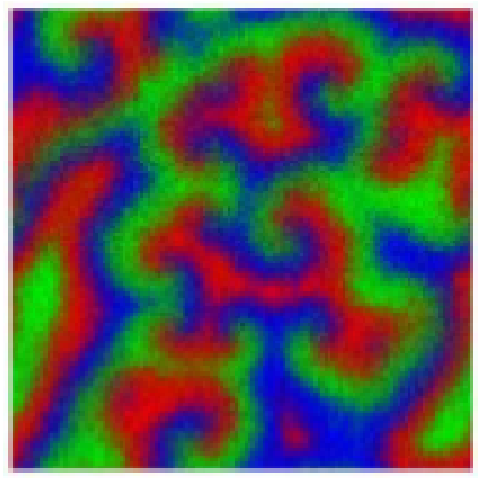}
	\includegraphics[width=0.225\linewidth]{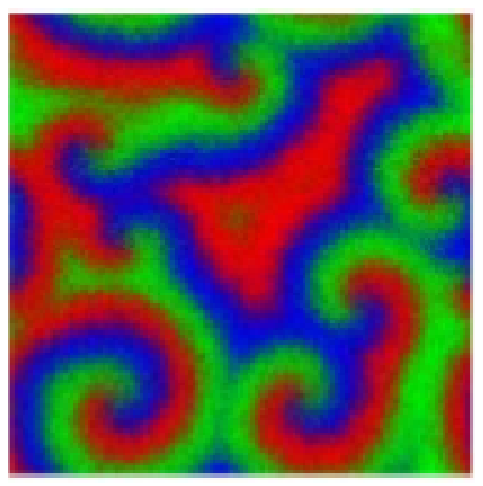}
	\caption{Top: Phase diagram of the two--dimensional RPS system around the Hopf bifurcation
	with contours of $c = (c_{{\rm AI}}, c_{{\rm EI}}, c_{{\rm BS}})$ in the $\sigma-\zeta$
	plane, see text. We distinguish four phases: spiral waves are unstable in AI, EI and SA phases,
	while they are stable in BS phase. The boundaries between the phases have been obtained using the
	 parameter $c$ given by equation~(\ref{c3}). Adapted from Ref.~\cite{SMR2}.
	Bottom: Typical long-time snapshots in the AI (left), EI (middle) and BS (right) phases
from stochastic simulations  of the metapopulation model (\ref{dom-rem})-(\ref{hop}) at low mutation rate, see text. Each color represents one species (dark dots are regions of low density).
 The parameters are $L=128$, $N=64$, $(\beta,\sigma,\mu,\delta_D,\delta_E)=(1,1,0.001,1,1)$, and $\zeta=1.8$ (left),
$1.2$ (middle), and $\zeta=0.6$ (right). In all panels, the initial condition is a random
perturbation of $\bm{s}^*$.
	Adapted from Ref.~\cite{MRS16}.}
 	\label{phasesCGLE}
\end{figure}

The two-dimensional versions of the generic model (\ref{dom-rem})-(\ref{hop})
are often characterized by the emergence of long-lived spiraling patterns that can rotate 
 clockwise or  anticlockwise  whose vortices can be considered as particles 
 that can be annihilated and created in pairs, see Figs.~\ref{Fig1RPS} and \ref{phasesCGLE}. 
A satisfactory description of these coherent structures can be obtained in terms
of equations~(\ref{RPSpde}) and from the  complex Ginzburg--Landau equation associated to them. The latter is
naturally derived within the metapopulation formulation
of Refs.~\cite{SMR,FigshareMovies,SMR2,cycl-rev,BS,MRS16} consisting of an
array of $L\times L$ patches each of which comprises a well-mixed sub-population
of constant size $N$. The composition of each patch changes according to the reactions (\ref{dom-rem})-(\ref{mut})
and individuals can migrate between two  neighboring patches 
${\bm \ell}$ and ${\bm \ell}'$ according to the processes (\ref{exch})-(\ref{hop}). Furthermore, when mutations occur with a rate $\mu>0$,
there are no absorbing states and the coexistence state ${\bm s}^*$ is no longer metastable but a proper reactive steady state.
%Within this metapopulation approach, stochastic effects can be captured by performing a
%size expansion  in the inverse of the carrying capacity $N$ of the underlying master equation~\cite{Gardiner}.
The partial differential equations (\ref{RPSpde}) are obtained at lowest order of a continuum limit $1/N$-expansion of the master
equation~\cite{Gardiner,VanKampen92,PaulEliSid,Tauber14},
 while a Fokker-Planck equation can be obtained to next order~\cite{Lugo08,Butler09,SMR2}.
The comparison with stochastic simulations have shown that equations~(\ref{RPSpde}) accurately capture the properties of the
lattice metapopulation model when $N\gtrsim 20$~\cite{SMR,SMR2,cycl-rev,BS}.
Moreover, the system of  partial differential equations (\ref{RPSpde}) can be well approximated by a complex Ginzburg Landau 
equation~\cite{CGLE} derived  by  performing a multiscale expansion about
the ensuing Hopf bifurcation~\cite{SMR,SMR2,BS}. Such an expansion is performed in terms of
 the `slow' variables $({\bm X}, T)=(\epsilon{\bm x},~\epsilon^2 t)$,
where \mbox{$\epsilon=\sqrt{3(\mu_H-\mu)}$} is the small parameter~\cite{Miller}.
The model's  complex Ginzburg Landau equation for the modulated amplitude $\mathcal{A}({\bm X}, T)$, which is a linear combination
of the rescaled species
densities~\cite{SMR,SMR2,cycl-rev}, thus reads:
	\begin{equation}
	\label{CGLE}
		\partial_{T} \mathcal{A} =
		D \Delta_{\bm X} \mathcal{A} + \mathcal{A} - (1 + i c) |\mathcal{A}|^2 \mathcal{A},
	\end{equation}
where  $\Delta_{\bm X}= \partial_{X_1}^2+\partial_{X_2}^2=\epsilon^{-2}(\partial_{x_1}^2+\partial_{x_2}^2)$ and
$\partial_{T}=\epsilon^{-2}\partial_{t}$, and after having rescaled $\mathcal{A}$ by a constant, the
parameters are
	\begin{equation}
	\label{c3}
		c = \frac
		{12\zeta (6\beta - \sigma)(\sigma + \zeta) + \sigma^2 (24\beta - \sigma)}
		{3\sqrt{3} \sigma (6\beta + \sigma)(\sigma + 2\zeta)} \quad \text{and} \quad
		D= \frac{3\beta D_E +\sigma D_D}{3\beta +\sigma}
	\end{equation}
equation~(\ref{CGLE}) is thus a {\it controlled approximation}  of the   partial differential equations
(\ref{RPSpde}) about the Hopf bifurcation that, in turn, provides a
reliable description of the metapopulation stochastic model when $N\gg 1$. The
complex Ginzburg Landau equation (\ref{CGLE}) thus allows us to accurately characterize the spatio-temporal spiraling patterns
when $\epsilon\ll 1$ ({\em i.e.}, $\mu\lesssim \mu_H$) and $N\gg 1$
by using
the phase diagram of the two-dimensional complex Ginzburg Landau equation, see, {\em e.g.},~\cite{CGLE}. While this approach
is valid when $\epsilon \ll 1$, it also provides significant insight into the system's spatio--temporal properties away from the Hopf bifurcation 
(when $\mu\ll \mu_H$)~\cite{SMR,SMR2,cycl-rev,BS}:
\begin{itemize}
\item For $\mu\lesssim \mu_H$ (close to Hopf bifurcation)~\cite{SMR,SMR2,cycl-rev}: Migration yields linear diffusion, 
which does not alter the
stability of the patterns, but $D\to \alpha D \;(\alpha>0)$ only rescales the space and the spiral wavelength
as $\lambda \to \lambda/\sqrt{\alpha}$.
As shown in the diagram of Figure~\ref{phasesCGLE} (top), there are four phases separated by the three critical values
$(c_{{\rm AI}}, c_{{\rm EI}}, c_{{\rm BS}})\approx (1.75, 1.25, 0.845)$, see Movies at~\cite{FigshareMovies}: No spiral waves can be sustained in the
 `absolute instability (AI) phase' ($c>c_{{\rm AI}}\approx 1.75$); spiral waves are  convectively unstable
in the Eckhaus instability (EI) phase with $c_{{\rm EI}}\approx 1.25<c<c_{{\rm AI}}$; stable spiral waves
 are found in
the bound state (BS) phase ($c_{{\rm BS}}\approx 0.845<c<c_{{\rm EI}}$); while 
the vortices corresponding to 
spiral waves rotating clockwise and anticlockwise can be considered as particles and 
antiparticles that  annihilate in pairs when they collide in the spiral annihilation (SA) phase
 when $0<c<c_{{\rm BS}}$.

\item For $\mu\ll \mu_H$ (far from Hopf bifurcation)~\cite{SMR2,cycl-rev,MRS16,Alastair16}:
Away from the Hopf bifurcation, the AI, EI and BS phases are still present and their boundaries appear to be essentially
the same as in the vicinity of the bifurcation~\cite{SMR2,BS,MRS16}. However, at low mutation rate, there
is no spiral annihilation and the SA phase is replaced by an extended BS phase, see Fig.~\ref{phasesCGLE} (bottom)
with  far-field breakup of the spiral waves when $\sigma \gg \zeta$~\cite{Jiang11,SMR2}.
This analysis confirms that the dynamics does not sustain stable spiral waves
when $\zeta\gg \sigma$, and hence corroborates the fact no stable spiraling patterns have been found 
in the two-dimensional CLVM (with $\zeta>0$ and $\sigma=\mu=0$)~\cite{Matti,He10}.
When $\mu\ll \mu_H$, non-linear diffusion matters and affects the stability of the spiral waves. In particular,
it was shown that when the hopping rate is increased with all the other rates kept constant, a far-field breakup of the spirals
occurs when $\delta_D\gg \delta_E$~\cite{SMR,SMR2,BS}.
\end{itemize}

The  complex Ginzburg Landau equation (\ref{CGLE}) permits an accurate prediction of the wavelength $\lambda$ 
of the spiraling patterns in the BS and EI phases near the Hopf bifurcation where $\mu \lesssim \mu_H$ 
($\epsilon \ll 1$)~\cite{SMR,SMR2,BS,MRS16}, and to infer an estimate of the wavelength when $\mu< \mu_H$ ($\epsilon = {\cal O}(1)$). 
It has indeed been shown that $\lambda$ decreases linearly with the mutation rate when $\mu<\mu_H$ is 
lowered~\cite{SMR2,MRS16}. Knowing the functional dependence of  $\lambda$, it has been possible to unravel the
 resolution issues arising when the  patterns predicted by  equations~(\ref{RPSpde}) are compared with those found 
in two-dimensional lattice simulations. This is achieved by determining the range of the diffusion coefficient within which 
the wavelength of the ensuing patterns is neither too small 
(of the order of the lattice-space) nor too large (outfitting the domain) for spiral arms to be observable 
in stochastic simulations on a finite grid~\cite{MRS16}. In Ref.~\cite{Alastair16}, the  characterization and stability of the spiraling patterns arising in the model (\ref{dom-rem})-(\ref{rep}) with linear diffusion
and no mutations ($\mu=0$) have been obtained directly from equations~(\ref{RPSpde}) with linear diffusion ($D_D= D_E$).
Furthermore, the `bosonic' formulation of the MLM (with $S_i+S_i \to S_i$ reactions, see Secs. 2.4 and 2.5)  
has been considered in Ref.~\cite{Serrao17} where a noisy complex Ginzburg--Landau 
equation that accounts for weak fluctuations near the Hopf bifurcation has been 
derived within the Doi--Peliti path integral formalism. With intrinsic reaction noise thus properly taken into account, 
the mapping to an effectively two-variable stochastic system is however only valid in a rather restricted range
of parameter space; in general three dynamical degrees of freedom are required.

\subsubsection{The dynamics of cyclic dominance in one dimension and on complex networks.}
The discussion in this section has so far focused on the biologically relevant case of  two-dimensional
systems. However, the analysis can, at least in principle,  be readily extended to one and three dimensions where we would
respectively expect traveling and scroll waves instead of spiral waves. In particular, the one-dimensional dynamics of
the CLVM and MLM with migration has recently attracted significant interest.
In Ref.~\cite{He10}, the mean extinction time of the one-dimensional CLVM with hopping and rates
$(\sigma_i, \zeta_i, \beta_i, \mu_i,\delta_E, \delta_D)=(0, \zeta/\delta_D, 0, 0, 0, 1)$
was computed and the power-law dependence $T_{\rm ex}\sim L^{2\gamma}$ with $\gamma\approx 1.5-1.8$ was obtained, while
the dynamics was found to be characterized by coarsening and the formation of growing domains
with a scaling similar to that of Refs.~\cite{Frachebourg96,Frachebourg96b}. The influence of the symmetry of the
pair-exchange ($\delta_D=\delta_E$) and reaction rates on the dynamics of the one-dimensional CVLM has been studied in Ref.~\cite{Venkat10},
while the effect of mutations on the coarsening dynamics and reactive steady state of the one-dimensional CLVM (without migration)
was investigated in Ref.~\cite{Winkler10}.
Interestingly, particularly rich dynamics  has been found for the one-dimensional MLM with pair-exchange, {\em i.e.}, for rates
$(\sigma_i, \zeta_i, \beta_i, \mu_i,\delta_E, \delta_D)=(\sigma, 0, \beta, 0, \delta, \delta)$: On
a one-dimensional array of patches having a very large carrying capacity, extinction was found to occur via
coarsening (rapid extinction),
or to be driven by heteroclinic orbits, or through the formation of  traveling waves~\cite{Rulands13}.
Besides lattice systems, the dynamics of cyclic dominance between three species has also been studied on
 random and complex networks,  which are settings of particular relevance in the context of evolutionary games and
behavioral sciences, see, {\it e.g.}, Refs.~\cite{Szabo07,OnNetworks2,cycl-rev} for reviews.
It would be interesting to investigate whether the theoretical approaches reviewed here, and devised for lattice systems,
could help shed further light on the dynamics of RPS-like systems on complex random topologies, such as small-world networks
on which intriguing oscillating patterns have been found~\cite{Szolnoki04}.

\section{Multiple species competition networks}
\label{sec:mulsp}
As discussed in the previous section, the many studies of the different variants of the rock-paper-scissors game \cite{Frey10}
have revealed the emergence of intriguing phenomena when adding stochastic effects and spatial dependence to the simple cyclic
interaction between three species. However, it is important to note that the cyclic three-species game is a very special situation,
where all the species interact with each other in a symmetric way. It is obvious that not many of the lessons learned for this
special case will be useful when discussing more complex situations (one exception is the fact that spirals will always form
when considering an odd number of species where each species attacks only a single other species;
however, in contrast to the three-species game these spirals are not composed of individuals belonging to a single species, due to the fact
that neutral species partially mix \cite{Roman13,Feng13}). This is especially true when aiming at an understanding
of realistic ecologies, as these ecologies are endowed with complex interaction networks that cannot be captured fully by only
considering symmetric networks. As such it is important to study more complex situations and develop theoretical
approaches that allow to comprehend the dynamics of more general networks than the cyclic three-species case, with the aim
of obtaining a more complete understanding of biodiversity, correlations, and spatio-temporal patterns \cite{May73}. 

In recent years, an increasing number of papers have focused on these more complex situations, ranging from four and more
species with symmetric interactions (see \cite{Szabo07} for a review of some early results)
to general interaction networks with an arbitrary number of species. Whereas the former
situation is now fairly well understood, results for more general cases are still rather scarce.

\subsection{Symmetric networks}
\begin{figure} [ht]
\includegraphics[width=0.28\columnwidth]{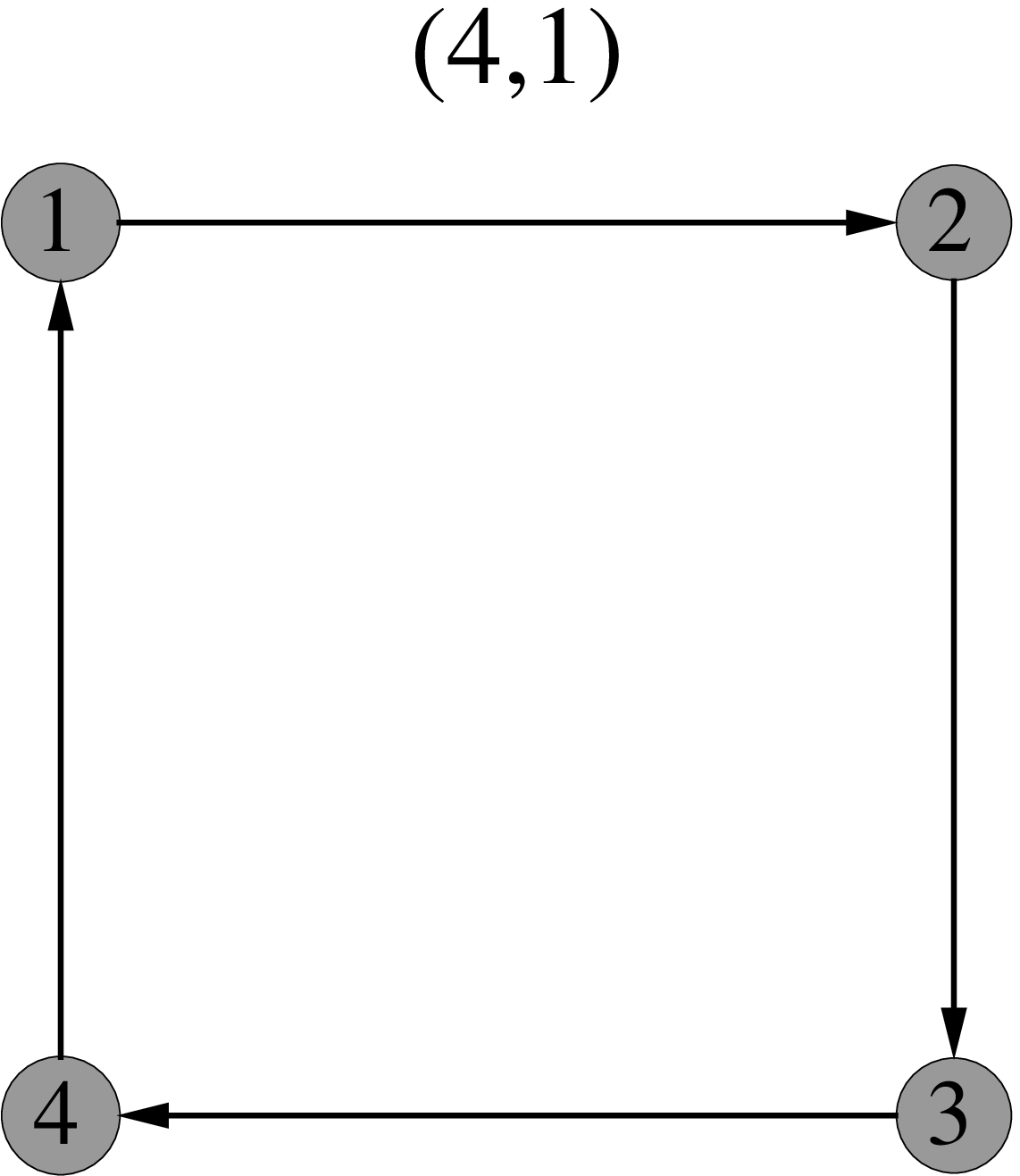}~~
\includegraphics[width=0.28\columnwidth]{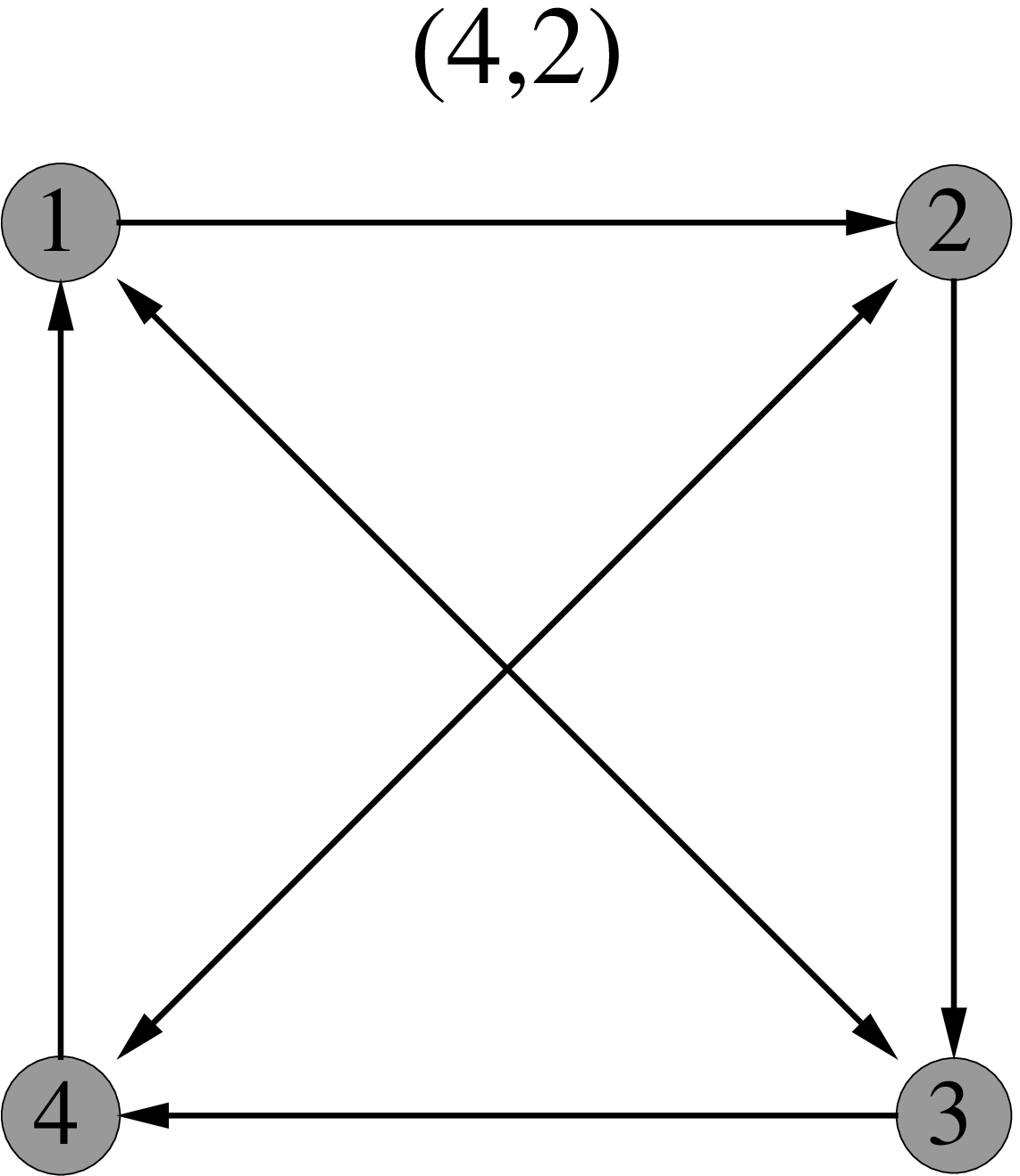}~~
\includegraphics[width=0.28\columnwidth]{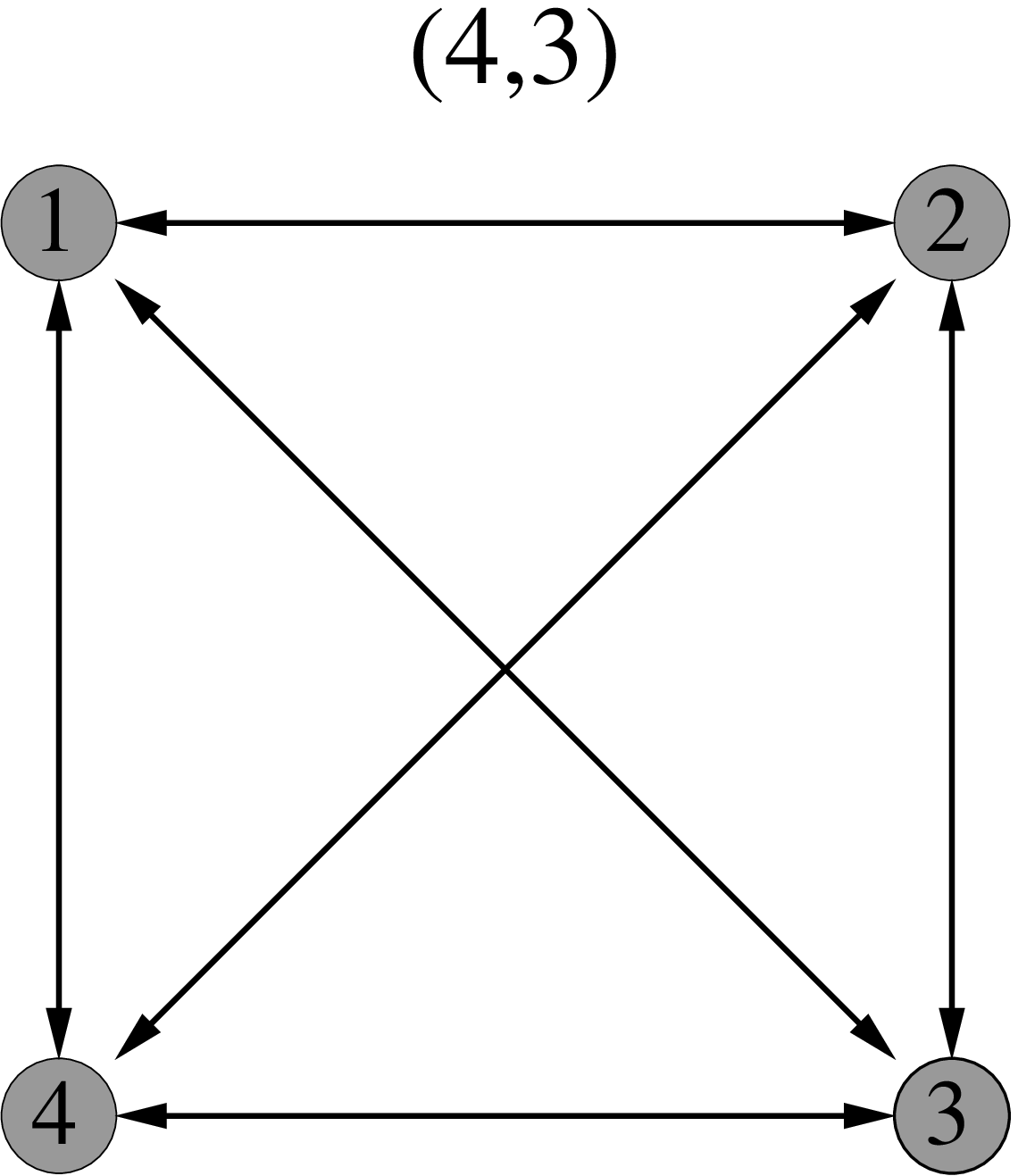}
\caption{The possible symmetric reaction schemes for a system with four species. The label $({\cal N},r)$
indicates the number of species ${\cal N}$ and the number of prey $r$ for every species.\label{fig1_ch4}
}
\end{figure}

When discussing symmetric networks of competing species, it is useful to introduce the following notation \cite{Roman13}.
Consider a system composed of ${\cal N}$ different species. We call model $({\cal N},r)$, $r < {\cal N}$, the model where each of
the ${\cal N}$ species preys on $r$ other species.
This is done in a cyclic way, {\em i.e.}, species $i$ preys on species $i+1$, $i+2$, $\cdots$,
$i+r$ (this has to be understood modulo ${\cal N}$). $({\cal N},1)$ is therefore identical to the cyclic
${\cal N}$-species game where every species preys on a single species and at the
same time is the prey of a different unique predator. On the opposite end $({\cal N},{\cal N}+1)$ describes the
situation where every species attacks every other species.
Figure \ref{fig1_ch4} shows for illustration the different symmetric predation
schemes that are possible for the simple case of four species.

%%%%%%%%%%%%%%%%%%%%%%%%%%%%%%%%%%%%%%%%%%%FIG 1.%%%%%%%%%%%%%%%%%%%%%%%%%%%%%%%%%%%%%%%%%%%%%%%%%%%%%%
%%%%%%%%%%%%%%%%%%%%%%%%%%%%%%%%%%%%%%%%%%%FIG 1.%%%%%%%%%%%%%%%%%%%%%%%%%%%%%%%%%%%%%%%%%%%%%%%%%%%%%%

\subsubsection{The cyclic case with one prey and one predator}
Already the earliest studies of cyclic cases with four or more species, where every species is preying on one other species while
being at the same time the prey of another species, pointed to new effects emerging
when going beyond the simple three-species situation. Frachebourg and co-workers considered cyclic Lotka--Volterra
systems, where immobile individuals belonging to ${\cal N}$ species interact in a cyclic way \cite{Frachebourg96,Frachebourg96b,Frachebourg98}.
The coarsening taking place in that situation yields the segregation into single-species domains. 
At a critical number ${\cal N}_c$ of species fixation sets in, yielding a frozen state composed of neighboring domains with non-interacting
species. This critical number is dependent on the dimensionality of the system, with ${\cal N}_c$ increasing from 5 in one space dimension
to 23 for the three-dimensional lattice. Another early observation concerns the marked differences between situations with an
even or odd number of species \cite{Kobayashi97,Sato02}. This parity law has a huge impact on the properties of a system,
yielding specific space-time patterns and extinction scenarios when the number of species is even.

In a study of the four-species cyclic game on a square lattice, Szab\'{o} and Sznaider \cite{Szabo04} observed the formation of a defensive
alliance where individuals from two non-interacting species (for example, species 1 and 3 for the case (4,1) in figure \ref{fig1_ch4})
mix in order to fight off the other alliance. When allowing for mobility of the individuals through jumping to empty
neighboring sites, a symmetry breaking ordering is encountered above a critical concentration of empty sites which results
in the formation of domains composed by two neutral species. These domains then undergo a coarsening process that stops when one
alliance completely fills the lattice. This formation of defensive alliances is a generic property in systems with an even
number of species that is very robust to modifications of the model \cite{He05,Szabo07b,Szabo08}.

\begin{figure} [ht]
\includegraphics[width=0.34\columnwidth,clip=true]{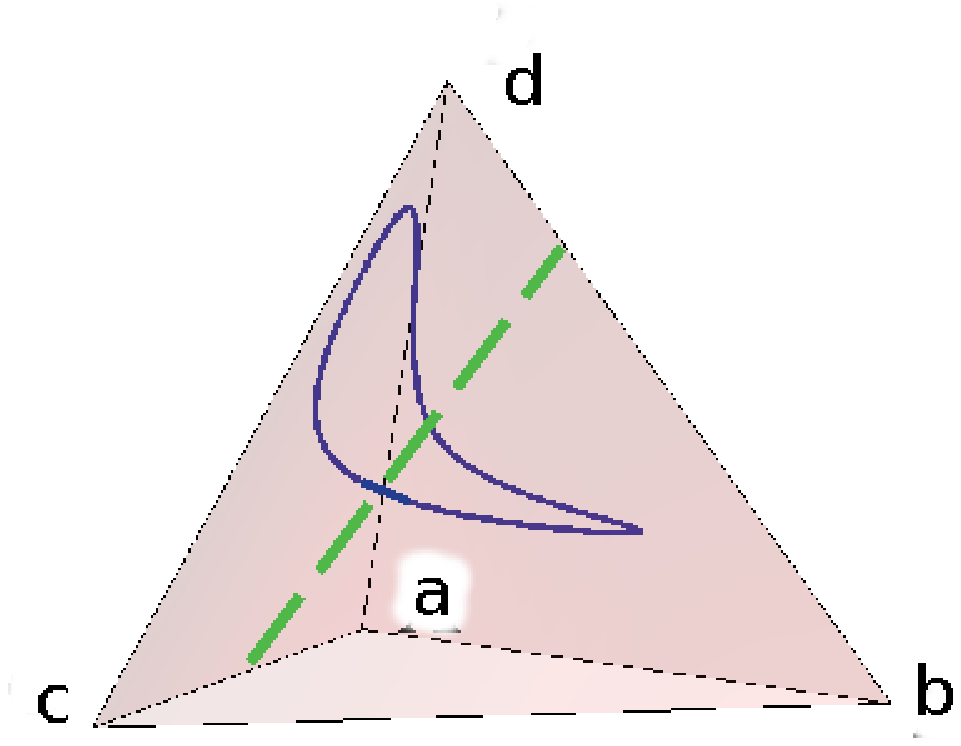}~~~~~~~~~~~~~~~
\includegraphics[width=0.42\columnwidth,clip=true]{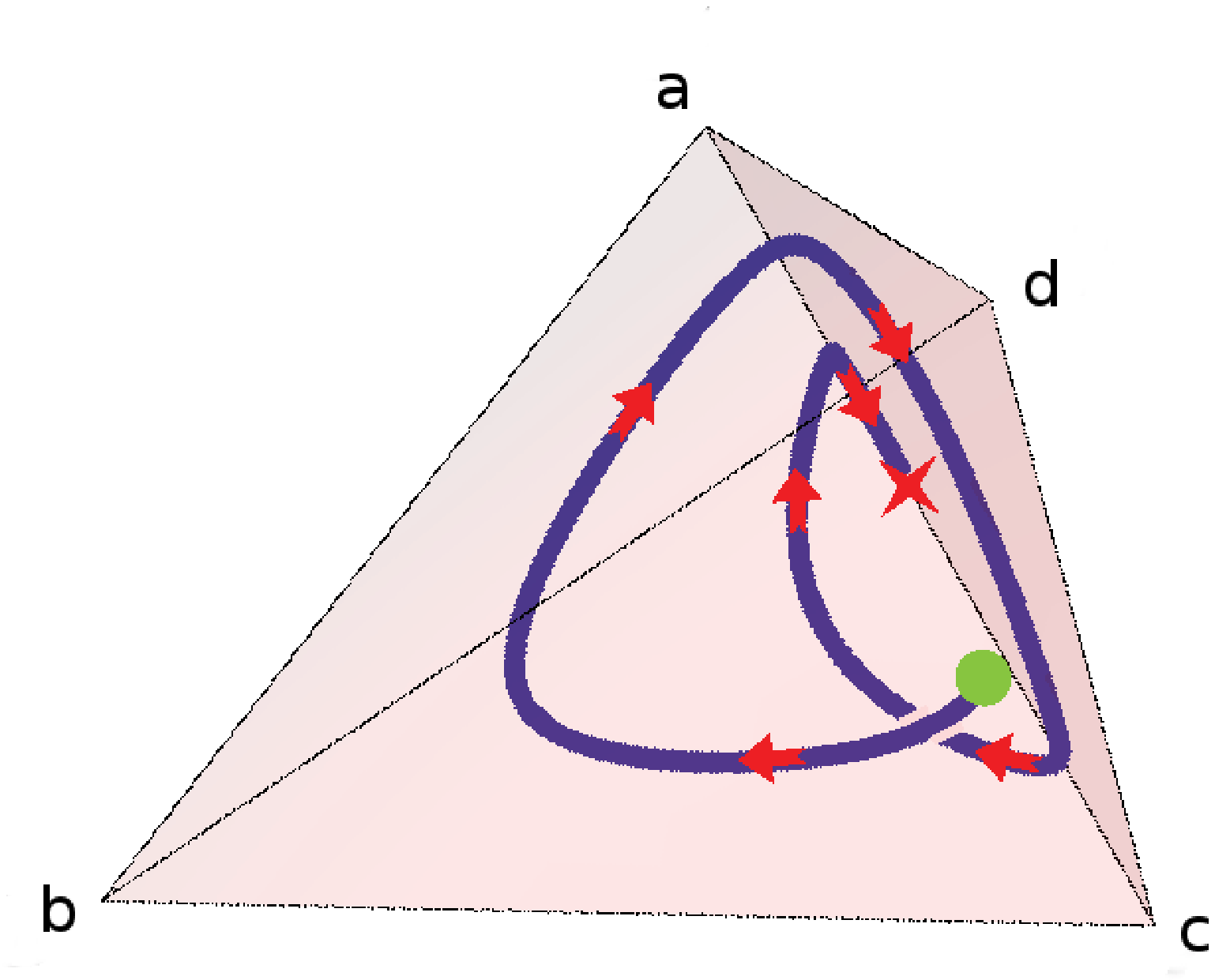}~~
\caption{(Left) Example of a closed loop (solid
curve) in the tetrahedron forming the configuration space for the four species cyclic game,
encircling the line of fixed points (green dashed line). (Right) Typical orbit for $\lambda > 0$ that starts at
the solid green circle and spirals towards an absorbing state,
indicated by the red cross on the a-c edge.
\label{fig2_ch4}
}
\end{figure}

The cyclic four-species case has been the subject of a series of recent papers
\cite{Case10,Durney11,Durney12,Roman12,Dobrinevski12,Intoy13,Guisoni13}
that have yielded a rather complete understanding for this case. One emphasis of these studies
was on the time evolution of the system and the exploration of the surprisingly rich variety of extinction scenarios.
In the well-mixed situation the mean-field rate equations and the correponding deterministic trajectories in configuration
space of the population fractions provide a convenient starting point in order to explore not only the end state but also
the evolution of the system towards this final state. Inspection of the mean-field equations \cite{Case10,Durney11}
\begin{eqnarray}
&& \partial_t a = a \left[k_A b - k_D d \right]~~;~~ \partial_t b = b \left[k_B c - k_A a \right] \nonumber \\
&& \partial_t c = c \left[k_C d - k_B b \right]~~;~~ \partial_t d = d \left[k_D a - k_C c \right]
\end{eqnarray}
where $a$ denotes the average population fraction of species $A$ and $k_A$ the rate at which species $A$ attacks
species $B$, reveals that the final state of the system (coexistence or survival of one of the alliances) is completely
determined by the key control parameter $\lambda = k_A k_C - k_B k_D$.
Indeed the quantity
\begin{equation}
Q = \frac{a^{k_B + k_C} c^{k_D + k_A}}{b^{k_C + k_D}d^{k_A + k_B}}
\end{equation}
evolves in the extremely simple manner $Q(t) = Q(0) e^{\lambda t}$. For $\lambda = 0$, $Q$ becomes an invariant, and
neither pair goes extinct. The system evolves along periodic, closed loops in configuration space that resemble the
rim of a saddle, see figure \ref{fig2_ch4}a. If, on the other hand, $k_A k_C \ne k_B k_D$, then $Q$ decays or grows
exponentially, which means that either $bd$ or $ac$ vanishes in the large time limit. Typically, the trajectory
in configuration space will be a rather complicated open orbit that spirals toward an absorbing state,
see figure \ref{fig2_ch4}b for an example. One can also
find limiting cases where the trajectory is a straight line that connects the initial state with the final, absorbing state.

%%%%%%%%%%%%%%%%%%%%%%%%%%%%%%%%%%%%%%%%%%%FIG 2.%%%%%%%%%%%%%%%%%%%%%%%%%%%%%%%%%%%%%%%%%%%%%%%%%%%%%%
%%%%%%%%%%%%%%%%%%%%%%%%%%%%%%%%%%%%%%%%%%%FIG 2.%%%%%%%%%%%%%%%%%%%%%%%%%%%%%%%%%%%%%%%%%%%%%%%%%%%%%%

Many of these mean-field features survive when considering the stochastic evolution of finite populations \cite{Case10,Durney12}:
the stochastic evolution is found to closely follow the mean-field trajectory, with notable deviations only appearing
in the vicinity of extinction events. For $\lambda$ close to 0, stochastic effects get more and more important, yielding
an increased probability that the system does not end up in the stationary state predicted by mean-field theory.
The probability distribution of the domination time $\tau$ \cite{Intoy13}, {\em i.e.}, the time needed for one alliance to fill the system,
shows for $\lambda=0$ an exponential tail that is a consequence of the fact that the system essentially performs an
unbiased random walk in configuration space. The extinction of two species forming an alliance can be viewed as a
Poisson process described by an exponential distribution.

Stochastic effects play a very important role when studying the cyclic four-species game on a lattice \cite{Roman12,Intoy13,Guisoni13}.
For example, the domination time probability distribution reveals the presence of different routes to extinction \cite{Intoy13}.
In the presence of neutral swappings this probability distribution exhibits for $\lambda = 0$ a crossover between two different exponential
decays. The earlier regime corresponds to extinctions taking place at very early stages of the coarsening process where small
domains contain mainly a single species. The second regime is characterized by very broad tails. These tails result
from very rare extinctions of one of the alliances in extremely long-lived states that are due to a stalemate between domains
where members of one alliance are well mixed. This transition is encountered in one- and two-dimensional lattices as well as
in systems with a fractal dimension.

As already mentioned, mobile individuals in two dimensions may yield a coarsening process where each domain only contains individuals
of one alliance. This is of course reminiscent of the well-studied coarsening process taking place in the two-dimensional
Ising model when quenching a system prepared in a disordered initial state to temperatures below the critical point \cite{Henkel10}.
Indeed, when viewing all individuals within one alliance as belonging to the same `type', we end up with two two kinds of domains,
in complete analogy to the domains in the Ising model that are formed by a majority of up or down spins. This of course neglects
the motion of individuals within domains as well as the preying events that take place at the interface between two domains.
In addition, the Ising model is governed by an energy term that via the Boltzmann factor determines the probability to
go from one configuration to the next, whereas the four-species model is a non-equilibrium model that breaks detailed balance.
As shown in \cite{Roman12}, in both models time-dependent quantities do display the same asymptotic behavior.
For example, the correlation length $L$ extracted from the space-time correlation displays as asymptotic growth regime
a square-root growth $L(t) \sim t^{1/2}$ similar to the Ising model. Interface fluctuations, which can be measured by setting up
a system composed of two halves separated by a straight line, with each of the halves containing only individuals from one of the alliances,
undergo a roughening process characterized by the same roughening exponents as an Ising interface.

Some variations of the basic scheme with four species that have also been studied include situations where the predation rate
is not the same for all species \cite{Szabo08} or is spatially variable \cite{Guisoni13} as well as cases where an
individual changes its character
following a time-dependent probability distribution \cite{Intoy15}.

Many of the results obtained for this simple cyclic four-species case with one prey and one predator remain valid
when considering a larger even number of species with the same interaction scheme \cite{Szabo07b,Durney11,Zia11}. In all cases,
two alliances, each composed of half of the species, are competing against each other, yielding similar phenomena as for the
four-species case, especially when all the predation rates are the same. As shown in \cite{Durney11}, at the mean-field
level all the conclusions reached for the four-species game with arbitrary values of the rates
are recovered when considering a larger even number of species.

\subsubsection{Other symmetric interaction networks}
\begin{figure} [!ht]
 \begin{center}

~~~~~~(a)~~~~~~~~~~~~~~~~~~~~~~~~~~~~~~~~~~~(b)~~~~~~~~~\\
\includegraphics[width=0.35\columnwidth]{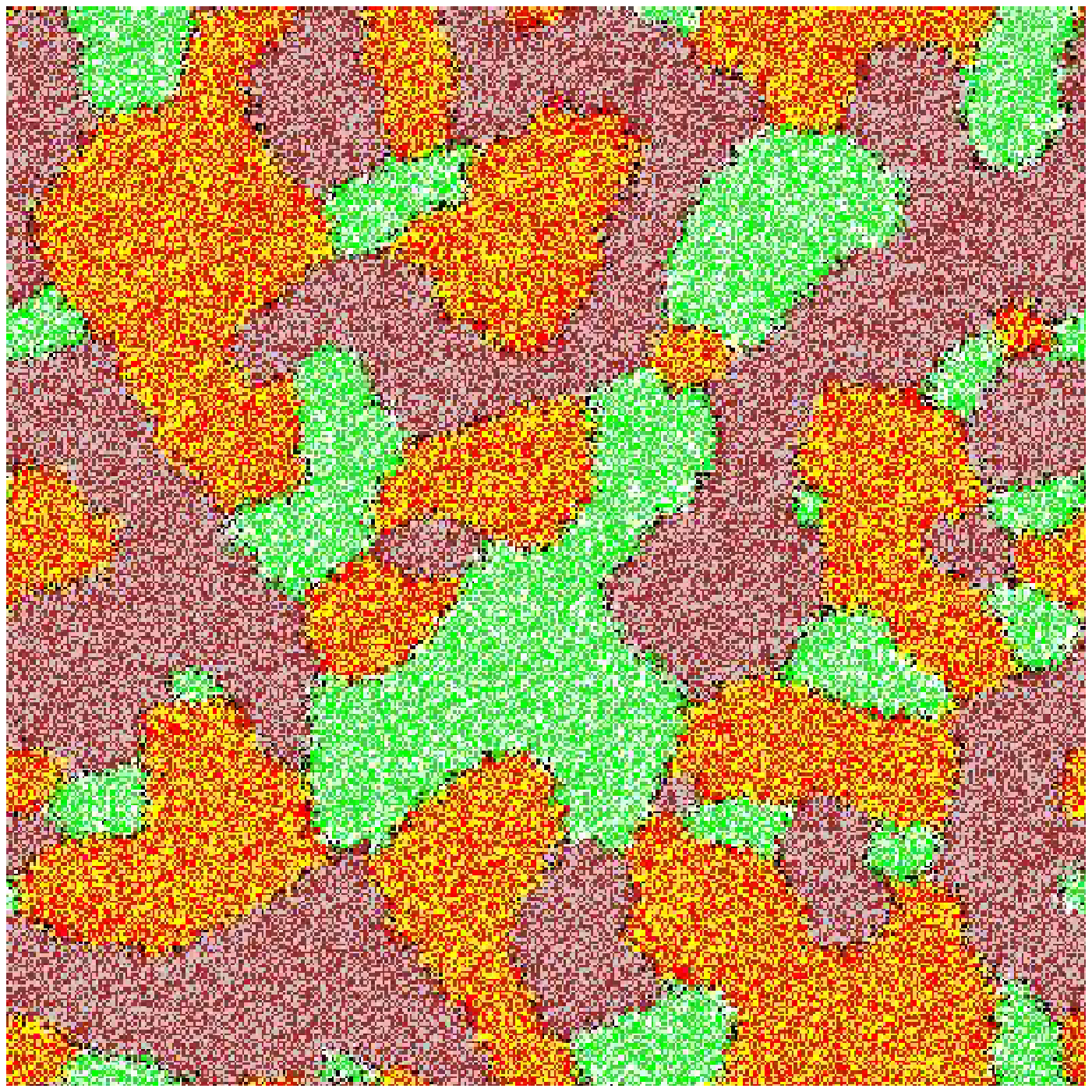}~~~
\includegraphics[width=0.35\columnwidth]{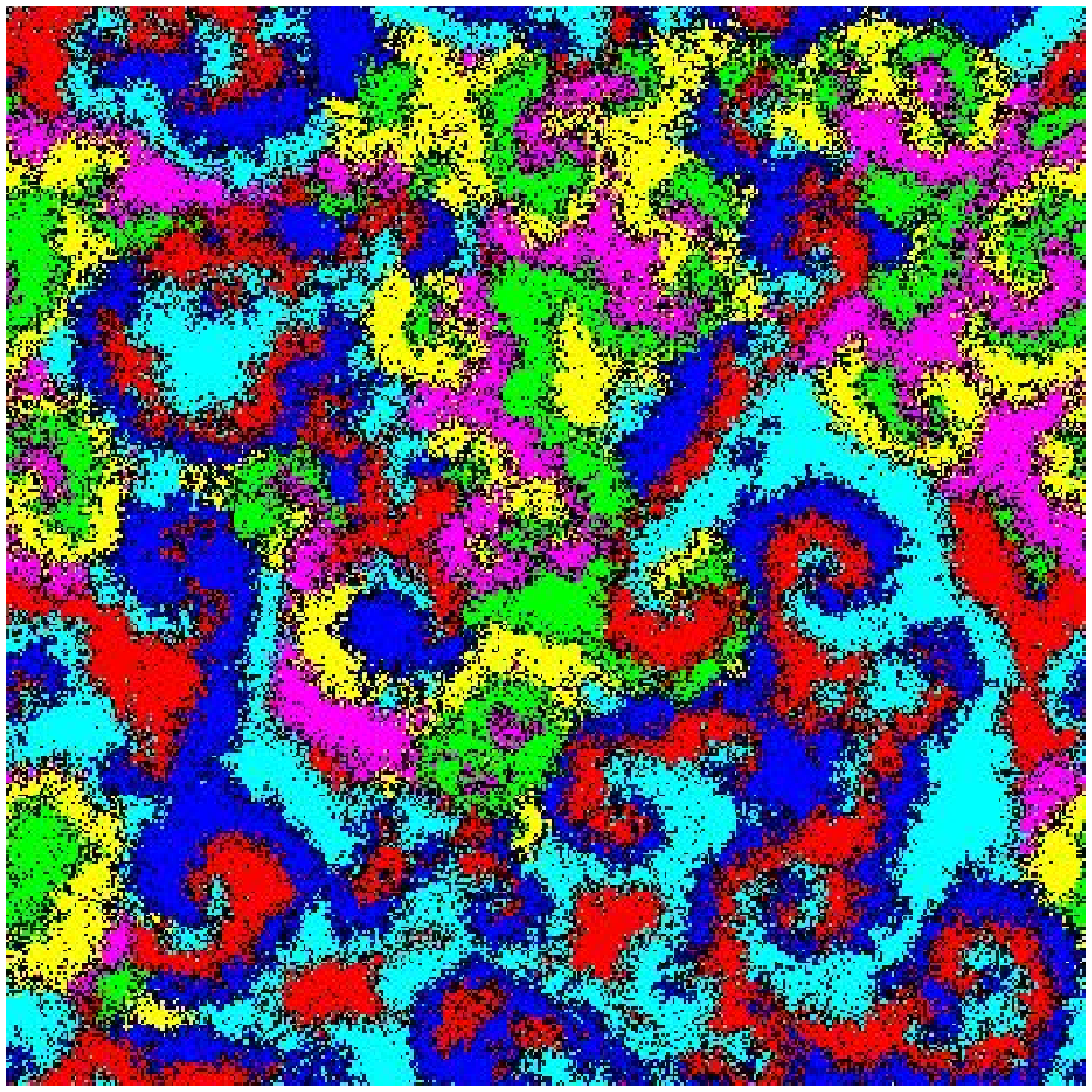}~~~\\[0.2cm]
~~~~~~(c)~~~~~~~~~~~~~~~~~~~~~~~~~~~~~~~~~~~(d)~~~~~~~~~\\
\includegraphics[width=0.35\columnwidth]{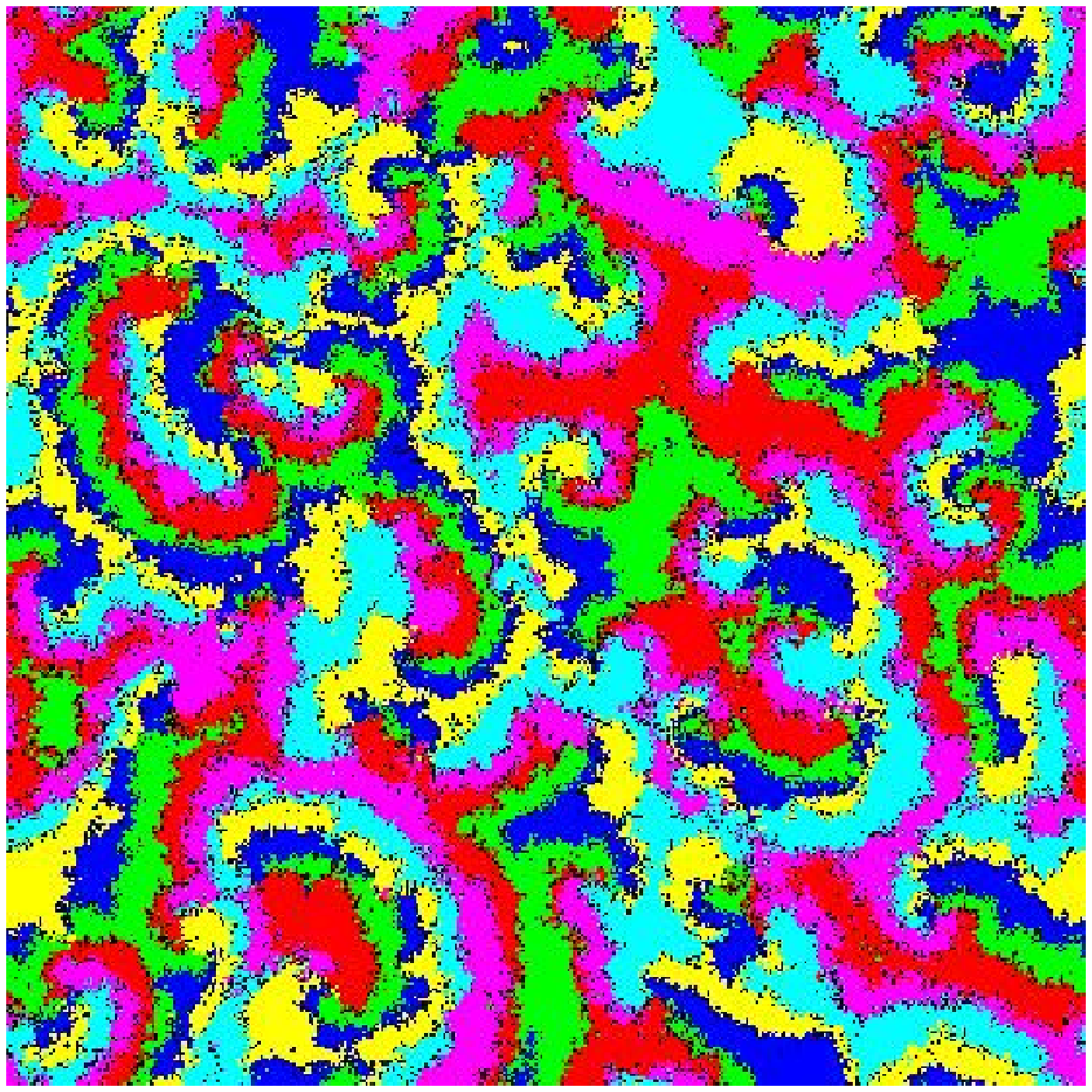}~~~
\includegraphics[width=0.35\columnwidth]{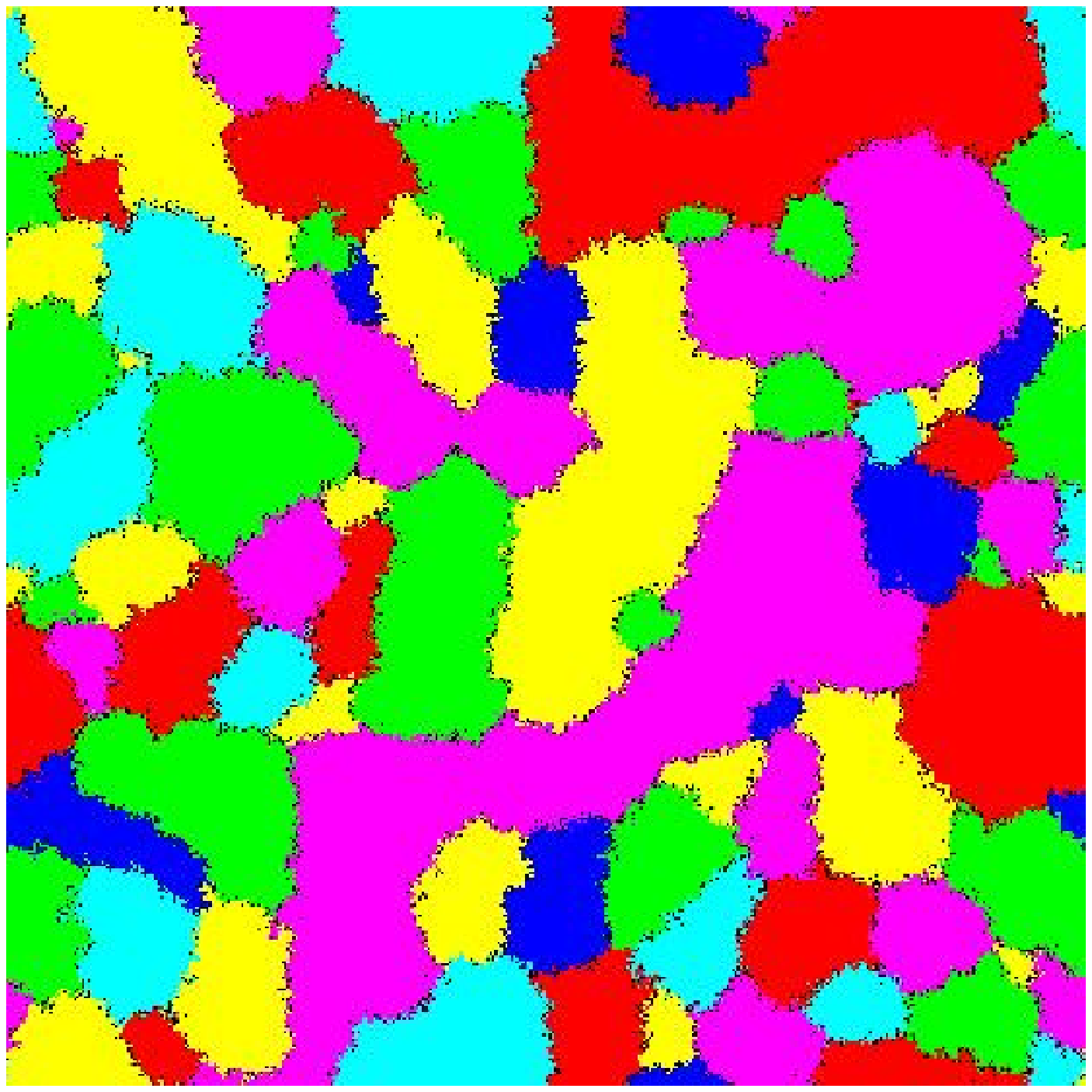}~~~
 \end{center}
\caption{(Color online) Space-time patterns emerging when changing the number of prey in the case of six species:
(a) coarsening of three different types of domains for (6,2), (b) two coarsening domain types with a non-trivial rock-paper-scissors
game within the domains for (6,3), (c) spirals and propagating wave fronts for (6,4), (d) segregation and formation of pure domains
for (6,5). All rates have been chosen equal to 1. For (a) different colors than for the other three cases were chosen for clarity.
\label{fig3_ch4}
}
\end{figure}

A straightforward way to generalize the simple cyclic case with a unique prey for each species is to consider instead $r > 1$ prey
which yields the already mentioned $({\cal N},r)$ models when done in a cyclic way \cite{Roman13,Mowlaei14,Roman16}. Increasing the number
of prey yields a rich variety of different space-time patterns as illustrated in figure \ref{fig3_ch4} for the six-species case.
Whereas for (6,1) two teams composed each of three non-interacting species yield the already described coarsening process with two
types of domains, the (6,2) model results in a coarsening process with three different types of domains, each domain being formed by
two mutually neutral species, see figure \ref{fig3_ch4}a. The case (6,3) shown in figure \ref{fig3_ch4}b is a very interesting one: within each
of the two types of domains three species undergo a cyclic rock-paper-scissors game \cite{Brown17a}. The (6,4) game in figure \ref{fig3_ch4}c is characterized
by spirals and propagating wave fronts. Finally, for (6,5) every species attacks every other species which results in segregation and
the formation of coarsening clusters that only contain individuals from one species, see figure \ref{fig3_ch4}d.

%%%%%%%%%%%%%%%%%%%%%%%%%%%%%%%%%%%%%%%%%%%FIG 3.%%%%%%%%%%%%%%%%%%%%%%%%%%%%%%%%%%%%%%%%%%%%%%%%%%%%%%
%%%%%%%%%%%%%%%%%%%%%%%%%%%%%%%%%%%%%%%%%%%FIG 3.%%%%%%%%%%%%%%%%%%%%%%%%%%%%%%%%%%%%%%%%%%%%%%%%%%%%%%

These patterns can be predicted by considering the square $\underline{B} = \underline{A}^2$ of the adjacancy matrix $\underline{A}$ \cite{Roman13}.
This matrix $\underline{B}$ contains information about preferred partnership formation. Indeed, when representing the game by a directed
graph like those shown in figure \ref{fig1_ch4}, element $b_{ij}$ then counts the number of paths of length 2 from vertex $i$ to vertex $j$, {\em i.e.},
paths of the form $i \longrightarrow k \longrightarrow j$ where $k$ is not equal to $i$ or $j$. Following the maxim that {\it the enemy of my
enemy is my friend}, species $j$ then has a preference to ally with the species that preys on most of its predators, this preferred ally of $j$
being identified by the condition $\max_i b_{ij}$. Whereas analysis of $\underline{B}$ is enough to fully characterize $({\cal N},r)$ games
with identical rates for all predation events, it can fail in the classification of extinction scenarios when rates are not identical.
In these situations the analysis of the Pfaffian of the interaction matrix has been shown to yield insights into the conditions for coexistence \cite{Knebel13}.

It is remarkable that some of the space-time patterns are characterized by a coarsening process with a non-trivial dynamics inside the
domains. One example can be found in figure~\ref{fig3_ch4}b which shows that for the (6,3) system a rock-paper-scissors game between the three species
of a team is sustained. This emergence of spirals within coarsening dynamics give rise to non-trivial internal dynamics.
This non-trivial dynamics inside the domains affects the coarsening process as well as the properties of the interfaces separating different domains,
yielding sets of exponents that differ markedly from those usually encountered in systems with curvature driven coarsening \cite{Brown17a}.
Whereas this appearance of spirals within coarsening domains
has originally been found for the case with site restriction where a site is occupied by at most one individual \cite{Roman13},
this intriguing space-time pattern is also observed in a “bosonic” implementation without a hard constraint on the occupation number per site \cite{Labavic16}.

The papers \cite{Perc07,Szabo08b} discuss the spatial (6,2) game in presence of predation rates that are not homogeneous. Assuming that
species $i$ replaces species $i+1$ with rate $\alpha$ and species $i+2$ with rate $\gamma$, the increase of mobility yields a transition
between a steady state where a three-species cyclic alliance prevails and a steady state where after the end of the coarsening process
shown in figure \ref{fig3_ch4}a one of the two-species neutral alliances fill the system. If $\gamma \neq \alpha$, then the three neutral alliances
do not undergo the coarsening process shown in figure \ref{fig3_ch4}a, but instead they play a spatial rock-paper-scissors game \cite{Szabo08b}.
This behavior is not predicted by the analysis of the square of the adjacency matrix, thus highlighting that for non-homogeneous rates
the approach of \cite{Roman13} does not allow to reliably predict the fate of an ecology. Introducing one more parameter by allowing for
alliance-specific heterogeneous predation rates, an even more complex behavior is observed, resulting in a non-monotonic dependence
of alliance survival on the difference of alliance-specific predation rates \cite{Perc07}.

A broader class of May--Leonard type systems that contains the $({\cal N},r)$ models as special cases has been studied by Avelino and co-workers in a series of papers
\cite{Avelino12a,Avelino12b,Avelino14a,Avelino14b,Avelino16}. Labeling the ${\cal N}$ different species by $i$ and making the cyclic identification
$i = i + k {\cal N}$, with $k$ an integer, these authors consider both right- and left-handed predation where a species $i$ attacks up to $\alpha_R$ species
to their right along the cycle and up to $\alpha_L$ species to their left along the cycle. $({\cal N},r)$ models are obtained when $\alpha_R = r$ and
$\alpha_L = 0$. In their analysis Avelino {\em et al.} focus on the emerging (interface) string network of empty sites and the corresponding junctions between
these strings. Depending on the number of species and
the chosen interactions, different types of junctions, associated to regions with a high concentration of empty sites, are identified.

A spatial five-species game with two prey and two predators,
where species $i$ replaces species $i+1$ with rate $p_1$ and species $i-2$ with rate $p_2$, was the topic
of some recent papers \cite{Kang13,Vukov13,Cheng14}. This model is an analogy
of the rock-paper-scissors-lizard-Spock game. Interestingly, this model presents a special ratio of the two rates (coined the golden point rule in
\cite{Kang13}) $q = p_1/p_2 = \frac{\sqrt{5}-1}{2}$ where two of the five eigenvalues of the interaction matrix vanish. This results in a zero-frequency
mode whose presence yields a vanishing dominance between any pair of mean-field solutions.
Monte Carlo simulations on the two-dimensional lattice show that this dominance vanishing also holds beyond the mean-field approximation
and yields a divergence of the species density fluctuations \cite{Vukov13}. Interesting results have also been obtained for the ratio $q=1$ where
one encounters the emergence of various local groups of three species each in different spatial regions \cite{Cheng14}. Changing the mobility
results in transitions between  different steady states.

\subsection{General competition and food networks}
An increasing, albeit still small, number of recent papers have focused on non-symmetric interaction networks. One of the dominant research thrust
in this context is the question how biodiversity and extinction scenarios change when going from a fully cyclic (non-transitive)
situation to a hierarchical (transitive) network by adding or modifying directed links. Examples include three-species cycles where one link is
reversed \cite{Daly15}, four-species cycles where three species are engaged in cyclic competition, whereas the fourth species interacts with the other three in
various ways \cite{Lutz13,Rulquin14,Dobrinevski14,Daly15}, a five-species cycle with additional links that yields five different levels of hierarchy \cite{Kang16},
six-species games where the condition of two prey and two predators is imposed in various ways (the symmetric version being the (6,2) game)
\cite{Szabo01b,Szabo05}, as well as nine-species cycles with complex interactions supposed to mimic the biochemical war among bacteria capable
of producing at most two different toxins \cite{Szabo01,Szabo07c}.

Whereas these different studies point to the amazingly rich properties of non-symmetric games,
it is difficult to draw general conclusions from specific case studies. As already mentioned, it was proposed in \cite{Knebel13} to use the
Pfaffian of the interaction matrix in order to understand the conditions for biodiversity of complex systems. This has been successfully applied
to a transitive four-species game as well as to a five-species cycle with heterogeneous predation rates. However, no systematic studies beyond these
specific cases have been published. An interesting path for further progress is provided by Szab\'{o} et al \cite{Szabo17} who show that cyclic
dominance of ${\cal N}$ species can be decomposed into $\left( {\cal N} -1 \right) \left( {\cal N} -2 \right) / 2$ RPS-type independent components.

In \cite{Roman16} Roman {\em et al.} build on the work \cite{Roman13} and propose different matrices, derived from the adjacency matrix, that allow to
fully characterize cases where the predation rates are homogeneous. Of course, in real ecologies the condition of homogeneity of predation rates
is not fulfilled. Still, restricting oneself to this situation already yields important insights in the properties of general food networks,
especially when classifying the possible inter-species relationships. The additional matrices needed for general interaction schemes encompass the {\it alliance}
matrix whose elements provide information on the best possible ally for each species, the {\it prey-allies} and {\it neutral allies} matrices that
distinguish for each species between allies hunted by that species and allies towards which that species has a neutral approach, as well as one
additional matrix that allows to identify neutral intermediaries between different species. As illustrated in \cite{Roman16} these matrices reveal
the full inter-species relationships for the most complex predator-prey systems as well as the possible extinction scenarios.

As a final remark we point out that one can also consider time-dependent rates and/or adjacency matrices in order to mimic various perturbations
to an ecological system, ranging from seasonal changes to the introduction of new species (for example through migration). In \cite{Botta14}
new species are introduced with some probability at empty sites, whereas interactions between the new species and already existing species
are formed randomly.
In \cite{Brown17} different perturbations are applied to the spatial (6,3) game (one perturbation involves changing the interaction scheme to (6,2)
during the coarsening process shown in figure \ref{fig3_ch4}b) and their effects are studied through the analysis of various
time-dependent quantities. These few studies provide some of the possible starting points for a systematic investigation of
the effects perturbations can have on predator-prey systems.

\section{Conclusion and outlook}
\label{sec:concl}

In this topical review, we have focused on stochastic predator-prey population dynamics in
spatially extended systems, and the investigation of dynamical correlations and fluctuations
beyond the realm of the standard mean-field rate equation analysis, which often turns out
inadequate in this context. We have specifically demonstrated how the transfer of both
analytical as well as numerical simulation tools from non-equilibrium statistical physics
has led to a host of unexpected novel and intriguing phenomena in these simple paradigmatic
model systems. These range from persistent population oscillations stabilized by intrinsic
demographic reaction noise and strong correlation-induced renormalizations of the associated
kinetic parameters to the emergence of genuine continuous out-of-equilibrium phase 
transitions, as well as the spontaneous formation of remarkably rich spatio-temporal 
patterns. We have discussed how the availability of spatial degrees of freedom can 
drastically extend extinction times through the emergence of such noise-stabilized 
structures, and hence promote ecological stability and species diversity. We have also
elucidated how spontaneous pattern formation and coarsening kinetics in multi-species 
competition networks can be understood and classified on the basis of mean-field theory.
Detailed investigations over the past two decades of stochastic spatial predator-prey 
dynamics have thus enriched our grasp and characterization of strongly out-of-equilibrium
systems. As a striking example, let us mention the recently uncovered intimate connection
of the directed percolation active-absorbing transition in predator-prey systems with the
long-unresolved problem of the onset of turbulence in shear flows~\cite{Shih15}.

We hope that the fields of ecology, population genetics, and epidemiology will in turn 
benefit from this much improved theoretical understanding of complex stochastic interacting 
and reacting particle models. This is of course not limited to the predator-prey type models
that are the subject of this overview. Non-equilibrium statistical physics has had a 
similarly strong impact on the study of stochastic and spatially extended single-'species' 
ecological systems; an up-to-date brief review is presented in Ref.~\cite{Pigolotti17}. Yet
of course the very simplified idealized models studied here cannot possibly capture the 
full complexity encountered in natural ecosystems. Nevertheless, we believe that the 
distinct physics approach of first isolating fundamental phenomena and identifying basic
quantitative features in reduced paradigmatic models and then re-synthesizing these into 
much more complicated systems should prove fruitful in ecology and population biology as 
well. Experimental verification of the relevance of correlation and fluctuation effects in 
simple artificial ecosystems whose dynamical evolution is fully controllable would naturally
constitute a crucial step towards validation of this assertion. We cannot provide an 
exhaustive list of attempts along this direction here, but merely mention two representative
and quite promising recent developments, namely (i) the construction of predator-prey 
molecular ecosystems in appropriately tailored DNA strands~\cite{Fujii13}; and (ii) the
genetic programming of \textit{E.~coli} bacteria to display various desired ecological 
features~\cite{Datla17}. We trust these and other efforts to construct controlled synthetic
ecosystems in the laboratory will turn out fruitful in the near future and provide major
novel avenues for both experimental and theoretical research in the area of stochastic
spatially extended population dynamics. 

\ack
The authors are indebted to Eli Ben-Naim, Bart Brown, Nicholas Butzin, Sara Case, 
Sheng Chen, Debanjan Dasgupta, Udaya Sree Datla, Clinton Durney, Shadi Esmaeili, Erwin Frey,
Ivan Georgiev, Nigel Goldenfeld, Qian He, Bassel Heiba, Ben Intoy, Luo-Luo Jiang, 
Alan McKane, David Konrad, Darka Labavi\'c, Will Mather, Hildegard Meyer-Ortmanns, 
Shahir Mowlaei, Tim Newman, Matjaz Perc, Tobias Reichenbach, Ahmed Roman, 
Alastair Rucklidge, Shannon Serrao, Max Shafer, Bartosz Szczesny, Attila Szolnoki, 
Sid Venkat, Mark Washenberger, Robert West, and Royce Zia for fruitful collbarations and/or 
insightful discussions. The work by M P on population dynamics is supported by the U S 
National Science Foundation through grants DMR-1205309 and DMR-1606814. M M is grateful for 
the support from the Alexander von Humboldt Foundation (Grant No. GBR/1119205 STP) and the 
hospitality of the LS Frey at the Arnold Sommerfeld Centre, Ludwig Maximilians University 
(LMU) in Munich, where part of this work was done.

\appendix

\end{document}